\documentclass[1p,10pt,times,sortyear&compress]{elsarticle}
\usepackage[T1]{fontenc}
\usepackage{amsbsy}
\usepackage[colorlinks]{hyperref}
\usepackage{amsmath}
\usepackage{mathrsfs}
\usepackage{color}
\usepackage{multirow}
\usepackage{booktabs}
\usepackage{subfig,morefloats}
\usepackage{float}
\journal{International Journal of Engineering Science}
\biboptions{authoryear}

\newcommand{\ii}[0]{\rm{i}}
\newcommand{\mm}[0]{\rm{m}}
\newcommand{\eq}[0]{\rm{eq}}
\newcommand{\pp}[0]{\rm{p}}
\newcommand{\ee}[0]{\rm{e}}
\newcommand{\aff}[0]{\rm{aff}}
\begin{document}
\begin{frontmatter}
\title{
	Anisotropic effect of regular particle distribution in elastic-plastic composites: the modified tangent cluster model and numerical homogenization}
	\author[IPPT]{K. Bieniek}
\author[PJTK]{M. Majewski}
\author[IPPT]{P. Ho{\l}obut}
\author[IPPT]{K. Kowalczyk-Gajewska\corref{mycorrespondingauthor}}
\ead{kkowalcz@ippt.pan.pl}
\address[IPPT]{Institute of Fundamental Technological Research, Polish Academy of Sciences,\\
	Pawi\'{n}skiego 5B, 02-106 Warsaw, Poland}
\address[PJTK]{Polish-Japanese Academy of Information Technology, Koszykowa 86, 02-008 Warsaw, Poland}
\cortext[mycorrespondingauthor]{Corresponding author, fax: +4822 8269815}
\begin{abstract}
{Estimation of macroscopic properties of heterogeneous materials has always posed significant problems.} Procedures based on numerical homogenization, although very flexible, consume a lot of time and computing power. Thus, many attempts have been made to develop analytical models that could provide robust and computationally efficient tools for this purpose. 
The goal of this paper is to develop a reliable analytical approach to finding the effective elastic-plastic response of metal matrix composites (MMC) and porous metals (PM) with a predefined particle or void distribution, as well as to examine the anisotropy induced by regular inhomogeneity arrangements. The proposed framework is based on the idea of Molinari \& El Mouden (1996) to improve classical mean-field models of thermoelastic media by taking into account the interactions between each pair of inhomogeneities within the material volume, known as a cluster model.
Both elastic and elasto-plastic regimes are examined. A new extension of the original formulation, aimed to account for the non-linear plastic regime, is performed with the use of the modified tangent linearization of the metal matrix constitutive law. The model uses the second stress moment to track the accumulated plastic strain in the matrix. In the examples, arrangements of spherical inhomogeneities in three Bravais lattices of cubic symmetry (Regular Cubic, Body-Centred Cubic and Face-Centred Cubic) are considered for two basic material scenarios: ``hard-in-soft'' (MMC) and ``soft-in-hard'' (PM). 
As a means of verification, the results of micromechanical mean-field modelling are compared with those of numerical homogenization performed using the {Finite Element Method (FEM). In the elastic regime, a comparison is also made with several other micromechanical models dedicated to periodic composites.} Within both regimes, the results obtained by the cluster model are qualitatively and quantitatively consistent with FEM calculations, especially for volume fractions of inclusions up to ~40\%.
\end{abstract}
\end{frontmatter}

\section{Introduction\label{Sec:Intro}}
Heterogeneity has always been a complicated topic in science and engineering, compared with the much simpler treatment of homogeneous materials. Steel, concrete and aluminium, for example, are widely used in many industries. Even though their real microstructures are heterogeneous, in the fields such as civil engineering (including official normatives and guidebooks) these materials are generally treated as homogeneous, which massively simplifies the process of design and makes it easier and more time-effective.
However, with the wider use of composites, polycrystalline and porous materials, with the possibilities of tailoring and designing microstructures opened by additive manufacturing, as well as with the demand for more accurate assessments of the parameters of materials, much more sophisticated theories and models were proposed in the last decades, dedicated to the description of the influence of material heterogeneity. 

While the use of numerical homogenization, in which the actual microstructure of a material can be directly subjected to mechanical analysis, seems to be the obvious choice at the present stage of development of computer technology, such a `brute force'  approach is still relatively resource-inefficient and time-consuming, which is the reason for the continuing development of analytical solutions, such as the mean-field models. The use of quick analytical computational tools allows for testing many scenarios (e.g. modifying parameters of phases, spatial distribution of inclusions or their volume fraction), understanding of the leading trends in the microstructure-property relationship and finding optimal solutions depending on the foreseen applications of a given composite. Thanks to short calculation times and ease of automation, many different scenarios may be tested efficiently. Examples are e.g. finding the lowest weight of an element along with a sufficient level of stiffness, or choosing the best arrangement of inclusions for the highest level of thermal insulation. More complex approaches may also include applying an external load while overseeing stresses in different phases or include inclusion failure.  

Many different micromechanical mean-field models aim to describe the connection between microstructure of a composite material and its macroscopic properties. There are various existing micromechanical models, many of them based on the Eshelby solution. Eshelby's problem assumes a single ellipsoidal inclusion in an infinite \emph{linearly elastic} medium, subjected to a homogeneous external load located at infinity. Among mean-field models originating from this fundamental solution, the Mori-Tanaka method is most widely used for composite materials. In this model, the medium material is the matrix material, while the far-field quantities are replaced with the average fields in the matrix phase. The model assumes that a single inclusion interacts only with the surrounding matrix while direct interactions between inclusions are neglected, which limits the applicability of the model to relatively small volume fractions. The other possibilities are the self-consistent and generalized self-consistent models, or differential and incremental schemes, cf. \citep{NematNasser99}. 

Unfortunately, the above-mentioned classical models fail to describe properly the scale effects related to the inclusion size as well as their spatial distribution within the sample, which may be an important source of a material's anisotropy even if it is composed of isotropic phases. As an answer to these issues, more sophisticated models were proposed.
\cite{Vilchevskaya21} distinguished two classes of approaches to dealing with the effect of regular particle distribution: i. homogenization models based on the one-particle approximation; and ii. analytical or semi-analytical solutions for periodic arrangements of identical inhomogeneities. 
As concerns the second class, the effect of regular particle distribution was first studied in the context of thermal conductivity. One of the very first contributions was given by \cite{ray92} and concerns a regular cubic lattice of spheres. Subsequently the issue was studied by  \cite{mc78,mcm78} and  \cite{sac82} who derived the overall thermal conductivity of a periodic array of spheres. \cite{lu99} and \cite{mercier00}  additionally analyzed the effect of the aspect ratio of the unit cell.
{As far as elastic properties are concerned, equivalent estimates of the effective properties of composite materials with regular inclusion arrangements were found analytically using different methods by \cite{NematNasser82}, \cite{Nunan84}, \cite{SANGANI19871}, \cite{Rodin93}, \cite{Kushch97}, \cite{Cohen04}, \cite{schj05} and \cite{Kushch11}.} 

Within the first class of methods, variational estimates accounting for the spatial distribution of inclusions were formulated by \cite{PonteCastaneda95}. The authors provided Hashin-Shtrikman-type estimates which, using two-point correlation functions, enable one to independently account for the shape and spatial distribution of inclusions. The results were restricted to the spatial correlations of an `ellipsoidal' form. \cite{HuWeng00} demonstrated that the proposed method delivers results similar to the ones obtained by the double inclusion model by \cite{HoriNematNasser93} when the outer shape of the double inclusion is assumed to carry information about the inhomogeneity arrangements. A similar concept was recently discussed by \cite{Kanaun08} who introduced the so-called `spatial correlation zone' surrounding each inhomogeneity. Additionally, \cite{Sevostianov14} and \cite{Sevostianov19} demonstrated that the \cite{PonteCastaneda95} approach coincides with the Maxwell model in which the `effective inclusion' is identified with the correlation zone. In all the above approaches the shape of this zone must be ellipsoidal, which limits the anisotropy of the overall properties, that is induced solely by the particles' spatial distribution, to three symmetry classes corresponding to the symmetry group of the second order tensor describing the shape of the effective ellipsoid: orthotropy, transverse isotropy and isotropy. Thus, for example, the cubic symmetry of elastic properties, which is induced by a regular array of spherical particles, cannot be accounted for by such approaches. This may also lead to the limited capability of those models to describe periodic composites, as concluded by \cite{Vilchevskaya21}.
The packing effect (but not the spatial distribution effect) can also be accounted for by the so-called morphologically representative pattern (MRP) approach developed in \citep{Bornert96,Marcadon07} and verified through comparison with numerical homogenization in \citep{Majewski17} and \citep{Majewski20} for elastic and elastic-plastic composites, respectively. An attempt at experimental validation of the effect was presented in \citep{Kowalczyk24} where samples with regular pores were produced using a 3D printing technique and tested in the regime of small strain to assess the elastic stiffness.  A more extensive review of the mentioned papers can be found in \citep{Kowalczyk21} or \citep{Vilchevskaya21}.

In this paper we selected and developed a different approach, originally proposed
in \citep{MolinariElMouden96} -- the cluster interaction model, which includes the effect of morphological and spatial distribution of particles.  The proposal was based on the Lippman-Schwinger-Dyson equation derived in the context of heterogeneous elasticity. The main idea was to directly take into account the interaction between inclusions. The model was later extended to coated inclusions
\citep{ElMoudenCherkaoui98} and thermoelasticity  \citep{ElMoudenMolinari00} (assuming a uniform temperature in the medium). It is worth observing that the cluster model was adopted to derive the thermal conductivity of a composite with coated inclusions in \citep{mercier00} referenced above. Recently, an extension of the approach to viscoplastic and elastic-viscoplastic composites was formulated by combining the scheme with the tangent additive law \citep{Kowalczyk21}. The proposal was validated by finite element calculations in the linear viscosity case. It can be noted here that \cite{ma04} also incorporated direct interactions between inclusions, however, without introducing the concept of the cluster. Using the developed method \cite{ma04} analyzed elastic-plastic composites employing modified secant linearization. Examples considered only random inclusion arrangement without its periodic extension to fill the whole space. 

The focus of this paper is on the rate-independent elastic-plastic composites for which \emph{the constitutive law is non-linear}. Thus, similarly to other classical mean-field models which are based on analytical solutions obtained in the frame of linear elasticity, {some additional steps must be taken to extend the cluster scheme to the elastic-plastic case} \citep{Suquet97}. First, the constitutive law of the matrix has to be effectively \textbf{linearized} and next the selected linearized stiffness, which is in general spatially non-uniform within the matrix, needs to be approximated by some \textbf{uniform} values. There are several possibilities for the linearization and `uniformization' steps available in the literature: the oldest model proposed by \cite{Kroner61}, the classical incremental tangent method due to \cite{Hill65} and the non-incremental secant method of \cite{Tandon88} or its modified version employing the second moments for uniformization proposed by \cite{Suquet95}.   Much less explored in this context is the affine (non-incremental tangent) linearization, which is well developed for non-linear visoplasticity \citep{Molinari87,Molinari04}, however, only conceptually signalled for elastic-plastic materials by \citep{Masson00}. Another related class of approaches, employing algorithmic incremental tangent or secant stiffness and originating from the finite element implementations of the elastic-plastic constitutive model, together with the use of the second moments of stress, were proposed in the series of papers by \citep{Doghri05,DoghriBrassart11,ElghezalDoghri18,NailiDoghri23} or by \cite{SongPeng20}.  There is also a family of variational approaches  which nevertheless bear some connections with the modified secant approach, cf. \citep{PonteCastaneda98,Lahellec13,LucchettaKondo21}. Some of these methods, when combined with the mean-field models neglecting interactions between inclusions, were subjected to numerical verification in \citep{Gonzalez04,Kursa18} and other papers mentioned above. 

It has been observed that most of the mean-field models formulated for elastic-plastic heterogeneous materials fail to predict correctly the plasticity developed in the processes with pure hydrostatic stress or even with high stress triaxialities, in particular for porous materials. It is especially true for the approaches which are solely based on the mean stress or strain per phase (the so-called first moments). An extensive discussion on this issue can be found in \cite{ElghezalDoghri18,NailiDoghri23}. It was demonstrated that the problem can be overcome, at least to some extent, when using the second stress moments in the `uniformization' step. 

Following the problem review above, the goal of this research is two-fold: (i) the extension of the cluster model to the case of rate-independent elastic-plastic composites and its broad verification, and (ii) analysis of the elastic and plastic anisotropy degree stemming from the spatial distribution of inclusions. In regard to the first problem, the main issue which needs to be solved is the selection of the linearization procedure which provides correct predictions of the composite material response for different loading scenarios, in particular, for both deviatoric and hydrostatic processes. 
In this paper we show that the first goal is satisfactorily achieved by the cluster model employing modified tangent linearization. The anisotropy degree assessed by the mean-field cluster approach is qualitatively and quantitatively compared with the results of more accurate analytical models in the elastic regime and those obtained by full-field numerical homogenization for both elastic and elastic-plastic composites. An additional paradigm of the present research is to make the model's formulation as computationally efficient as possible in order to make it a good candidate for implementation with a view to large-scale finite element calculations, cf. \citep{Sadowski17b}, so whenever possible the closed-form relations are derived. To our best knowledge, a mean-field model for elastic-plastic metal matrix composites with comparable predictive capabilities is not available in the existing literature. 

The paper consists of five sections. After the present introductory part, the formulation of the cluster scheme in the elastic regime and its extension to the elastic-plastic case are provided in Section~\ref{Sec:ClMod}. Section~\ref{Sec:Numer} discusses details of the numerical finite element calculations performed to validate the approach. In Section~\ref{Sec:Results} the anisotropy degree, as predicted by the cluster model and numerical homogenization, and resulting from the spatial distribution of inclusions, is analysed in the linear and non-linear regime of material behaviour. The predictions of both approaches are compared for three unit cells of cubic symmetry: RC (simple cubic), BCC (body centered cubic) and FCC (face centered cubic). The paper closes with conclusions. Additionally, two appendices provide the details of the calculation algorithm for the cluster model and discuss mathematical properties of the fourth-order tensor describing the two-particle interaction.

\section{Formulation of the cluster model\label{Sec:ClMod}}
\subsection{Cluster model in the elastic regime}\label{SubSec:ClusterModel}
In the cluster model \citep{MolinariElMouden96}, an elementary cubic volume (a unit cell) containing a finite {number $M$ of inclusions} is created, which is then periodically reproduced to fill the whole space. {This periodically filled space will be denoted by $\mathscr{P}$.} Within the unit cell we allow any spatial layout of inclusions, be it regular or random, provided that it is periodic at the boundaries of the cell, thus allowing the assumed periodic extension over the whole space. To each inclusion of the unit cell we may assign a set of inclusions which are {its periodic images in $\mathscr{P}$} and are supposed to carry the same mechanical fields. This set of inclusions is called \emph{a family}. If all inclusions within the unit cell are symmetrically equivalent like e.g. in the RC, BCC or FCC cubic lattices of inclusions of the same size, then we deal with the one-family case. However, when the inclusions within the cell are not symmetrically equivalent, in which case they may exhibit different mean strains and stresses, like for example in random arrangements, then multiple families must be considered. In this paper we restrict ourselves to the single-family situation, focusing on the overall material anisotropy induced by the spatial distribution of inclusions. The random unit cells are to be analysed in the next paper, in which we would like to concentrate our attention on the stress and strain heterogeneity among different particles within the cell stemming from their location with respect to the neighbouring particles.      

\begin{figure}[ht]
\centering
\begin{tabular}{ccccc}
(a) & (b)\\
\includegraphics[width=7cm]{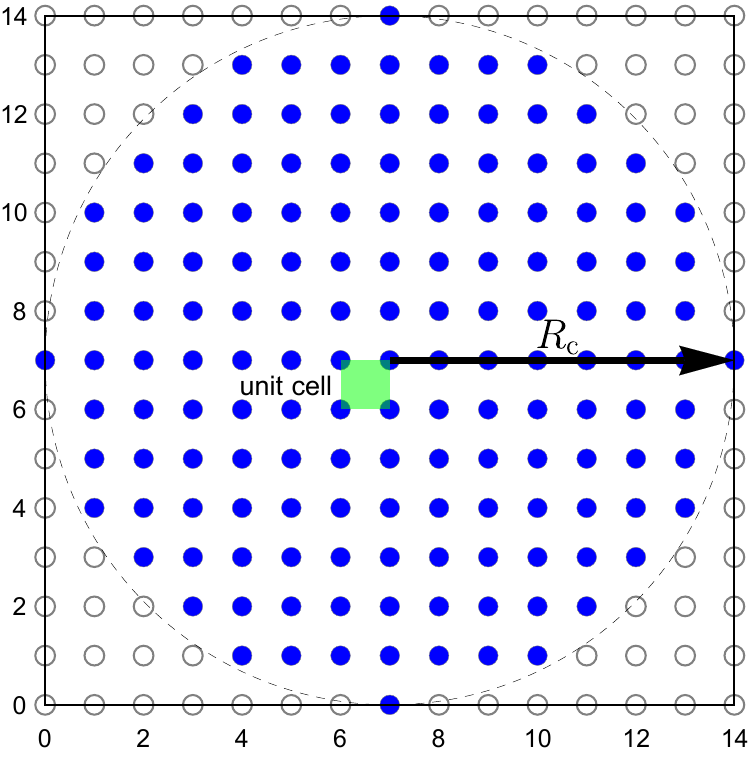}&
\includegraphics[width=7cm]{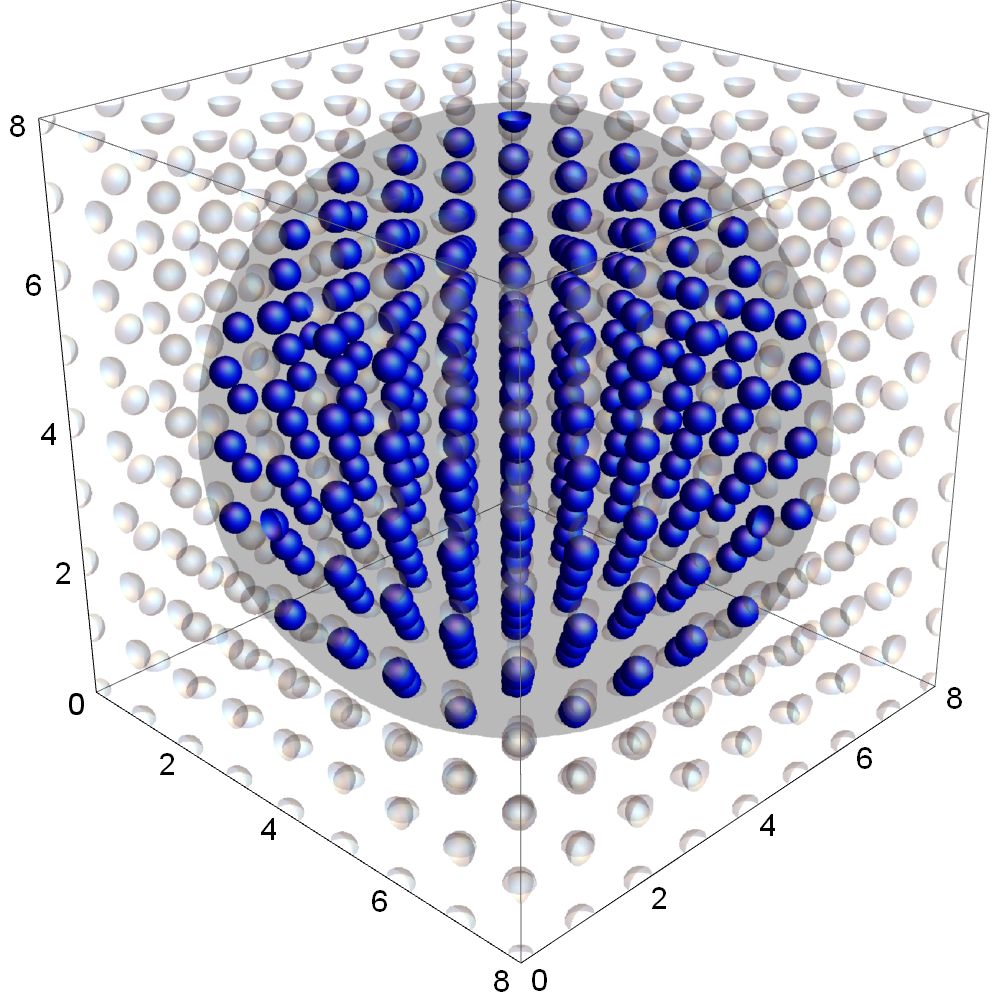}
\end{tabular}
\caption{
(a) {A 2D cluster composed of the inclusions, in blue,  whose centers are on or inside the circle of radius $R_{\rm{c}}$.} 
In this case of the RC ``net'', the unit cell size, shown in green,  is $d=1$ and the cluster radius is {$R_{\rm{c}}=7$}. 
(b) {A 3D cluster of radius $R_{\rm{c}}=4$ and unit cell size $d=1$}.
}
\label{Fig:clusterradius}
\end{figure}

The core idea of \cite{MolinariElMouden96} is to augment classical mean-field models by taking into account interactions between each pair of inclusions within the analyzed space, in addition to interactions between the inclusions and the matrix. Due to the character of the fundamental solution by Green's function technique (see \cite{Berveiller87}), such interactions decay when the distance between inclusions {increases. Therefore, we may restrict the analysis only to pairs of inclusions contained in some material sub-volume, which are called \emph{a cluster}.} In order to establish {a cluster}, a reference inclusion is chosen within the basic unit cell (usually, the centermost one) and then a sphere {$\mathscr{K}$ of a chosen radius is set up in $\mathscr{P}$ around the center of the reference inclusion.} In calculations, only {those inclusions of $\mathscr{P}$ whose centers lie on or inside $\mathscr{K}$ are considered}, as depicted in Fig.~\ref{Fig:clusterradius}(a). {This set of $N$ inclusions is a cluster. Its size is defined as the radius of the smallest $\mathscr{K}$ that determines it. For the purposes of this paper, we shall also define the unit cell size as the distance between consecutive inclusions of the RC ``net'' of the cluster (BCC and FCC contain extra inclusions inside the RC ``net'').}

In brief, the input to the model is the material parameters for each phase, geometry and external (macroscopic) stresses or strains. The results are homogenized macroscopic parameters of the entire sample, such as: elastic stiffness moduli, thermal conductivity, the coefficient of thermal expansion or yield stress, as well as stresses and strains for each phase. These stresses and strains are assumed to be uniform within each phase. {Although only mechanical parameters are considered in this article,} we present below the set of relations valid for thermoelastic composites since they are required for the extension of the model to elasto-plasticity presented in the next subsection. 

The starting point is the constitutive equations for the thermoelastic material, locally relating the stress $\boldsymbol{\sigma}$ and strain $\boldsymbol{\varepsilon}$ in each phase $r$, {which are also valid for the respective mean values $\boldsymbol{\sigma}_r=\left<\boldsymbol{\sigma}\right>_r$ and $\boldsymbol{\varepsilon}_r=\left<\boldsymbol{\varepsilon}\right>_r$ over volume $V_r$ occupied by phase $r$:}
\begin{equation}\label{Eq:tel}
\boldsymbol{\sigma}=\mathbb{C}_r\cdot\boldsymbol{\varepsilon}-\boldsymbol{\beta}_r\theta\rightarrow \boldsymbol{\sigma}_r=\mathbb{C}_r\cdot\boldsymbol{\varepsilon}_r-\boldsymbol{\beta}_r\theta\,,
\end{equation}
where $<.>_r=1/V_r\int_{V_r}(.)\,\rm{d}V$ denotes the operation of volume averaging, $\mathbb{C}_r$ is the fourth-order stiffness tensor for phase $r$, and $\boldsymbol{\beta}_r\theta$ is the thermal stress at temperature $\theta$ uniform in the composite ($\boldsymbol{\beta}_r={\mathbb{C}_r}\cdot\boldsymbol{\alpha}_r$, and $\boldsymbol{\alpha}_r$ is the linear thermal expansion tensor).

As in any mean-field model, the effective stiffness tensor $\bar{\mathbb{C}}$ and effective thermal stress $\bar{\boldsymbol{\beta}}\theta$ for the composite representative volume $V$, such that
\begin{equation}\label{Eq:tot}
\bar{\boldsymbol{\sigma}}=\sum_rf_r\boldsymbol{\sigma}_r=	\bar{\mathbb{C}}\cdot\bar{\boldsymbol{\varepsilon}}-\bar{\boldsymbol{\beta}}\theta\,,\quad
\bar{\boldsymbol{\varepsilon}}=\sum_rf_r\boldsymbol{\varepsilon}_r\,,
\end{equation}
 are found as a weighted sum of respective mean quantities for each phase, where the weights are the relevant strain localization tensors  $\mathbb{A}_r$:
\begin{equation}\label{Eq:C}
\bar{\mathbb{C}}={\sum_r} f_r\mathbb{C}_r\mathbb{A}_r\,,\quad \bar{\boldsymbol{\beta}}={\sum_r} f_r\boldsymbol{\beta}_r\cdot\mathbb{A}_r\,,
\end{equation}
and $f_r$ is the volume fraction of phase $r$. The strain localization relations take a standard form 
\begin{equation}\label{Eq:loc}
\boldsymbol{\varepsilon}_r=\mathbb{A}_r\cdot\bar{\boldsymbol{\varepsilon}}+\mathbf{e}_r\theta\,.
\end{equation}
When the strain localization tensors $\mathbb{A}_r$ and $\mathbf{e}_r$ are known, the set of relations (\ref{Eq:tel}--\ref{Eq:loc}) enables one to find the mean stresses and strains per phase and macroscopic stress or strain, depending on which of these two quantities is prescribed by the boundary conditions.

It has been demonstrated by \cite{Kowalczyk21} that for the single-family case the strain localization tensor and the free term {$\mathbf{e}_r$}  of the cluster model are calculated differently for the inclusions and for the matrix:
\begin{equation}\label{Eq:Ai}
\mathbb{A}_{\ii}=\left(\mathbb{I}+\left((1-f_{\ii})\mathbb{P}_0-\bar{\boldsymbol{\Gamma}}\right)(\mathbb{C}_{\ii}-\mathbb{C}_{\mm})\right)^{-1}\,,\quad \mathbf{e}_{\ii}=\mathbb{A}_{\ii}\left((1-f_{\ii})\mathbb{P}_0-\bar{\boldsymbol{\Gamma}}\right)\cdot\left(\boldsymbol{\beta}_{\ii}-\boldsymbol{\beta}_{\mm}\right)\,,
\end{equation}
\begin{equation}\label{Eq:Am}
\mathbb{A}_{\mm}=\frac{1}{1-f_{\ii}}\left(\mathbb{I}-f_{\ii}\mathbb{A}_{\ii}\right)\,,\quad {\mathbf{e}_{\mm}=-\frac{f_{\ii}}{1-f_{\ii}}\mathbf{e}_{\ii}}\,,
\end{equation}
where $\bar{\boldsymbol{\Gamma}}$ is the interaction tensor,
$\mathbb{P}_0$ is the polarization tensor (specified in \ref{SubSec:AppStepByStep}) and $\mathbb{I}$ is the fourth-order identity tensor. Subscript ``m'' denotes the respective parameters for the matrix, while subscript ``i'' for the inclusions. After substituting Eqs.~(\ref{Eq:Ai}--\ref{Eq:Am}) into Eq.~(\ref{Eq:C}) one obtains
\begin{equation}
\label{Eq:C1F}
\bar{\mathbb{C}}=\mathbb{C}_{\mm}+f_{\ii}\left[(1-f_{\ii})\mathbb{P}_0-\bar{\boldsymbol{\Gamma}}+(\mathbb{C}_{\ii}-\mathbb{C}_{\mm})^{-1}\right]^{-1}\,,\quad
\bar{\boldsymbol{\beta}}=\boldsymbol{\beta}_{\mm}+f_{\ii}(\boldsymbol{\beta}_{\ii}-\boldsymbol{\beta}_{\mm})\cdot\mathbb{A}_{\ii}\,.
\end{equation}

The interaction tensor $\bar{\boldsymbol{\Gamma}}$ is the core of the cluster model. It is a fourth-order tensor that accounts for the interactions between the inclusions and is specified as:  
\begin{equation}
\bar{\boldsymbol{\Gamma}}
=
\sum_{J\neq{I}} \boldsymbol{\Gamma}^{IJ}\,.
\end{equation}
The symmetry group of this tensor results from the symmetry of the unit cell. 
$\boldsymbol{\Gamma}^{IJ}$ is calculated as an interaction tensor between the reference inclusion $I$ and every other inclusion $J$ within the cluster (a separate tensor for each pair $I-J$, see Fig.~\ref{Fig:cluster_interaction}). The procedure of finding effective values is described in detail in \ref{SubSec:AppStepByStep}, as a step-by-step algorithm for the readers' convenience, if they wish to implement the model or test it by themselves. Note that, contrary to the \citep{PonteCastaneda95} approach and related ones, the effect of the spatial distribution is described in the cluster approach by a fourth order tensor (not a second order tensor), so there are eight symmetry classes of anisotropy which could be accounted for within the cluster approach, including the cubic symmetry class \citep{Forte96}. It is worth mentioning then when $\bar{\boldsymbol{\Gamma}}$ is set to zero in the formulas above, the classical Mori-Tanaka model for a two-phase composite is recovered. 

As already mentioned, the cluster model allows one to perform calculations for different regular and irregular arrangements. This provides the flexibility of adjusting inclusion sizes, their locations and sample proportions.
{The} most time-consuming part of the cluster model is the process of calculating the interaction tensor $\bar{\boldsymbol{\Gamma}}$. As it is a fourth-order tensor that needs to be calculated for each inclusion and then summed over all possible pairs, {this can pose a significant efficiency problem in large layouts.} To solve this problem, let us observe that the $\bar{\boldsymbol{\Gamma}}$ tensor depends only on the inclusion size, cell geometry and material parameters of the matrix (see \ref{Sec:PropGammaIJ}). Thus, as long as the inclusions are spherical, their layout remains the same, and the same material for the matrix is used, the $\bar{\boldsymbol{\Gamma}}$ tensor will always be the same. This opens the way to optimizing this most resource-intensive part of the calculations. 

\begin{figure}[ht]
	\centering
	\includegraphics{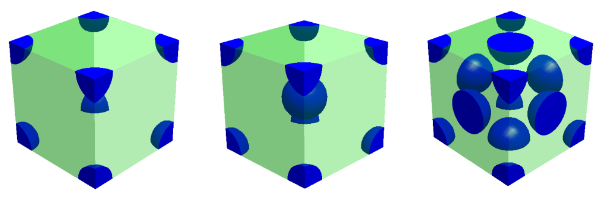}
	\caption{Unit cells for different arrangements: RC (left), BCC (middle) and FCC (right). Note: the size of inclusions is the same in all three cases, but their volume fractions are different.}
	\label{Fig:RC_BCC_FCC}
\end{figure}

In this publication we focus on regular arrangements with cubic symmetry: RC, BCC or FCC (Fig.~\ref{Fig:RC_BCC_FCC}), composed of spherical inclusions. Therefore, we intend to optimize the calculations for these three common regular arrangements.
For this purpose, the Mathematica software was used to automatically perform the summation for a specified set of inclusions. The results obtained for a cluster of radius $R_{\rm{c}}=20d$, which was sufficient to reach the required convergence of the $\bar{\boldsymbol{\Gamma}}$ components, are given by the following formulas (note that $\bar{\boldsymbol{\Gamma}}$ is of cubic symmetry in the considered cases and the components below are given in the frame of material symmetry):

\begin{equation}\label{Eq:GammaClosed}
\bar{\Gamma}_{1111}=
\bar{\Gamma}_{2222}=
\bar{\Gamma}_{3333}=
\bar{\Gamma}_{arrangement}\,,
\end{equation}
\begin{equation}
\bar{\Gamma}_{1122}=
\bar{\Gamma}_{1133}=
\bar{\Gamma}_{2233}=
-\frac{1}{2}\bar{\Gamma}_{arrangement}\,,
\end{equation}
\begin{equation}
\bar{\Gamma}_{1212}=
\bar{\Gamma}_{1313}=
\bar{\Gamma}_{2323}=
-\frac{1}{2}\bar{\Gamma}_{arrangement}\,,
\end{equation}
with $\bar{\Gamma}_{arrangement}$ being described by the following expressions, depending on the sample arrangement (RC, BCC, FCC):
\begin{equation}\label{Eq:XRC}
\bar{\Gamma}_{\rm RC}=
\frac{\left(r/d\right)^{3}}{3G_{\mm}\left(1-\nu_{\mm}\right)}     \left(2.3322-7.4597\left(r/d\right)^{2}\right)\,,
\end{equation}
\begin{equation}\label{Eq:XBCC}
\bar{\Gamma}_{\rm BCC}=
\frac{\left(r/d\right)^{3}}{3G_{\mm}\left(1-\nu_{\mm}\right)}     \left(-1.4414+7.4555\left(r/d\right)^{2}\right)\,,
\end{equation}
\begin{equation}\label{Eq:XFCC}
\bar{\Gamma}_{\rm FCC}=
\frac{\left(r/d\right)^{3}}{3G_{\mm}\left(1-\nu_{\mm}\right)}     \left(-2.5643+18.0617\left(r/d\right)^{2}\right)\,,
\end{equation}
where $r$ is the radius of inclusions, $d$ is the unit cell size, $G_{\mm}$ is the shear modulus of the matrix and $\nu_{\mm}$ is the Poissons's ratio of the matrix. Note that $\bar{\boldsymbol{\Gamma}}$ carries information about the shape and spatial distribution of heterogeneities, but it is independent of the inclusions' material properties, so, in particular, the same expressions are to be used for the MMC and PM cases. The ratio $r/d$ in the above formulas can be alternatively expressed by the volume fraction of inclusions $f_{\ii}$ according to the relation
\begin{equation}\label{Eq:rodfromfin}
\left(\frac{r}{d}\right)^3=\frac{3f_{\ii}}{4\pi n}\,,
\end{equation}
where $n$ is the number of inclusions in the respective unit cell, i.e. $n=1, 2$ and 4 for RC, BCC and FCC, respectively. 

The above closed-form formulas are compact and consume much less calculation time when implemented. When tested and compared with the full-calculation approach, a significant time saving was achieved. This shows that pre-defined functions can serve as an efficient way of calculating effective material parameters for known regular arrangements and speed up the analysis even further, which is essential when implementing the cluster model in large scale FEM computations \citep{Sadowski17b,Sadowski21}.

\subsection{Extension of the cluster model for elastic-plastic composites}
\label{SubSec:Linearisation}
The cluster model presented above was derived in the framework of linear thermoelasticity of the matrix material. So, in particular, {it was assumed in the derivation that} the matrix stiffness tensor $\mathbb{C}_{\mm}$ and the thermal stress tensor $\boldsymbol{\beta}_{\mm}\theta$ were constant within the matrix phase. The goal of this section is to provide a possible extension of the cluster model for elastic-plastic composites. Due to the involved non-linearity of the constitutive law of the metal matrix, the extension of any mean-field model for this case requires additional assumptions. {The general idea is to present the constitutive equations for the matrix in a linearized form, with stress- and strain-dependent parameters, which can be substituted into the cluster model. Then the response of the entire composite is computed by an incremental solution procedure using the cluster model based on the linearized equations of the matrix. This cluster model is updated at each integration step according to the current stress and strain state of the material.} Below, we first present the local model of the matrix phase and next explain and specify all the necessary steps required for the cluster model extension for elasto-plasticity.

Similarly to \cite{Majewski20}, in the analysis we assume {that in the plastic regime the metal matrix is locally described} by the Huber-von Mises yield function with the associated flow rule (the so-called isotropic J2 plasticity) and the power-law isotropic hardening (see Table~\ref{Tab:ElasticPlasticModel} for the respective equations). 

\begin{table}[ht]
	\centering
	\begin{tabular}{lcc}
		\hline
		\multicolumn{3}{c}{Elastic-plastic model of the matrix material} 
		\\ 
		\hline
		\textup{Yield condition} & $f(\boldsymbol{\sigma})=\sigma_{\eq}-Y(\varepsilon_{\eq}^{\pp})\leq
		0$ & $\boldsymbol{\sigma}^{'}=\mathrm{dev}(\boldsymbol{\sigma})$, $\sigma_{\eq}=\sqrt{\frac{3}{2}\boldsymbol{\sigma}^{'}\cdot\boldsymbol{\sigma}^{'}}$
		\\
		\textup{Flow rule} & $\dot{\boldsymbol{\varepsilon}}^{\pp}=\lambda\frac{3\boldsymbol{\sigma}^{'}}{2 Y}$ & $\dot{\boldsymbol{\varepsilon}}^{\pp}=\dot{\boldsymbol{\varepsilon}}-\dot{\boldsymbol{\varepsilon}}^{\ee}$
		\\
		\textup{Consistency conditions} & $\lambda \, f = 0
		\quad \wedge \quad
		\lambda \, \dot{f} = 0$ & 
		\\
		\textup{Hardening law} & $Y \left(\varepsilon^{\pp}_{\eq} \right) =
		Y_0 +
		h \, \left(\varepsilon^{\pp}_{\eq} \right)^n$ & ${\dot{\varepsilon}^{\pp}_{\eq}=\sqrt{\frac{2}{3} {\boldsymbol{\dot{\varepsilon}}}^{\pp} \cdot {\boldsymbol{\dot{\varepsilon}}}^{\pp}} = \lambda}$ 
		\\
		\hline
	\end{tabular}
	\caption{Local constitutive relations for the elastic-plastic metal matrix. $Y_0,n,h$ - matrix material parameters to be specified.} \label{Tab:ElasticPlasticModel}
\end{table}

In the model, the strain and strain rate are additively decomposed into the elastic and plastic parts and the stress or stress rate can be obtained from the linear Hooke's law for the elastic part of the strain or strain rate, namely:
\begin{equation}\label{Eq:elpl}
\dot{\boldsymbol{\sigma}}=\mathbb{C}^{\ee}_{\mm}\cdot(\dot{\boldsymbol{\varepsilon}}-{\dot{\boldsymbol{\varepsilon}}^{\pp}(\boldsymbol{\sigma},\varepsilon^{\pp}_{\eq})})\,,\quad {\boldsymbol{\sigma}}=\mathbb{C}^{\ee}_{\mm}\cdot({\boldsymbol{\varepsilon}}-{\boldsymbol{\varepsilon}}^{\pp}(\boldsymbol{\sigma},\varepsilon^{\pp}_{\eq}))\,.
\end{equation}
Moreover, in the plastic regime ($\lambda(\boldsymbol{\sigma},\varepsilon^{\pp}_{\eq})>0$) a one-to-one relation exists between the equivalent platic strain $\varepsilon_{\eq}^{\pp}$ and the equivalent von Mises stress, namely:
\begin{equation}\label{Eq:seq-eeq}
\sigma_{\eq}=Y_0 +
h \, \left(\varepsilon^{\pp}_{\eq} \right)^n 	\quad\rightarrow\quad
\varepsilon^{\pp}_{\eq}({\sigma_{\eq}})=\left(\frac{\sigma_{\eq}-Y_0}{h}\right)^{1/n}\,.
\end{equation}

The above equations cannot be directly used to find the respective cluster model formula of the first section as the plastic strain evolves in a non-linear way and is spatially non-uniform within the matrix. Thus, some additional steps must be taken to incorporate those equations into the cluster model framework. First, the constitutive law of the matrix, in either of the two forms \eqref{Eq:elpl}, has to be \textbf{linearized} and next the obtained linearized stiffness tensor and thermal-like stress, which are in general spatially non-uniform within the matrix, need to be approximated by some \textbf{uniform} values.

Thus in \textbf{the first linearization step}  the elastic-plastic relations \eqref{Eq:elpl} are written in the thermoelastic-like (affine) \underline{incremental} or \underline{non-incremental} form as follows:
\begin{eqnarray}\label{Eq:affine-inc}
\dot{\boldsymbol{\sigma}}&=&\mathbb{C}_{\mm}^{\rm{ep,aff}}(\boldsymbol{\sigma},\varepsilon^p_{\eq})\cdot\dot{\boldsymbol{\varepsilon}}-{\boldsymbol{\beta}}_{\mm}^{\aff}(\boldsymbol{\sigma},\varepsilon^{\pp}_{\eq})\\\nonumber
\textrm{or}&&\\ \label{Eq:affine} {\boldsymbol{\sigma}}&=&\mathbb{C}_{\mm}^{\rm{ep,aff}}(\boldsymbol{\sigma},\varepsilon^{\pp}_{\eq})\cdot{\boldsymbol{\varepsilon}}-{\boldsymbol{\beta}}_{\mm}^{\aff}(\boldsymbol{\sigma},\varepsilon^{\pp}_{\eq})\,,
\end{eqnarray}
where the definitions of $\mathbb{C}^{\rm{ep,aff}}_{\mm}(\boldsymbol{\sigma},\varepsilon^{\pp}_{\eq})$ and $\boldsymbol{\beta}_{\mm}^{\aff}(\boldsymbol{\sigma},\varepsilon^{\pp}_{\eq})$ depend on the applied linearization method and are not the same for incremental and non-incremental formulations\footnote{The same notation was used for both cases to avoid an unnecessary inflation of symbols.}. In \textbf{the  second `uniformization' step} the quantities $\boldsymbol{\sigma}$ and $\varepsilon^{\pp}_{\eq}$ are approximated by some values, uniform within the matrix material, denoted as $\hat{\boldsymbol{\sigma}}$ and $\hat{\varepsilon}^{\pp}_{\eq}$:
\begin{equation}
\mathbb{C}_{\mm}^{\rm{ep,aff}}(\boldsymbol{\sigma},\varepsilon^{\pp}_{\eq})\approx \mathbb{C}^{\rm{ep,aff}}_{\mm}(\hat{\boldsymbol{\sigma}},\hat{\varepsilon}^{\pp}_{\eq})\,,\quad
{\boldsymbol{\beta}}_{\mm}^{\aff}(\boldsymbol{\sigma},\varepsilon^{\pp}_{\eq})\approx{\boldsymbol{\beta}}_{\mm}^{\aff}(\hat{\boldsymbol{\sigma}},\hat{\varepsilon}^{\pp}_{\eq})\,.
\end{equation}

Additionally, in the case of the cluster model, to make use of the closed-form formulas and preserve computational robustness of the model, in \textbf{the third step} the elastic-plastic stiffness {$\mathbb{C}^{\rm{ep,aff}}_{\mm}(\hat{\boldsymbol{\sigma}},\hat{\varepsilon}^{\pp}_{\eq})$} is \textbf{isotropized}, so that equations \eqref{Eq:affine} are finally approximated as:
\begin{eqnarray}\label{Eq:affine-uni}
\dot{\boldsymbol{\sigma}}&=&\mathbb{C}^{\rm{ep,aff,iso}}_{\mm}(\hat{\boldsymbol{\sigma}},\hat{\varepsilon}^{\pp}_{\eq})\cdot\dot{\boldsymbol{\varepsilon}}-{\boldsymbol{\beta}}_{\mm}^{\aff*}(\hat{\boldsymbol{\sigma}},\hat{\varepsilon}^{\pp}_{\eq})\\ \nonumber
\textrm{or}&&\\ \label{Eq:affine-uni-non}	{\boldsymbol{\sigma}}&=&\mathbb{C}^{\rm{ep,aff,iso}}_{\mm}(\hat{\boldsymbol{\sigma}},\hat{\varepsilon}^{\pp}_{\eq})\cdot{\boldsymbol{\varepsilon}}-{\boldsymbol{\beta}}_{\mm}^{\aff*}(\hat{\boldsymbol{\sigma}},\hat{\varepsilon}^{\pp}_{\eq})\,,
\end{eqnarray}
where
\begin{equation}\label{Eq:affine-uni-iso}
\mathbb{C}^{\rm{ep,aff,iso}}_{\mm}(\hat{\boldsymbol{\sigma}},\hat{\varepsilon}^{\pp}_{\eq})=3K_{\mm}\mathbb{I}^{\rm{P}}+2G^{\rm{ep,aff,iso}}_{\mm}(\hat{\boldsymbol{\sigma}},\hat{\varepsilon}^{\pp}_{\eq})\mathbb{I}^{\rm{D}}\,,
\end{equation}
{${\boldsymbol{\beta}}_{\mm}^{\aff*}(\hat{\boldsymbol{\sigma}},\hat{\varepsilon}^{\pp}_{\eq})$ is the form of $\boldsymbol{\beta}_{\mm}^{\aff}(\hat{\boldsymbol{\sigma}},\hat{\varepsilon}^{\pp}_{\eq})$ after isotropization of $\mathbb{C}^{\rm{ep,aff}}_{\mm}(\hat{\boldsymbol{\sigma}},\hat{\varepsilon}^{\pp}_{\eq})$, $K_{\mm}$ is the constant elastic bulk modulus of the matrix, $G^{\rm{ep,aff,iso}}_{\mm}(\hat{\boldsymbol{\sigma}},\hat{\varepsilon}^{\pp}_{\eq})$ is the isotropic affine shear modulus of the matrix, and $\mathbb{I}^{\rm{P}}$, $\mathbb{I}^{\rm{D}}$ are defined in Eq.~(\ref{eq:IPD}).}

With such an approximation of the local matrix law we may proceed with applying the cluster model with the reservation that $\hat{\boldsymbol{\sigma}}_{\mm}$ and $\hat{\varepsilon}^{\pp}_{\eq}$ evolve as deformation advances so the solution must be obtained step-by-step. Note that the relation between the mean stress or stress rate and mean strain or strain rate in the matrix, resulting from Eq.~\eqref{Eq:affine-uni-iso}, takes the form equivalent to Eq.~\eqref{Eq:tel}$_2$ in Section~\ref{SubSec:ClusterModel}:
\begin{eqnarray}\label{Eq:affine-uni-mean}
\dot{\boldsymbol{\sigma}}_{\mm}&=&\mathbb{C}^{\rm{ep,aff,iso}}_{\mm}(\hat{\boldsymbol{\sigma}},\hat{\varepsilon}^{\pp}_{\eq})\cdot\dot{\boldsymbol{\varepsilon}}_{\mm}-{\boldsymbol{\beta}}_{\mm}^{\aff*}(\hat{\boldsymbol{\sigma}},\hat{\varepsilon}^{\pp}_{\eq})\\ \nonumber
\textrm{or}&&\\ \label{Eq:affine-uni-mean-non}
{\boldsymbol{\sigma}}_{\mm}&=&\mathbb{C}^{\rm{ep,aff,iso}}_{\mm}(\hat{\boldsymbol{\sigma}},\hat{\varepsilon}^{\pp}_{\eq})\cdot{\boldsymbol{\varepsilon}}_{\mm}-{\boldsymbol{\beta}}_{\mm}^{\aff*}(\hat{\boldsymbol{\sigma}},\hat{\varepsilon}^{\pp}_{\eq})\,,
\end{eqnarray}
where the argument $\hat{\boldsymbol{\sigma}}$ is not necessarily equal to ${\boldsymbol{\sigma}}_{\mm}$ and, similarly, the current $\hat{\varepsilon}^{\pp}_{\eq}$ is not necessarily calculated based on mean values of stress and strain tensors (which are also called the first stress and strain moments).

In this paper we focus our attention on three variants of the cluster model extension which employ tangent elastic-plastic moduli for the linearization step: incremental tangent, non-incremental tangent (affine) and modified tangent schemes. 
The tangent elastic-plastic moduli are defined using the rate form of the local relations \eqref{Eq:elpl}. The local flow rule and consistency relation lead to the following definition of the \underline{anisotropic} tangent elastic-plastic modulus 
\begin{equation}
\dot{\boldsymbol{\sigma}}=\mathbb{C}^{\rm{ep,t}}_{\mm}(\boldsymbol{\sigma},\varepsilon_{\eq}^{\pp})\cdot\dot{\boldsymbol{\varepsilon}}\,\quad \textrm{with}\quad\mathbb{C}^{\rm{ep,t}}_{\mm}(\boldsymbol{\sigma},\varepsilon_{\eq}^{\pp})=
\mathbb{C}^{\ee}_{\mm}\quad \rm{if}\quad \lambda(\boldsymbol{\sigma},\varepsilon^{\pp}_{\eq})=0
\end{equation}
and 
\begin{equation}\label{C-ep}
\mathbb{C}^{\rm{ep,t}}_{\mm}(\boldsymbol{\sigma},\varepsilon_{\eq}^{\pp})=3K_{\mm}\mathbb{I}^{\rm{P}}
+2G_{\mm}^{\rm{ep,t}}(\varepsilon_{\eq}^{\pp}) \mathbb{N}^{\rm{D}}(\boldsymbol{\sigma})+2G_{\mm}(\mathbb{I}^{\rm{D}}- \mathbb{N}^{\rm{D}}(\boldsymbol{\sigma}))\quad \rm{if}\quad\lambda(\boldsymbol{\sigma},\varepsilon^{\pp}_{\eq})>0\,,
\end{equation}
where the fourth order tensor  $\mathbb{N}^{\rm{D}}$ is
\begin{equation}
\mathbb{N}^{\rm{D}}(\boldsymbol{\sigma})=\mathbf{N}'(\boldsymbol{\sigma})\otimes\mathbf{N}'(\boldsymbol{\sigma})
\end{equation} 
and $\mathbf{N}'(\boldsymbol{\sigma})$ is the direction of the stress deviator $\boldsymbol{\sigma}'$ while the modulus $G^{\rm{ep,t}}_{\mm}(\varepsilon_{\eq}^{\pp})$ decreases with the accumulated plastic strain:
\begin{equation}\label{Eq:Gep-tan}
\mathbf{N}'=\frac{\boldsymbol{\sigma}'}{\sqrt{\boldsymbol{\sigma}'\cdot\boldsymbol{\sigma}'}}\,,\quad
G_{\mm}^{\rm{ep,t}}(\varepsilon_{\eq}^{\pp})=G_{\mm}\frac{Y'(\varepsilon_{\eq}^{\pp})}{Y'(\varepsilon_{\eq}^{\pp})+3G_{\mm}}=G_{\mm}\frac{nh(\varepsilon_{\eq}^{\pp})^{n-1}}{nh(\varepsilon_{\eq}^{\pp})^{n-1}+3G_{\mm}}\,,
\end{equation} 
where $Y'(\varepsilon_{\eq}^{\pp})$ denotes the derivative of $Y(\varepsilon_{\eq}^{\pp})$ with respect to $\varepsilon_{\eq}^{\pp}$ {and $G_{\mm}$ is the constant elastic shear modulus of the matrix}.

The three mentioned variants of tangent linearization and `uniformization' rely on the tangent modulus \eqref{C-ep} and specify $\mathbb{C}^{\rm{ep,aff,iso}}_{\mm}(\hat{\boldsymbol{\sigma}},\hat{\varepsilon}^{\pp}_{\eq})$ and ${\boldsymbol{\beta}}_{\mm}^{\aff*}(\hat{\boldsymbol{\sigma}},\hat{\varepsilon}^{\pp}_{\eq})$ as follows.

\begin{itemize}
		
	\item \textbf{The Tangent model.} 
 The affine local relation \eqref{Eq:affine-uni} is postulated with $G^{\rm{ep,aff,iso}}_{\mm}(\hat{\boldsymbol{\sigma}},\hat{\varepsilon}^{\pp}_{\eq})=G^{\rm{ep,t}}_{\mm}(\bar{\varepsilon}^{\pp}_{\eq})$ for $\lambda({\boldsymbol{\sigma}}_m,\bar{\varepsilon}^{\pp}_{\eq})>0$ , so it is calculated based on the mean values of the stress and strain field in the matrix (the so-called first moments). In particular, using relation \eqref{Eq:seq-eeq} we use the mean value of stress in the matrix phase to calculate $\hat{\varepsilon}^{\pp}_{\eq}$ ($\hat{\sigma}_{\eq}:=\bar{\sigma}_{\eq}$):
	\begin{equation}\label{Eq:Gep-tan-1st}
	{\varepsilon}^{\pp}_{\eq}\approx\hat{\varepsilon}^{\pp}_{\eq}:=\bar{\varepsilon}^{\pp}_{\eq}({\bar{\sigma}_{\eq}})=\left(\frac{\bar{\sigma}_{\eq}-Y_0}{h}\right)^{1/n}
	\quad\textrm{where}
	\quad {\bar{\sigma}_{\eq}=\sqrt{\frac{3}{2}\boldsymbol{\sigma}_{\mm}^{'}\cdot\boldsymbol{\sigma}_{\mm}^{'}}\,,}
	\end{equation}
	{with $\boldsymbol{\sigma}_{\mm}^{'}$ being the deviatoric part of $\boldsymbol{\sigma}_{\mm}$.} The free term ${\boldsymbol{\beta}}_{\mm}^{\aff*}(\hat{\boldsymbol{\sigma}},\hat{\varepsilon}^{\pp}_{\eq})$ in Eq.~\eqref{Eq:affine-uni} incorporates the anisotropic part of the tangent stiffness tensor and is specified as:
	\begin{equation}\label{Eq:beta-tan-1st}
	{\boldsymbol{\beta}}_{\mm}^{\aff*}(\hat{\boldsymbol{\sigma}},\hat{\varepsilon}^{\pp}_{\eq})\approx	{\boldsymbol{\beta}}_{\mm}^{\aff*}({\boldsymbol{\sigma}}_{\mm},\bar{\varepsilon}^{\pp}_{\eq})=2(G^{\rm{ep,t}}(\bar{\varepsilon}^{\pp}_{\eq})-G_{\mm})(\mathbb{I}^{\rm{D}}-\mathbb{N}^{\rm{D}}(\boldsymbol{\sigma}_{\mm}))\cdot\dot{\boldsymbol{\varepsilon}}_{\mm}\,,
	\end{equation}
	so again it is calculated using the mean values of the stress and strain fields in the matrix phase. 
	This isotropic approximation of the tangent stiffness is justified by the fact that for the proportional loading process, in which $\mathbf{N}'$ remains constant at a given material point of the matrix phase, we have
	\begin{equation}
	\dot{\boldsymbol{\sigma}}=\mathbb{C}^{\rm{ep,t}}_{\mm}(\boldsymbol{\sigma},\varepsilon_{\eq}^{\pp})\cdot\dot{\boldsymbol{\varepsilon}}=\mathbb{C}^{\rm{ep,t,iso}}_{\mm}(\varepsilon_{\eq}^{\pp})\cdot\dot{\boldsymbol{\varepsilon}}\,.
	\end{equation}
	It is worth mentioning that in the case of the proposed tangent linearization and unformization based on the first moments, the analogical equality for mean values, namely
	\begin{equation}
	\dot{\boldsymbol{\sigma}}_{\mm}=\mathbb{C}^{\rm{ep,t}}_{\mm}(\boldsymbol{\sigma}_{\mm},\bar{\varepsilon}_{\eq}^{\pp})\cdot\dot{\boldsymbol{\varepsilon}}_{\mm}=\mathbb{C}^{\rm{ep,t,iso}}_{\mm}(\bar{\varepsilon}_{\eq}^{\pp})\cdot\dot{\boldsymbol{\varepsilon}}_{\mm}\,,
	\end{equation}
	remains valid for most of the processes considered in the present study: hydrostatic extension and isochoric tension in the direction of an edge or diagonal of cubic RC, BCC and FCC unit cells, so in these cases ${\boldsymbol{\beta}}_{\mm}^{\aff*}(\hat{\boldsymbol{\sigma}},\hat{\varepsilon}^{\pp}_{\eq})=\mathbf{0}$. On the other hand, in general, the condition is not fulfilled, as for example for isochoric tension in an arbitrary direction within the cubic unit cell.

	\item \textbf{The Non-incremental Tangent (Affine) model}. In \cite{Masson00} it was proposed to consider the local law in the matrix in the non-incremental form, while employing the tangent modulus. Following this proposal, for the non-incremental relation \eqref{Eq:affine-uni-iso} we set $G^{\rm{ep,aff,iso}}_{\mm}(\hat{\boldsymbol{\sigma}},\hat{\varepsilon}^{\pp}_{\eq})=G^{\rm{ep,t}}_{\mm}(\bar{\varepsilon}^{\pp}_{\eq})$ where $G^{\rm{ep,t}}_{\mm}(\bar{\varepsilon}^{\pp}_{\eq})$ is given by Eq.~\eqref{Eq:Gep-tan}. Then the `thermal stress'-like term is given by
	\begin{equation}\label{Eq:beta-aff}
	{\boldsymbol{\beta}}_{\mm}^{\aff*}(\hat{\boldsymbol{\sigma}},\varepsilon^{\pp}_{\eq})=\mathbb{C}^{\rm{ep,t,iso}}_{\mm}(\bar{\varepsilon}_{\eq}^{\pp})\cdot{\boldsymbol{\varepsilon}}_{\mm}-{\boldsymbol{\sigma}}_{\mm}\,,
	\end{equation}
	so again `uniformization' is performed using the mean quantities (first moments) per phase. In practice, in the numerical step-by-step algorithm, the $\boldsymbol{\beta}$-term for step $t+\Delta t$ is calculated as:
	\begin{equation}\label{Eq:Affine-update}
	{\boldsymbol{\beta}}_{\mm}^{\aff*}(t+\Delta t)=\mathbb{C}^{\rm{ep,t,iso}}_{\mm}(t+\Delta t)\cdot{\boldsymbol{\varepsilon}}_{\mm}(t)-{\boldsymbol{\sigma}}_{\mm}(t)-(\mathbb{C}^{\rm{ep,t}}_{\mm}(t+\Delta t)-\mathbb{C}^{\rm{ep,t,iso}}_{\mm}(t+\Delta t))\cdot\Delta{\boldsymbol{\varepsilon}}_{\mm}\,,
	\end{equation}
	where $\Delta{\boldsymbol{\varepsilon}}_{\mm}={\boldsymbol{\varepsilon}}_{\mm}(t+\Delta t)-{\boldsymbol{\varepsilon}}_{\mm}(t)$, and the last term in Eq.~\eqref{Eq:Affine-update} vanishes for the proportional processes mentioned above.

	\item \textbf{The \underline{modified} Tangent model.} A less common method of `uniformization' of the affine moduli is based on the use of the so-called second moment of stress \citep{Suquet97}, namely:
	\begin{equation}\label{Eq:def-Sm}
	\bar{S}_{\mm}=\mathbb{I}^D\cdot\left<\boldsymbol{\sigma}\otimes\boldsymbol{\sigma}\right>_{\mm}=\left<\boldsymbol{\sigma}'\cdot\boldsymbol{\sigma}'\right>_{\mm}=\frac{2}{3}\left< \sigma_{\eq}^2 \right>_{\mm}\,,
	\end{equation}	
	so that
	\begin{equation}\label{Eq:bbar-sig}
	{\varepsilon}^{\pp}_{\eq}\approx\hat{\varepsilon}^{\pp}_{\eq}=\bar{\bar{\varepsilon}}^{\pp}_{\eq}({\bar{\bar{\sigma}}_{\eq}}) \quad\textrm{where}
	\quad \bar{\bar{\sigma}}_{\eq}=\sqrt{\left< \sigma_{\eq}^2 \right>_{\mm}}=\sqrt{\frac{3}{2}\bar{S}_{\mm}}
	\end{equation}
	and $\hat{\varepsilon}^{\pp}_{\eq}=\bar{\bar{\varepsilon}}^{\pp}_{\eq}$ is used to calculate the isotropic tangent shear modulus and the $\boldsymbol{\beta}$-term in Eq.~\eqref{Eq:affine-uni-iso}. Also the yield condition is checked based on $\bar{\bar{\sigma}}_{\eq}$. Because this value is independent of $\boldsymbol{\sigma}_{\mm}$ and $\boldsymbol{\varepsilon}_{\mm}$, which could be found using the set of equations of the cluster model, an additional relation is needed to track the value of $\bar{S}_{\mm}$ in the course of calculations.  To this end in \citep{Berbenni21}, in the context of elastic-viscoplastic two-phase materials and another mean-field approach, it was proposed to use the Hill-Mandel lemma \citep{Hill67} in the form:
	\begin{equation}
	\bar{\boldsymbol{\sigma}}\cdot\dot{\bar{\boldsymbol{\varepsilon}}}=
	\left<	\boldsymbol{\sigma}\cdot\dot{\boldsymbol{\varepsilon}}\right>={f_{\ii}\left<	\boldsymbol{\sigma}\cdot\dot{\boldsymbol{\varepsilon}}\right>_{\ii}+(1-f_{\ii})\left<	\boldsymbol{\sigma}\cdot\dot{\boldsymbol{\varepsilon}}\right>_{\mm}\,.}
	\end{equation}
	Next, observing that the heterogeneity of fields within the inclusion phase and the hydrostatic part of fields in the matrix are less pronounced, the formula was simplified as follows
	\begin{equation}\label{Eq:Hill-app}
	{\bar{\boldsymbol{\sigma}}\cdot\dot{\bar{\boldsymbol{\varepsilon}}}=f_{\ii}{\boldsymbol{\sigma}}_{\ii}\cdot\dot{{\boldsymbol{\varepsilon}}}_{\ii}+3(1-f_{\ii}){\sigma}_{\mm}^{\rm{h}}\dot{{\varepsilon}}^{\rm{h}}_{\mm}+(1-f_{\ii})\left< \boldsymbol{\sigma}'\cdot\dot{\boldsymbol{\varepsilon}}'\right>_{\mm}\,,}
	\end{equation} 
	{where $\sigma_{\mm}^{\rm{h}}\mathbf{I}$ and $\varepsilon^{\rm{h}}_{\mm}\mathbf{I}$ are the hydrostatic parts of $\boldsymbol{\sigma}_{\mm}$ and $\boldsymbol{\varepsilon}_{\mm}$, respectively, while $\boldsymbol{\varepsilon}'$ is the deviatoric part of $\boldsymbol{\varepsilon}$. As it is seen, following this simplification regarding the inclusion phase and the hydrostatic part of fields in the elastic-plastic matrix, only the mean values of fields per phase are taken into account.} The last term, employing the deviatoric part of the stress and strain rate fields in the matrix, is expanded and approximated by introducing Eq.~\eqref{Eq:affine-uni-iso} with
	\begin{eqnarray}
	\mathbb{C}^{\rm{ep,t,iso}}_{\mm}(\varepsilon_{\eq}^{\pp})&\approx&\mathbb{C}^{\rm{ep,t}}_{\mm}(\bar{\bar{\varepsilon}}_{\eq}^{\pp})=3K_{\mm}\mathbb{I}^{\rm{P}}
	+2G_{\mm}^{\rm{ep,t}}(\bar{\bar{\varepsilon}}_{\eq}^{\pp}) \mathbb{I}^{\rm{D}}\,,\\ \label{Eq:beta-tan-2nd}
	{\boldsymbol{\beta}}_{\mm}^{\aff*}(\hat{\boldsymbol{\sigma}},\hat{\varepsilon}^{\pp}_{\eq})&\approx&	{\boldsymbol{\beta}}_{\mm}^{\aff*}({\boldsymbol{\sigma}}_{\mm},\bar{\bar{\varepsilon}}^{\pp}_{\eq})=2(G^{\rm{ep,t}}_{\mm}(\bar{\bar{\varepsilon}}^{\pp}_{\eq})-G_{\mm})(\mathbb{I}^{\rm{D}}-\mathbb{N}^{\rm{D}}(\boldsymbol{\sigma}_{\mm}))\cdot\dot{\boldsymbol{\varepsilon}}_{\mm}\,,
	\end{eqnarray}
	after which it is obtained
	\begin{equation}
	\left< \boldsymbol{\sigma}'\cdot\dot{\boldsymbol{\varepsilon}}'\right>_{\mm}\approx
	\frac{1}{2G_{\mm}^{\rm{ep,t}}}\left< \boldsymbol{\sigma}'\cdot(\dot{\boldsymbol{\sigma}}'+{\boldsymbol{\beta}_{\mm}^{\aff*}})\right>_{\mm}=\\ \frac{1}{2G_{\mm}^{\rm{ep,t}}}\left( \left<\boldsymbol{\sigma}'\cdot\dot{\boldsymbol{\sigma}}'\right>_{\mm}+\boldsymbol{\sigma}'_{\mm}\cdot{\boldsymbol{\beta}_{\mm}^{\aff*}}\right)=\frac{1}{2G_{\mm}^{\rm{ep,t}}}\left<\boldsymbol{\sigma}'\cdot\dot{\boldsymbol{\sigma}}'\right>_{\mm}\,.
	\end{equation} 
	Based on the definition \eqref{Eq:def-Sm} we observe that
	\begin{equation}
	\dot{\bar{S}}_{\mm}=2\left<\boldsymbol{\sigma}'\cdot\dot{\boldsymbol{\sigma}}'\right>_{\mm}\,,
	\end{equation}
	so the following relation is obtained for the time rate of the second moment:
	\begin{equation}\label{Eq:dotSm}
	{\dot{\bar{S}}_{\mm}=  \frac{4G_{\mm}^{\rm{ep,t}}}{1-f_{\ii}}\left(\bar{\boldsymbol{\sigma}}\cdot\dot{\bar{\boldsymbol{\varepsilon}}}-f_{\ii}{\boldsymbol{\sigma}}_{\ii}\cdot\dot{{\boldsymbol{\varepsilon}}}_{\ii}-3(1-f_{\ii}){\sigma}_{\mm}^{\rm{h}}\dot{{\varepsilon}}^{\rm{h}}_{\mm}\right)\,.}
	\end{equation}
    In \citep{Berbenni21}, the current value of $\bar{S}_m$ was found from the evolution equation: $\bar{S}_{\mm}(t+\Delta t)=\bar{S}_{\mm}(t)+\dot{\bar{S}}_{\mm}(t)\Delta t$.
	However, since $\bar{S}_{\mm}$ is a quadratic function of $\boldsymbol{\sigma}$, it has been found in the present context of the elastic-plastic matrix that for a computationally efficient update of $\bar{S}_{\mm}$ we will also require the second time rate of $\bar{S}_{\mm}$. This quantity can be found using Hill's condition presented fully in rate form, namely
	\begin{equation}
	\dot{\bar{\boldsymbol{\sigma}}}\cdot\dot{\bar{\boldsymbol{\varepsilon}}}=
	\left<	\dot{\boldsymbol{\sigma}}\cdot\dot{\boldsymbol{\varepsilon}}\right>={f_{\ii}\left<	\dot{\boldsymbol{\sigma}}\cdot\dot{\boldsymbol{\varepsilon}}\right>_{\ii}+(1-f_{\ii})\left<	\dot{\boldsymbol{\sigma}}\cdot\dot{\boldsymbol{\varepsilon}}\right>_{\mm}}\,,
	\end{equation}
	which after the same simplifications as in Eq.~\eqref{Eq:Hill-app} reduces to
	\begin{equation}\label{Eq:Hill-app2}
	\dot{  \bar{\boldsymbol{\sigma}}}\cdot\dot{\bar{\boldsymbol{\varepsilon}}}={f_{\ii}{	\dot{\boldsymbol{\sigma}}}_{\ii}\cdot\dot{{\boldsymbol{\varepsilon}}}_{\ii}+3(1-f_{\ii}){\dot{\sigma}}_{\mm}^{\rm{h}}\dot{{\varepsilon}}^{\rm{h}}_{\mm}+(1-f_{\ii})\left<	\dot{\boldsymbol{\sigma}}'\cdot\dot{\boldsymbol{\varepsilon}}'\right>_{\mm}\,.}
	\end{equation} 
	After substituting Eq.~\eqref{Eq:affine-uni-iso}, the last term is found as
	\begin{equation}
	\left< \dot{\boldsymbol{\sigma}}'\cdot\dot{\boldsymbol{\varepsilon}}'\right>_{\mm}\approx
	\frac{1}{2G_{\mm}^{\rm{ep,t}}}\left< \dot{\boldsymbol{\sigma}}'\cdot(\dot{\boldsymbol{\sigma}}'+\boldsymbol{\beta}_{\mm}^{\aff*})\right>_{\mm}= \frac{1}{2G_{\mm}^{\rm{ep,t}}}\left( \left<\dot{\boldsymbol{\sigma}}'\cdot\dot{\boldsymbol{\sigma}}'\right>_{\mm}+\dot{\boldsymbol{\sigma}}'_{\mm}\cdot\boldsymbol{\beta}_{\mm}^{\aff*}\right)\,.
	\end{equation}
	On the other hand it is assumed that
	\begin{equation}\label{Eq:ddotSm}
	\ddot{\bar{S}}_{\mm}=2\left<{\boldsymbol{\sigma}}'\cdot\ddot{\boldsymbol{\sigma}}'+\dot{\boldsymbol{\sigma}}'\cdot\dot{\boldsymbol{\sigma}}'\right>_{\mm}\approx {2 \left<\dot{\boldsymbol{\sigma}}'\cdot\dot{\boldsymbol{\sigma}}'\right>_{\mm}\,,}
	\end{equation}
	that is we assume that the stress increment along $\Delta t$ is sufficiently smooth to neglect $\ddot{\boldsymbol{\sigma}}'$. Combing Eqs.~(\ref{Eq:Hill-app2}--\ref{Eq:ddotSm}) we find
	\begin{equation}\label{Eq:ddotSm2}
	\ddot{\bar{S}}_{\mm}=  {\frac{4G_{\mm}^{\rm{ep,t}}}{1-f_{\ii}}\left(\dot{\bar{\boldsymbol{\sigma}}}\cdot\dot{\bar{\boldsymbol{\varepsilon}}}-f_{\ii}{	\dot{\boldsymbol{\sigma}}}_{\ii}\cdot\dot{{\boldsymbol{\varepsilon}}}_{\ii}-3(1-f_{\ii})\dot{{\sigma}}_{\mm}^{\rm{h}}\dot{{\varepsilon}}^{\rm{h}}_{\mm}\right)-2\dot{\boldsymbol{\sigma}}'_{\mm}\cdot\boldsymbol{\beta}_{\mm}\,.}
	\end{equation}
	Using Eqs.~\eqref{Eq:dotSm} and \eqref{Eq:ddotSm}, the value of $\bar{S}_{\mm}$ is updated according to
	\begin{equation}\label{Eq:Sm-update}
	\bar{S}_{\mm}(t+\Delta t)=\bar{S}_{\mm}(t)+\dot{\bar{S}}_{\mm}\Delta t+\frac{1}{2}\ddot{\bar{S}}_{\mm} \Delta t^2
	\end{equation}
	and then used in Eq.~\eqref{Eq:bbar-sig} to find the current value of $\bar{\bar{\sigma}}_{\eq}$. Let us note that the modified (incremental) tangent scheme was also employed in \cite{DoghriBrassart11}; however, the authors calculated the second moments using the algorithmic tangent modulus for the elastic predictor step and the formula based on differentiation of the effective elastic stiffness with respect to the shear modulus. 
\end{itemize}

It should be noted that the cluster scheme extension to elastic-plastic materials can be also performed using the classical or modified secant scheme model. It has been decided not to follow this path since the definition of the secant elastic-plastic modulus is restricted to the proportional loading processes. By contrast, the tangent elastic-plastic modulus {$\mathbb{C}^{\rm{ep,t}}_{\mm}(\boldsymbol{\sigma},\varepsilon_{\eq}^{\pp})$} in Eq.~\eqref{C-ep} can be calculated for any deformation path. As mentioned above, the difficulty of the tangent linearization method as compared to the secant model is the fact that, even though the plasticity model is isotropic, the exact tensor {$\mathbb{C}^{\rm{ep,t}}_{\mm}(\boldsymbol{\sigma},\varepsilon_{\eq}^{\pp})$} 
 is orthotropic with the orthotropy directions induced by the principal axes of the plastic strain rate or equivalently stress in the matrix. Although there exist specifications of the polarization tensor $\mathbb{P}_0$ and the interaction tensor between two inclusions $\boldsymbol{\Gamma}^{IJ}$ for an anisotropic matrix material \citep{Berveiller87}, their application requires numerical integration to be performed and the closed-form relations given in \ref{SubSec:AppStepByStep} and, consequently, Eqs.~(\ref{Eq:XRC}--\ref{Eq:XFCC}) cannot be applied. This strongly affects the computational efficiency and robustness of the method, which are supposed to be the main advantages of the mean-field schemes making them suitable for use in large-scale FEM computations \citep{Sadowski17b}. As demonstrated above, the issue has been solved in the present study by using the isotropic $\mathbb{C}^{\rm{ep,aff,iso}}_{\mm}$ specified by Eq.~\eqref{Eq:affine-uni-iso} with $G^{\rm{ep,t}}(\hat{\varepsilon}^{\pp}_{\eq})$ as the shear modulus to preserve the relative algebraic simplicity of the scheme in the elastic regime. The assumed isotropic approximation corresponds to the so-called \emph{soft} isotropization proposed in \cite{Chaboche05}. 

It should be stressed that all three steps: linearization, uniformization and isotropization introduce additional approximations to the mean-field model as compared to the elastic regime, which may further affect the quality of predictions. Thus a selection of the most appropriate approach must be based on verification with respect to full-field solutions, obtained e.g. by numerical FEM homogenization.

\section{Numerical models\label{Sec:Numer}}
Throughout the paper, Finite Element Method (FEM) calculations are used as benchmarks for checking the correctness of predictions of the cluster model. Each numerical test using FEM consisted of generating a geometric representation of a periodic volume element (a unit cell) of the composite, then generating a periodic FEM mesh for the given geometry, and finally performing FEM computations. The volume elements were meshed in the NetGen software \citep{Schoberl97,Gangl2020} with second-order (10-node) tetrahedral finite elements. Periodicity of the mesh was enforced in the process, i.e. the opposite sides of a volume element were identically meshed. The sensitivity of results to the increasing mesh density was analyzed to decide upon a required mesh refinement. FEM computations were performed in the AceFEM software \citep{Korelc02}, using periodic displacement boundary conditions. Periodic boundary conditions imply that if any two FEM mesh nodes A and B lie on one side of a volume element, while nodes C and D are their periodic images, respectively, on the opposite side of the volume element, then
\begin{equation}\label{eq:periodic}
	\mathbf{u}_{\rm{A}}-\mathbf{u}_{\rm{C}} = \mathbf{u}_{\rm{B}}-\mathbf{u}_{\rm{D}}\,,
\end{equation}
where $\mathbf{u}_{\rm{A}}$, $\mathbf{u}_{\rm{B}}$, $\mathbf{u}_{\rm{C}}$ and $\mathbf{u}_{\rm{D}}$ are the displacements of the respective nodes. To implement the periodic displacement boundary conditions for a given macroscopic deformation tensor $\bar{\boldsymbol{\varepsilon}}$ of the unit volume element, the nodes at four vertices of the volume element are assigned displacements of the form
\begin{equation}\label{eq:deformation}
	\mathbf{u}_i = \bar{\boldsymbol{\varepsilon}}\cdot\mathbf{x}_i\ , \quad i\in\{0, 1, 2, 3\}\,,
\end{equation}
where $\mathbf{x}_0=(0,0,0)$, $\mathbf{x}_1=(1,0,0)$, $\mathbf{x}_2=(0,1,0)$ and $\mathbf{x}_3=(0,0,1)$, while all periodic relations in Eq.~\eqref{eq:periodic} are enforced during the FEM solution by taking node A as vertex 0 and C as vertex 1, 2 or 3, depending on the wall pairing. A more detailed description of the applied numerical homogenization procedures can be found in \citep{Majewski20,Majewski22}.

\begin{figure}[H]
	\centering
	\begin{tabular}{cccc}
		RC MMC&RC Voids&BCC MCC&BCC Voids\\
		\includegraphics[angle=0,width=0.24\textwidth]{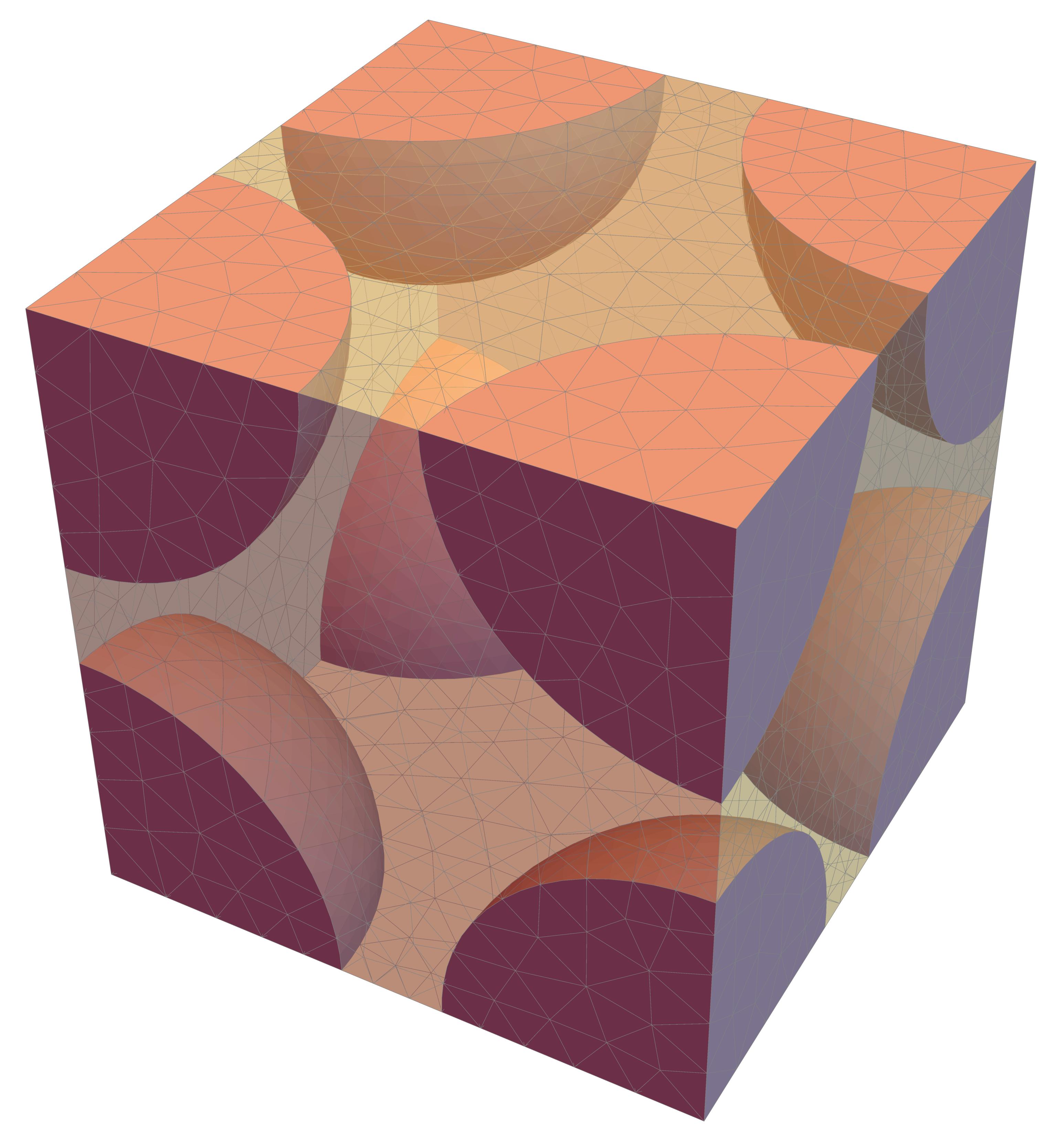}&
		\includegraphics[angle=0,width=0.24\textwidth]{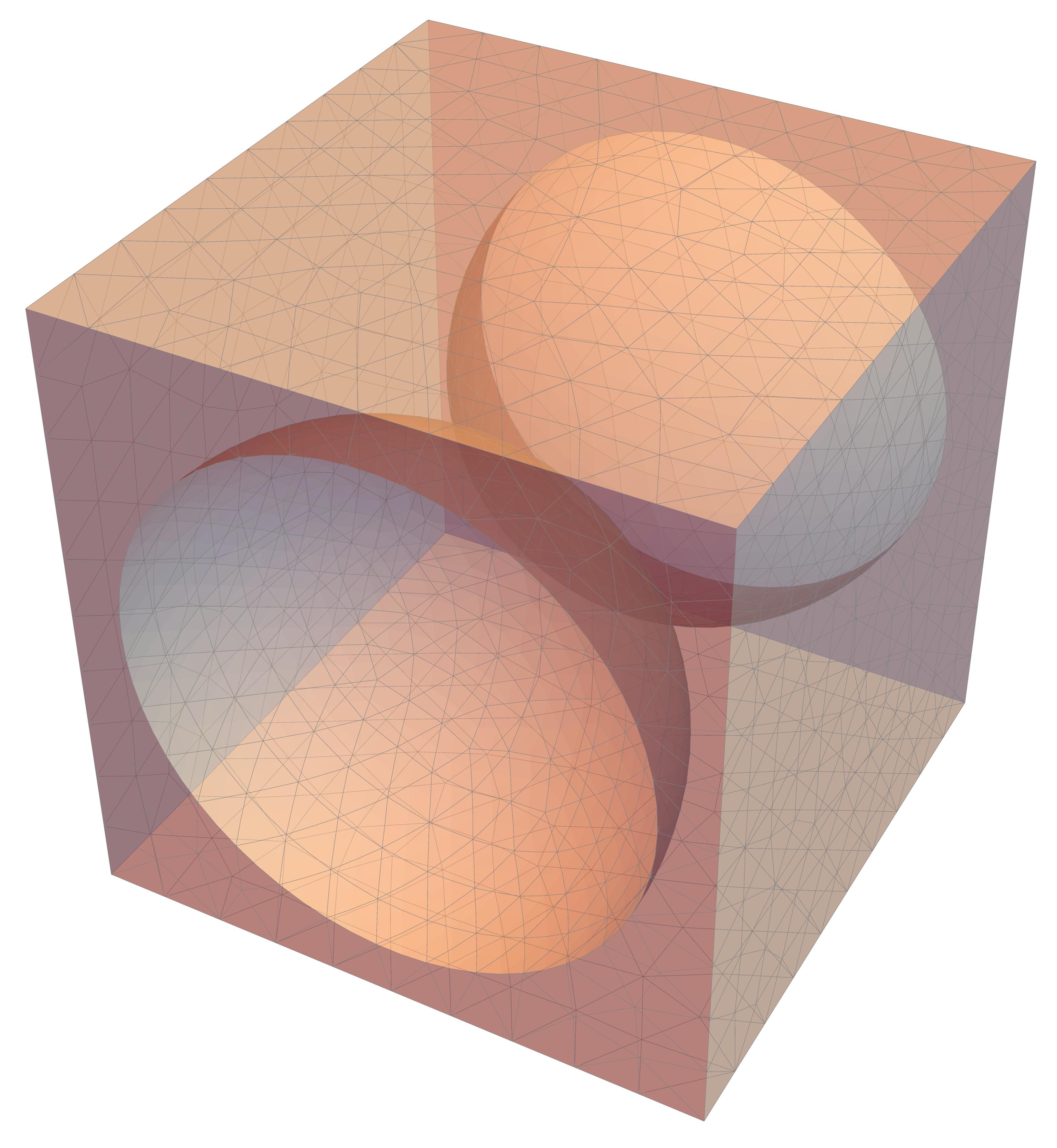}&
		\includegraphics[angle=0,width=0.24\textwidth]{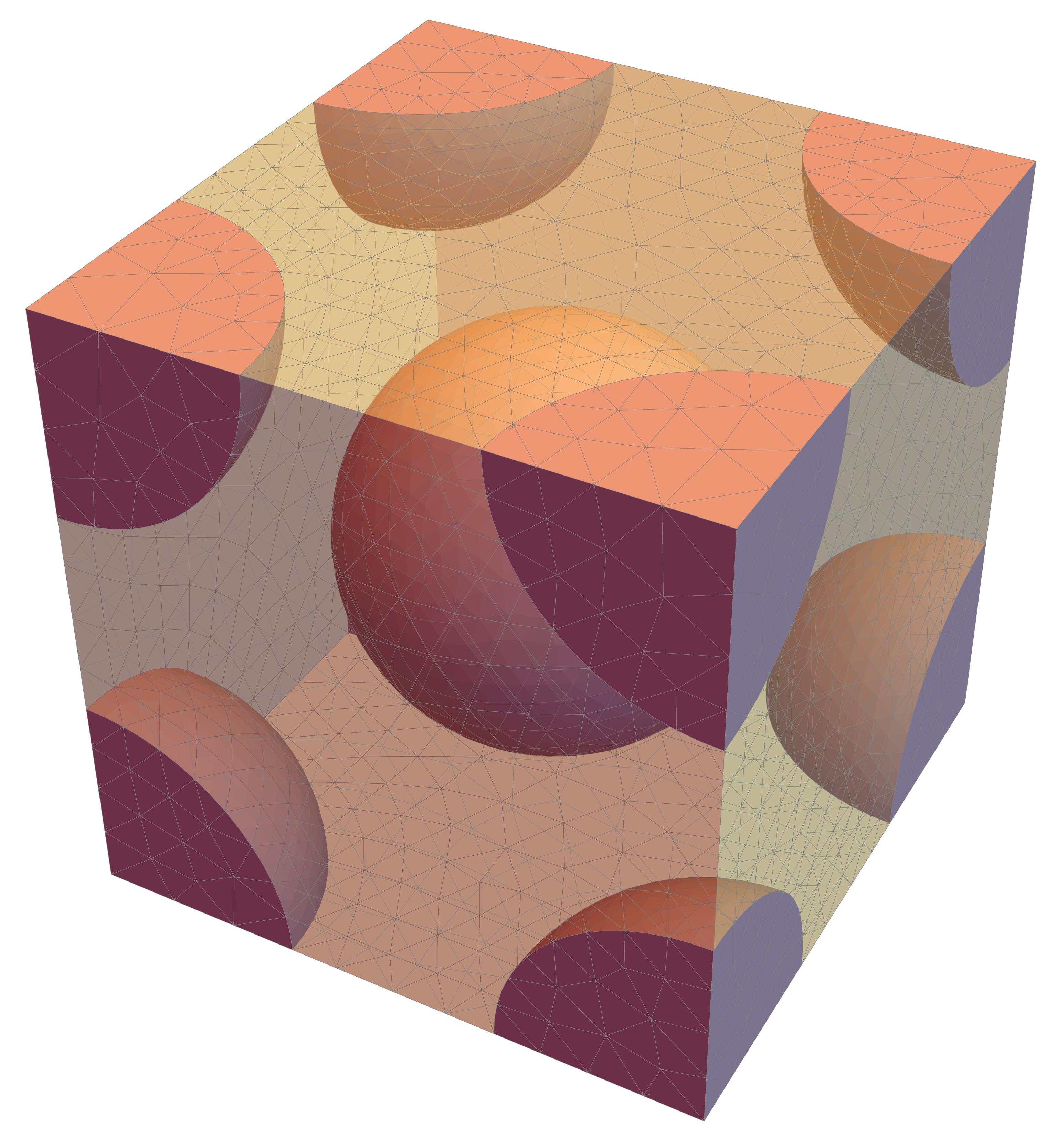}&
		\includegraphics[angle=0,width=0.24\textwidth]{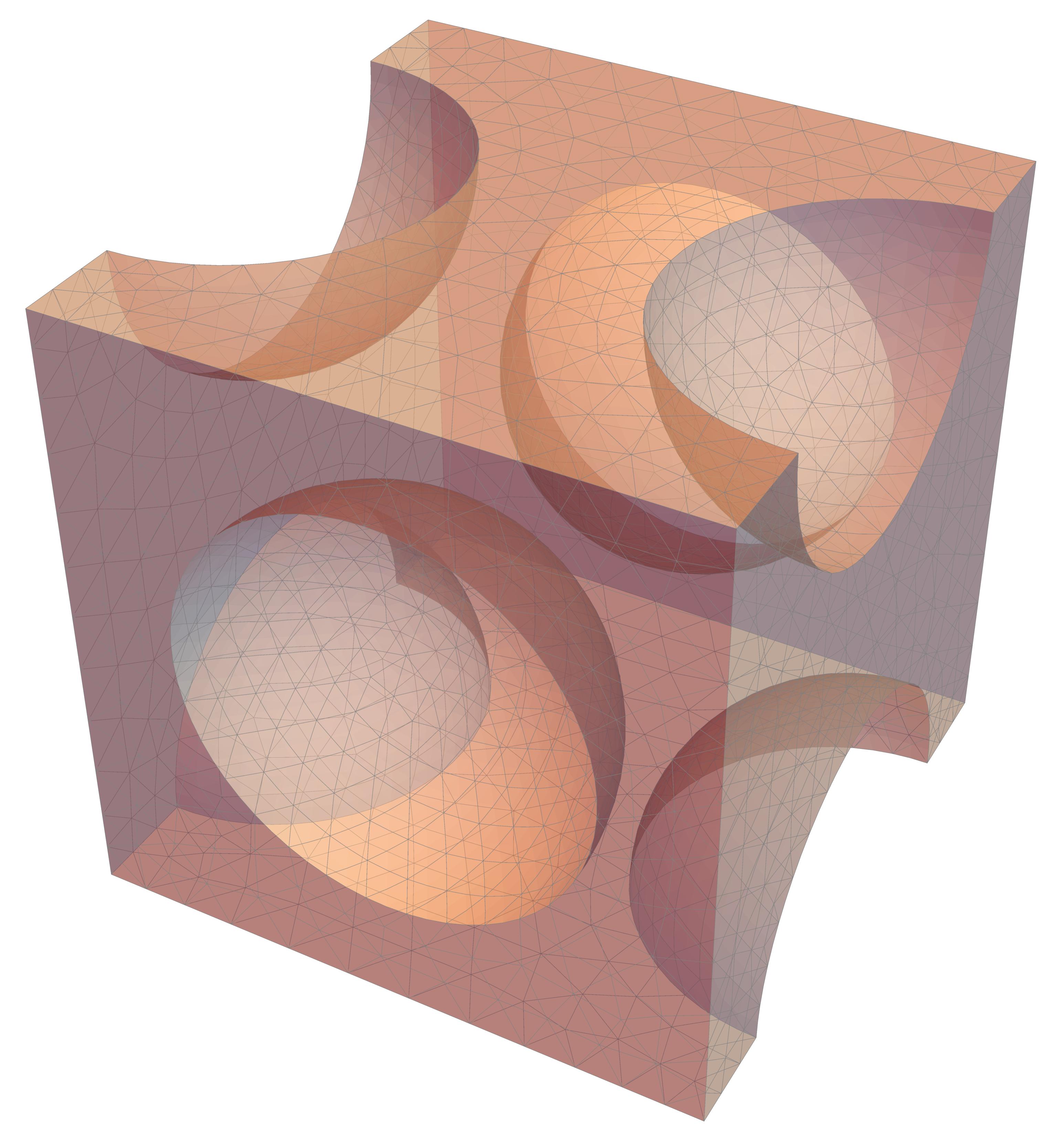}
	\end{tabular}
	\caption{Unit volume elements with $f_{\textup{i}}=30\%$ and FEM meshes shown. Regular arrangements of inclusions: Regular Cubic (RC) and Body-Centered Cubic (BCC), and two material configurations: Metal Matrix reinforced with Ceramic spheres (MMC) and metal matrix weakened by voids (PM). 
		\label{Fig:UVEMMCVoids}}
\end{figure}

The materials analysed in this paper, in terms of geometry, are composites with regularly distributed inhomogeneities. Fig.~\ref{Fig:UVEMMCVoids} presents unit volume elements with regular arrangements of inhomogeneities used in the FEM simulations: Regular Cubic (RC) and Body-Centered Cubic (BCC), with either spherical inclusions, for modelling a Metal Matrix reinforced with Ceramic spheres (MMC), or spherical voids, for modelling a porous metal matrix (PM).
It can be seen that, for the same regular systems, the positions of inclusions in the unit volume elements of the MMC and of the material with voids are different (e.g. Fig.~\ref{Fig:UVEMMCVoids}: RC MMC vs RC Voids). This results from the applied method of enforcing periodicity of the AceFEM solution, which requires that there exist FEM mesh nodes at the vertices of the volume element (if there were voids at the vertices, no mesh nodes would be there).

\section{Results\label{Sec:Results}}

\subsection{Elastic regime\label{SubSec:RegularElastic}}
{In the present section, the behaviour of the model is examined by observing how effective composite properties change with the volume fraction and arrangement of inclusions in the elastic regime.
In order to assess the behaviour of the model, theoretical materials with a large contrast between the matrix and inclusion phases are investigated as benchmark cases. Their parameters are listed in Table~\ref{Tab:TestMaterialParameters}.
\begin{table}[ht]
\centering
\begin{tabular}{lccc}
\hline
 &  & Ultra hard inclusions  & Ultra soft inclusions 
\\
Phase & Matrix 
material & (UHI) & (USI)
\\ 
\hline
\textup{Shear modulus,} $G$ [GPa] & 1 & 10000 & 0.0001
\\
\textup{Bulk modulus,} $K$ [GPa] & 2.1667 & 21667 & 0.00021667
\\
\textup{Poisson's ratio,} $\nu$ [-] & 0.3 & 0.3 & 0.3
\\
\hline
\end{tabular}
\caption{Theoretical material parameters used for testing the model in the elastic regime.} \label{Tab:TestMaterialParameters}
\end{table}
Two material configurations are considered: ultra hard inclusions (UHI) and ultra soft inclusions (USI) (Table~\ref{Tab:TestMaterialParameters}). The parameters do not correspond to any particular material, however, the first one is similar to e.g. a metal matrix with ceramic inclusions, especially in the case of an advanced plastic flow when the stiffness modulus of the matrix decreases. The second scenario is the opposite case, close to e.g. a matrix with voids (metal foams or similar) or a situation when ceramic inclusions have been damaged and no longer carry loads.

As explained before, three different arrangements of inclusions are considered: regular cubic (RC), body-centered cubic (BCC) and face-centered cubic (FCC). They are presented in Fig.~\ref{Fig:RC_BCC_FCC}. Since the cluster model does not allow for overlap of inclusions, the maximum possible volume fraction of inclusions is limited to: $\pi/6$ for the RC, $\sqrt{3}\pi/8$ for the BCC and $\sqrt{2}\pi/6$ for the FCC arrangement.

Predictions of the cluster model, obtained using the closed-form formulas in Eqs.~(\ref{Eq:GammaClosed}--\ref{Eq:rodfromfin}), along with the results of calculations by FEM, are shown in Figs.~\ref{Fig:elastic_RC_uhiusiK}--\ref{Fig:elastic_FCC_uhiusiG}. For comparison, Figs.~\ref{Fig:elastic_RC_uhiusiK}--\ref{Fig:elastic_FCC_uhiusiG} also present predictions of several other micromechanical models, whenever they are applicable. These models are referred to by the names of the first authors of the papers in which they were proposed, namely: \emph{Cohen} \citep{Cohen04}, \emph{Nemat-Nasser} \citep{NematNasser82}---derived for RC, \emph{Nunan} \citep{Nunan84}---derived for UHI and \emph{Rodin} \citep{Rodin93}. For each setup, the effective bulk/shear modulus was divided by the bulk/shear modulus of the matrix. Because of the cubic symmetry of arrangements, the effective properties of the composite also exhibit cubic symmetry. Cubic symmetry can be best described by two shear moduli $\bar{G_1}$ and $\bar{G_2}$ \citep{Majewski17}. These effective shear moduli, together with the effective bulk modulus $\bar{K}$, are calculated as follows:

\begin{equation}
3\bar{K}=
1/3\left(\bar{C}_{1111}+\bar{C}_{2222}+\bar{C}_{3333}\right)
+
2/3\left(\bar{C}_{1122}+\bar{C}_{1133}+\bar{C}_{2233}\right)\,,
\end{equation}
\begin{equation}
2\bar{G_1}=
1/3\left(\bar{C}_{1111}+\bar{C}_{2222}+\bar{C}_{3333}\right)
-
1/3\left(\bar{C}_{1122}+\bar{C}_{1133}+\bar{C}_{2233}\right)\,,
\end{equation}
\begin{equation}
\bar{G_2}=
1/3\left(\bar{C}_{2323}+\bar{C}_{1313}+\bar{C}_{1212}\right)\,.
\end{equation}
Note that for the stiffness tensor of perfect cubic symmetry we have $\bar{C}_{1111}=\bar{C}_{2222}=\bar{C}_{3333}$, $\bar{C}_{1122}=\bar{C}_{1133}=\bar{C}_{2233}$ and $\bar{C}_{2323}=\bar{C}_{1313}=\bar{C}_{1212}$.


\begin{figure}[ht]
\centering
\begin{tabular}{cc}
	(a)&(b)\\
\includegraphics[height=6cm]{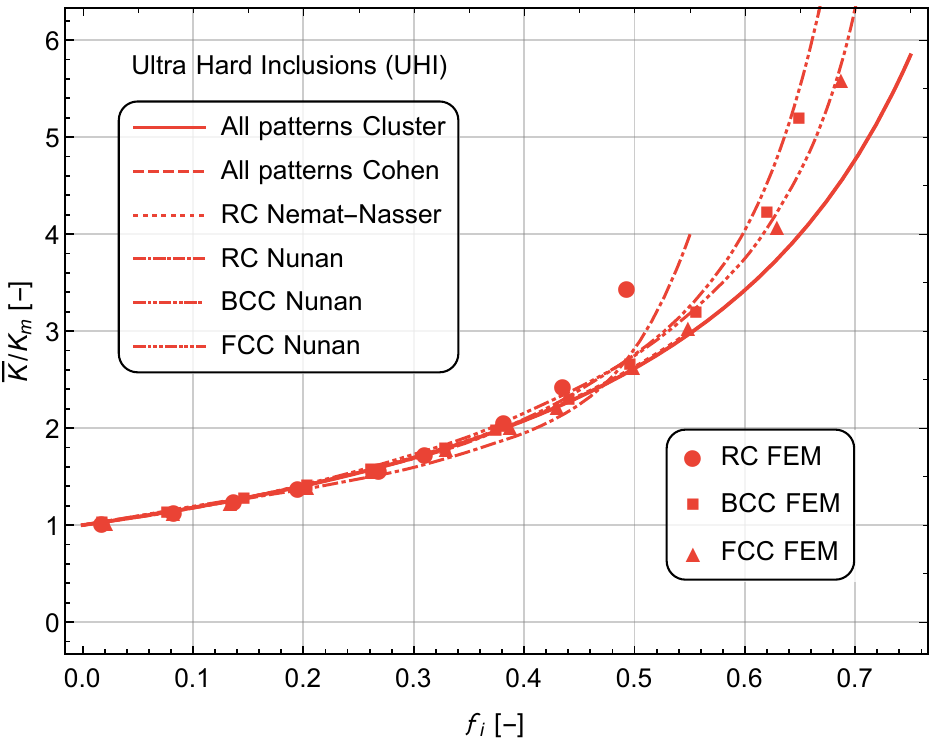} &
\includegraphics[height=6cm]{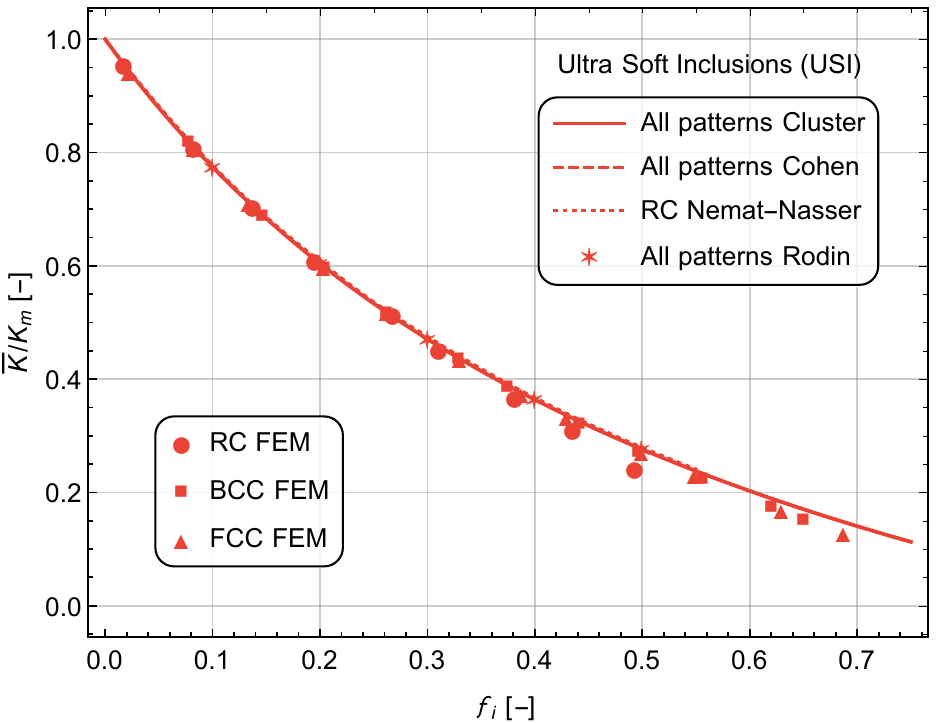}
\end{tabular}
\caption{Effective values of the bulk modulus relative to the matrix properties. All arrangements, a) ultra hard inclusions (UHI) and b) ultra soft inclusions (USI). The curves for the cluster, Cohen and Nemat-Nasser models nearly coincide.}
\label{Fig:elastic_RC_uhiusiK}
\end{figure}

\begin{figure}[ht]
\centering
\begin{tabular}{cc}
	(a)&(b)\\
\includegraphics[height=6cm]{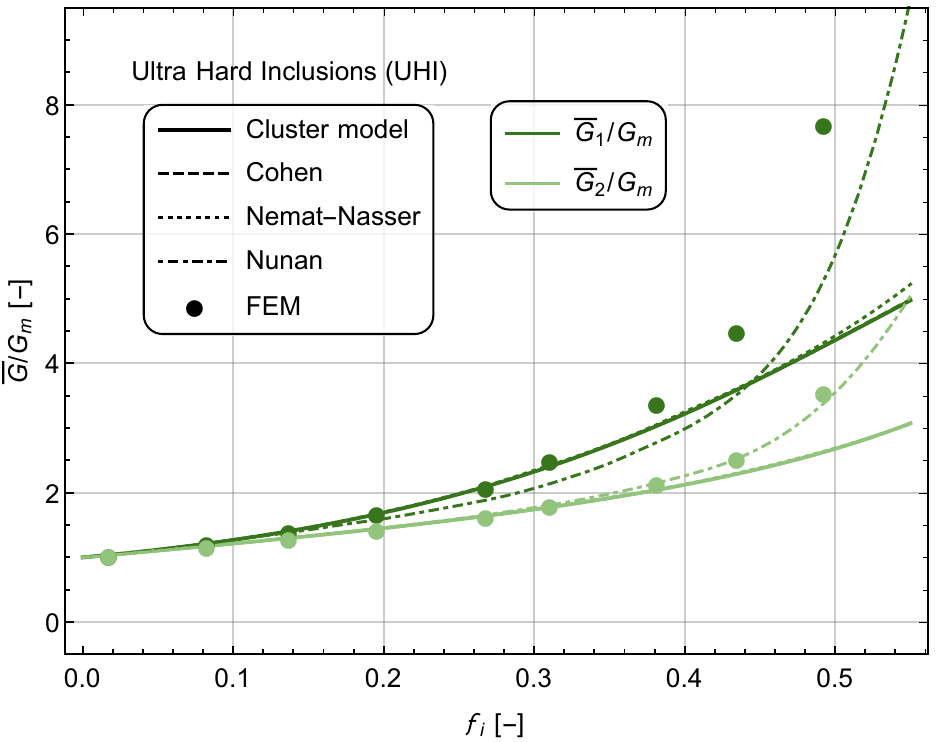}& 
\includegraphics[height=6cm]{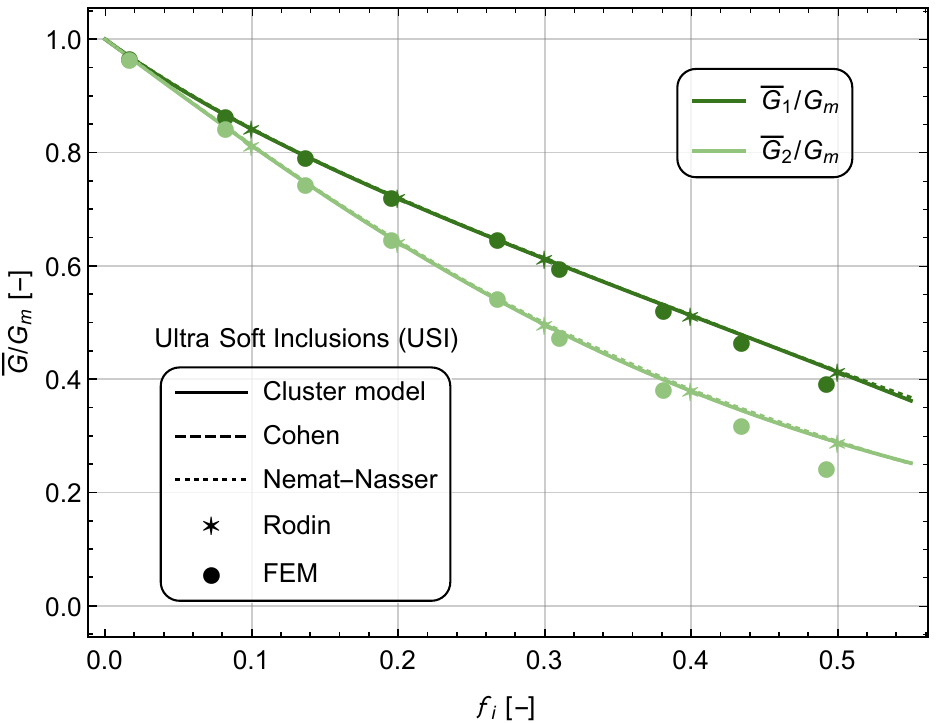}
\end{tabular}
\caption{Effective values of the shear moduli relative to the matrix properties. RC arrangement, a) ultra hard inclusions (UHI) and b) ultra soft inclusions (USI). The curves for the cluster, Cohen and Nemat-Nasser models nearly coincide, except for the Nemat-Nasser model for UHI at high volume fractions.}
\label{Fig:elastic_RC_uhiusiG}
\end{figure}

\begin{figure}[ht]
\centering
\begin{tabular}{cc}
	(a)&(b)\\
\includegraphics[height=6cm]{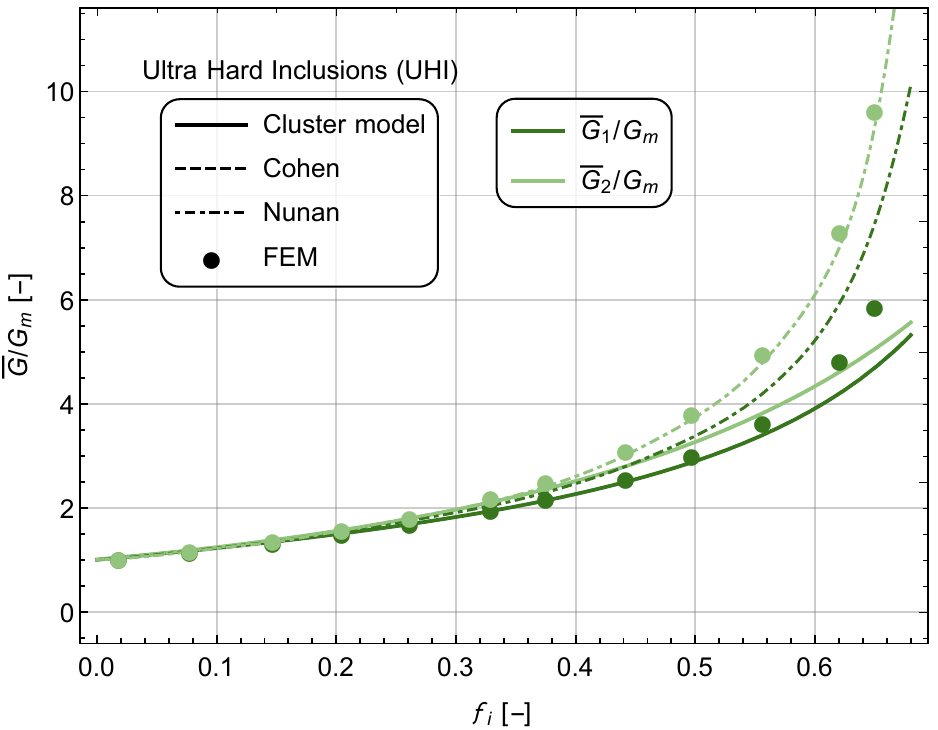} &
\includegraphics[height=6cm]{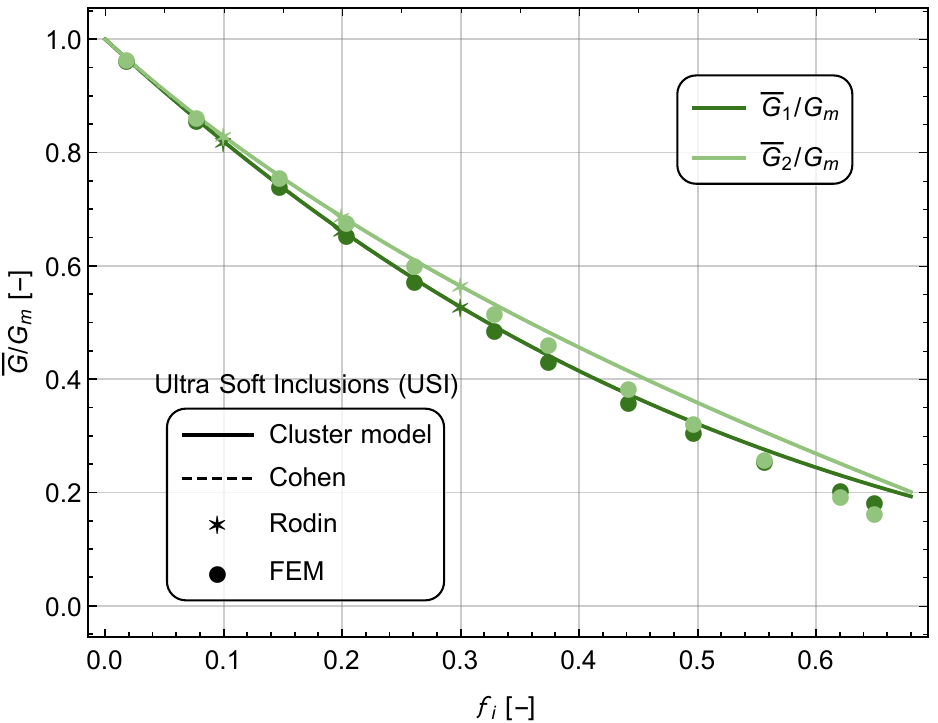}
\end{tabular}
\caption{Effective values of the shear moduli relative to the matrix properties. BCC arrangement, a) ultra hard inclusions (UHI) and b) ultra soft inclusions (USI). The curves for the cluster and Cohen models coincide.}
\label{Fig:elastic_BCC_uhiusiG}
\end{figure}

\begin{figure}[ht]
\centering
\begin{tabular}{cc}
	(a)&(b)\\
\includegraphics[height=6cm]{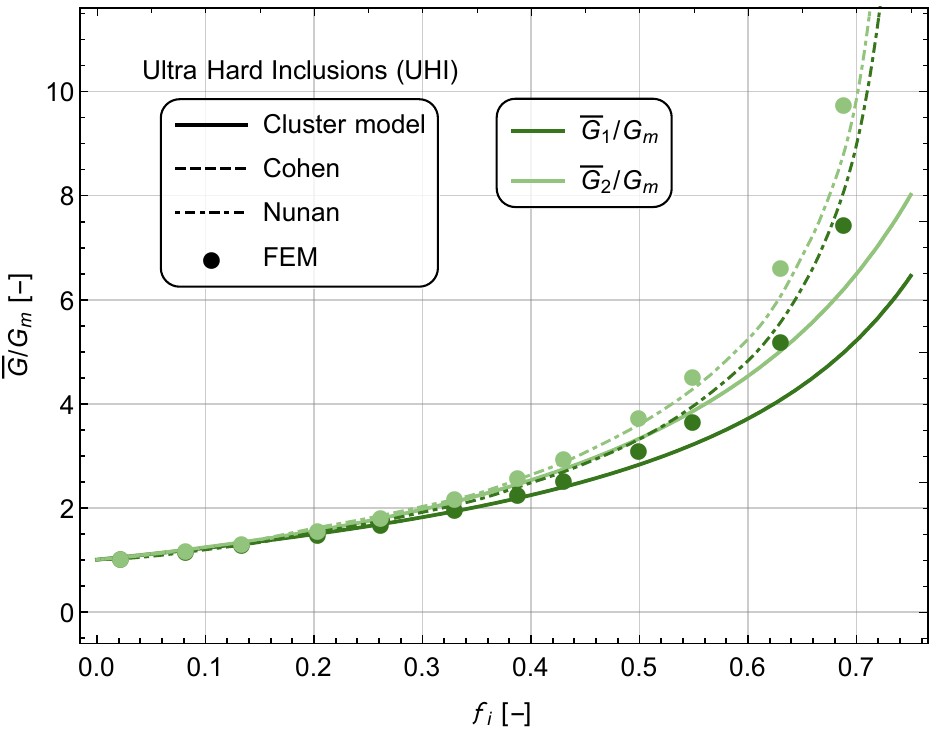} &
\includegraphics[height=6cm]{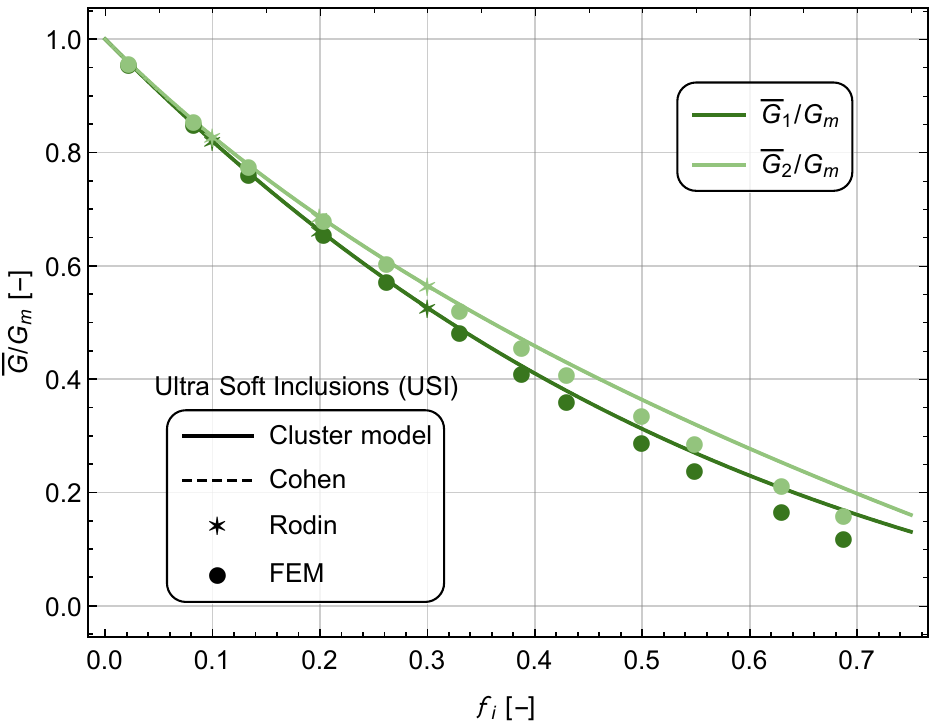}
\end{tabular}
\caption{Effective values of the shear moduli relative to the matrix properties. FCC arrangement, a) ultra hard inclusions (UHI) and b) ultra soft inclusions (USI). The curves for the cluster and Cohen models coincide.}
\label{Fig:elastic_FCC_uhiusiG}
\end{figure}

As can be demonstrated analytically, the effective bulk modulus obtained by the cluster model for a given volume fraction of inclusions is independent of the spatial arrangement of inclusions, be it RC, BCC or FCC. The same holds true for the Cohen and Rodin models, but not for the Nunan model. The respective plots of the bulk modulus assessments are presented in Fig.~\ref{Fig:elastic_RC_uhiusiK}. It can be seen that the curves for the cluster model and those for the Cohen and Nemat-Nasser methods almost coincide.

The cluster model, similarly to the Cohen, Nunan and Rodin ones, captures the difference between the arrangements with respect to the shear moduli (Figs.~\ref{Fig:elastic_RC_uhiusiG}--\ref{Fig:elastic_FCC_uhiusiG}). Again, the predictions of the cluster, Cohen, Nemat-Nasser and Rodin models are almost the same, while the Nunan model differs, especially at higher volume fractions. 
The strongest predicted anisotropy is visible for the RC arrangement. The BCC and FCC arrangements behave similarly, with a lower, but still visible, difference between the two shear moduli. For both the UHI and USI scenarios, $\bar{G_1}$ has higher values than $\bar{G_2}$ for the RC arrangement, while for the BCC and FCC it is the opposite: $\bar{G_2}$ is higher than $\bar{G_1}$. This means that the Zener anisotropy ratio $G_2/G_1$ is larger than one in the first case, while in the second case it is smaller than one, which demonstrates a qualitative difference between the estimated anisotropies. Moreover, an asymmetry of the inclusion content effect is found between hard and soft inclusions. Namely, for example, for the RC cell the UH inclusions cause a more substantial increase in $G_1$ than in $G_2$, while the US inclusions lead to a less substantial decrease in the $G_1$ than in the $G_2$ modulus. Such behaviour results from the fact that $\bar{\boldsymbol{\Gamma}}$ does not depend on the inclusions' material properties. It should be recalled that the difference between the $G_1$ and $G_2$ moduli, and so material anisotropy, is not captured by the classical Mori-Tanaka model whose predictions only take the volume content and the shape of inclusions into account. It is also not reproduced by the more elaborate models discussed in the introduction which use the concept of an effective ellipsoidal inhomogeneity to describe the spatial distribution effect, since for cubic arrangements this effective inclusion would have the shape of a sphere (compare \cite{Vilchevskaya21}).

It can be seen in Figs.~\ref{Fig:elastic_RC_uhiusiK}--\ref{Fig:elastic_FCC_uhiusiG} that the cluster model generally gives results which are very close to the ones obtained by the Finite Element Method for volume fractions of inclusions below $f_{\rm{i}}$=0.40, confirming good predictive capabilities of the model. However, there is a significant discrepancy between the cluster model and FEM for volume fractions of inclusions reaching their maximum possible values. This fact is the most visible in scenarios where inclusions are stiffer than the matrix and in the case of the shear moduli. The Nunan method is the only one which seems to better approximate the FEM results than the cluster model at high volume fractions of inclusions, while it also predicts differences between the effective bulk moduli for the RC, BCC and FCC arrangements. However, it has the drawback of being more computationally complex and designed only for rigid inclusions.

It is clear that the volume fraction of inclusions is the main factor influencing the effective properties of a composite. However, a comparison of results for different arrangements of inclusions provides further information about the secondary impact of the inclusions' spatial distribution. The results of such a comparison are shown in Figs.~\ref{Fig:RCvBCCvFCC_uhiusi_G1} and \ref{Fig:RCvBCCvFCC_uhiusi_G2}.
\begin{figure}[ht]
\centering
\begin{tabular}{cc}
	(a)&(b)\\
\includegraphics[width=7.5cm]{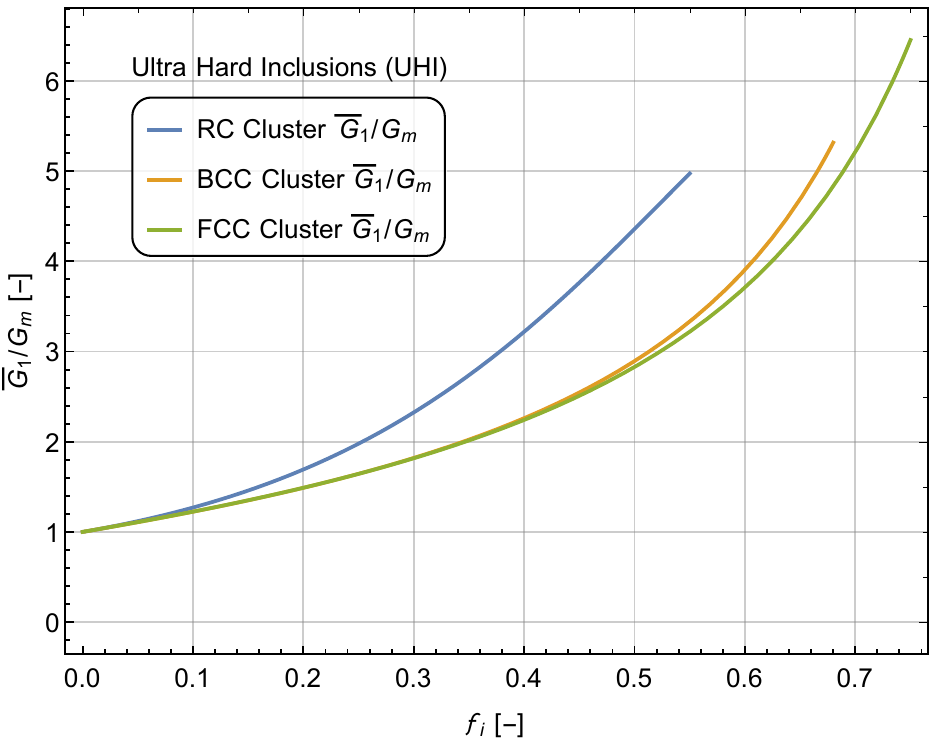}& 
\includegraphics[width=7.68cm]{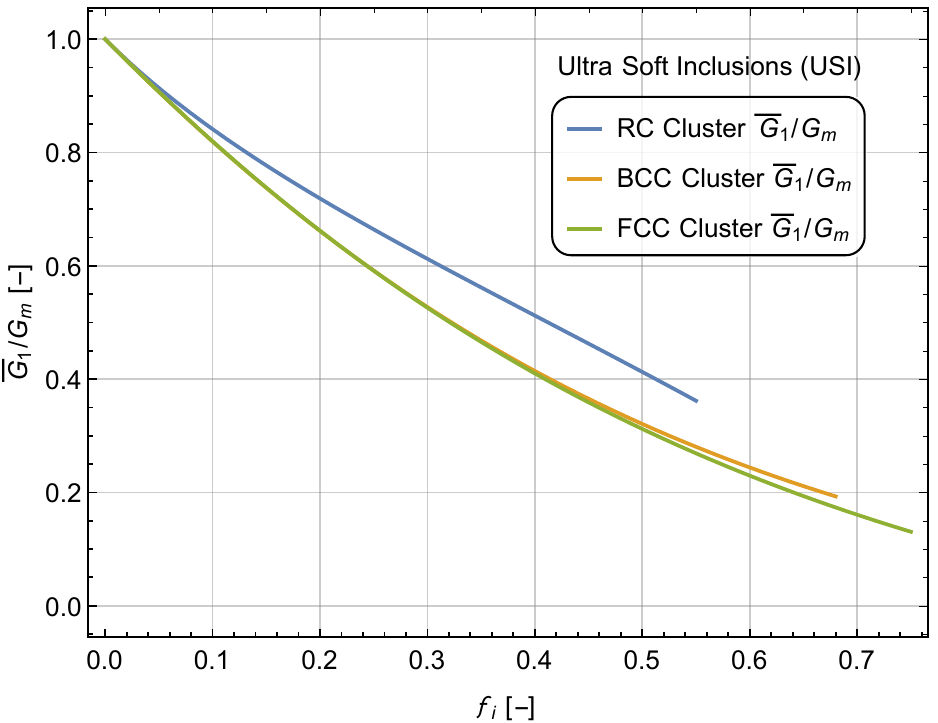}
\end{tabular}
\caption{Comparison of the effective first shear moduli of the RC, BCC and FCC arrangements for: a) ultra hard inclusions (UHI) and b) ultra soft inclusions (USI).}
\label{Fig:RCvBCCvFCC_uhiusi_G1}
\end{figure}

\begin{figure}[ht]
\centering
\begin{tabular}{cc}
	(a)&(b)\\
\includegraphics[width=7.5cm]{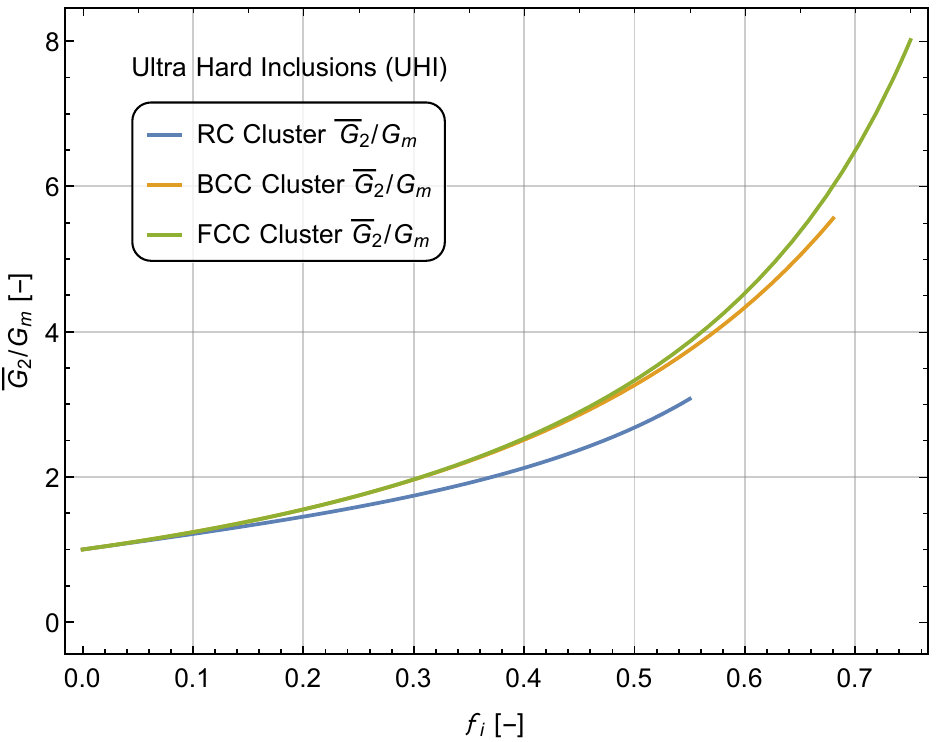} &
\includegraphics[width=7.68cm]{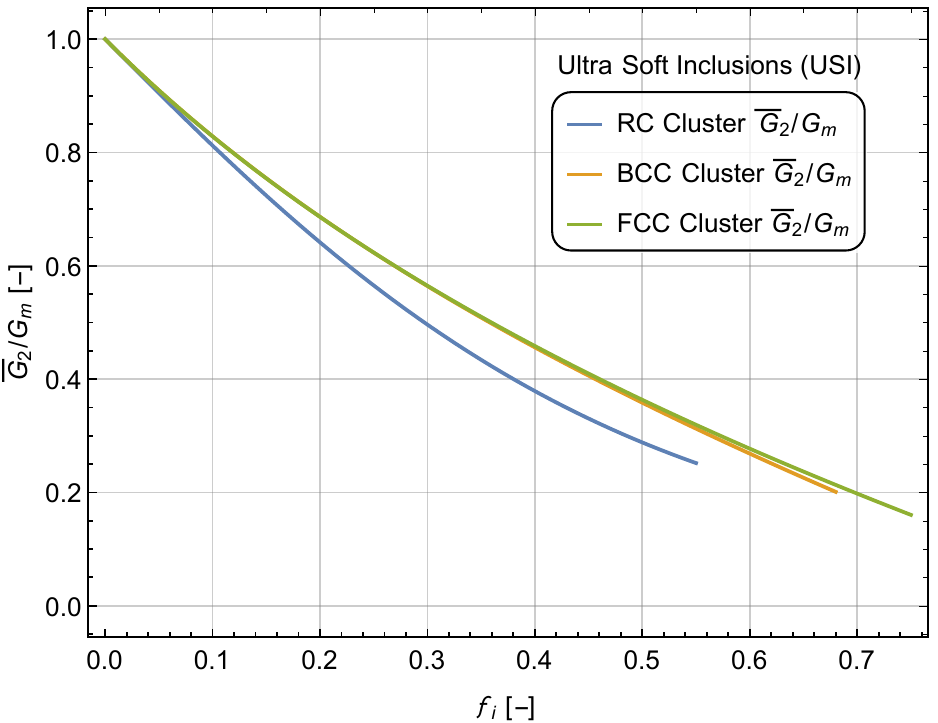}
\end{tabular}
\caption{Comparison of the effective second shear moduli of the RC, BCC and FCC arrangements for: a) ultra hard inclusions (UHI) and b) ultra soft inclusions (USI).}
\label{Fig:RCvBCCvFCC_uhiusi_G2}
\end{figure}

As previously discussed, the effective bulk modulus $\bar{K}$ obtained by the cluster model is identical for all three arrangements. On the other hand, the predicted shear moduli $\bar{G_1}$ and $\bar{G_2}$ show clear differences between the three arrangements, especially when RC is compared with BCC and FCC. The values for the two latter ones are relatively close to each other. For the same volume fraction of inclusions, the RC arrangement is stiffer than the other arrangements in terms of $\bar{G_1}$ and less stiff in terms of $\bar{G_2}$.  
}

\subsection{Elastic-plastic regime}\label{SubSec:RegularPlastic}
\paragraph{Deviatoric loading} In this part, the boundary conditions used in the elastic-plastic regime are described by the global deformation tensors $\bar{\boldsymbol{\varepsilon}}$ with the following representations in the basis coaxial with the unit cell's edges:
\begin{equation} \label{Eq:isochoric}
{{\bar{\varepsilon}}_{ij}^{(0,0,1)}=E_{\eq} \, 
\left(\begin{array}{ccc} -1/2 & 0& 0 \\ 0 & -1/2 & 0 \\ 0 & 0 & 1 \end{array}\right)
\,,\quad
{\bar{\varepsilon}}_{ij}^{(1,1,1)}=E_{\eq} \,  
\left(\begin{array}{ccc} 0 & 1/2& 1/2 \\ 1/2 & 0& 1/2  \\1/2 & 1/2& 0 \end{array}\right)
\,.}
\end{equation}
Note that such strain tensors correspond to isochoric extension in the {direction of a vector $\mathbf{v}$: $\bar{\boldsymbol{\varepsilon}}=\frac{1}{2}E_{\eq}\,(3\mathbf{n}\otimes\mathbf{n}-\mathbf{I})$, where $\mathbf{n}=\mathbf{v}/\|\mathbf{v}\|$.} For $\bar{\varepsilon}_{ij}^{(0,0,1)}$, $\mathbf{v}$ is oriented along the unit cell's edge, while for $\bar{\varepsilon}_{ij}^{(1,1,1)}$ along the cell's main diagonal. For the unit cell of cubic symmetry these strain processes result in a proportional deviatoric overall stress, namely: $\bar{\boldsymbol{\sigma}}\sim\bar{\boldsymbol{\varepsilon}}$. 

The results which follow concern a composite material made of a metal matrix reinforced with ceramic particles or weakened by spherical voids.
It is assumed that the metallic phase is elastic-plastic with a power law of isotropic hardening, while the ceramic phase is an ideally elastic material.
The material parameters of the composite's phases are listed in Table~\ref{Tab:MaterialParameters}.
\begin{table}[!h]
\centering
\begin{tabular}{lccc}
\hline
Phase & Binder : Metal matrix & Filler 1 : Ceramic & Filler 2 : Voids
\\
\hline
\textup{Young's modulus,} $E$ [GPa] & 75 & 400 & -
\\
\textup{Poisson's ratio,} $\nu$ & 0.3 & 0.2 & -
\\
\textup{Initial yield stress,} $Y_0$ [MPa] & 75 & - & -
\\
\textup{Plastic modulus,} $h$ [MPa] & 416 & - & -
\\
\textup{Exponent}, $n$ & 0.3895 & - & -
\\
\hline
\end{tabular}
\caption{The assumed material parameters of an elastic-plastic metal matrix reinforced with elastic ceramic particles (see also \citep{PonteCastaneda98}) or weakened by voids. \label{Tab:MaterialParameters}}
\end{table}
\begin{figure}[H]
	\centering
	\begin{tabular}{cc}
		(a)&(b)\\
		\includegraphics[angle=0,width=0.45\textwidth]{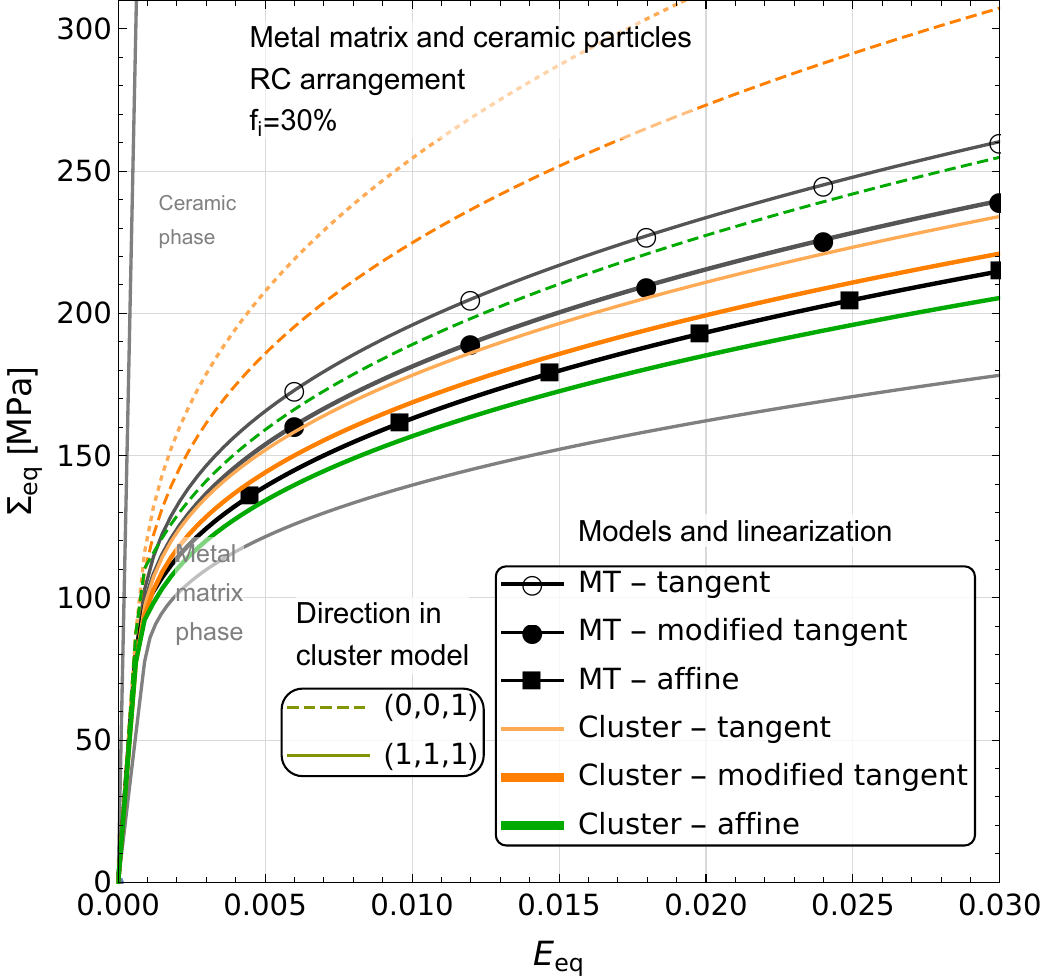}&
		\includegraphics[angle=0,width=0.45\textwidth]{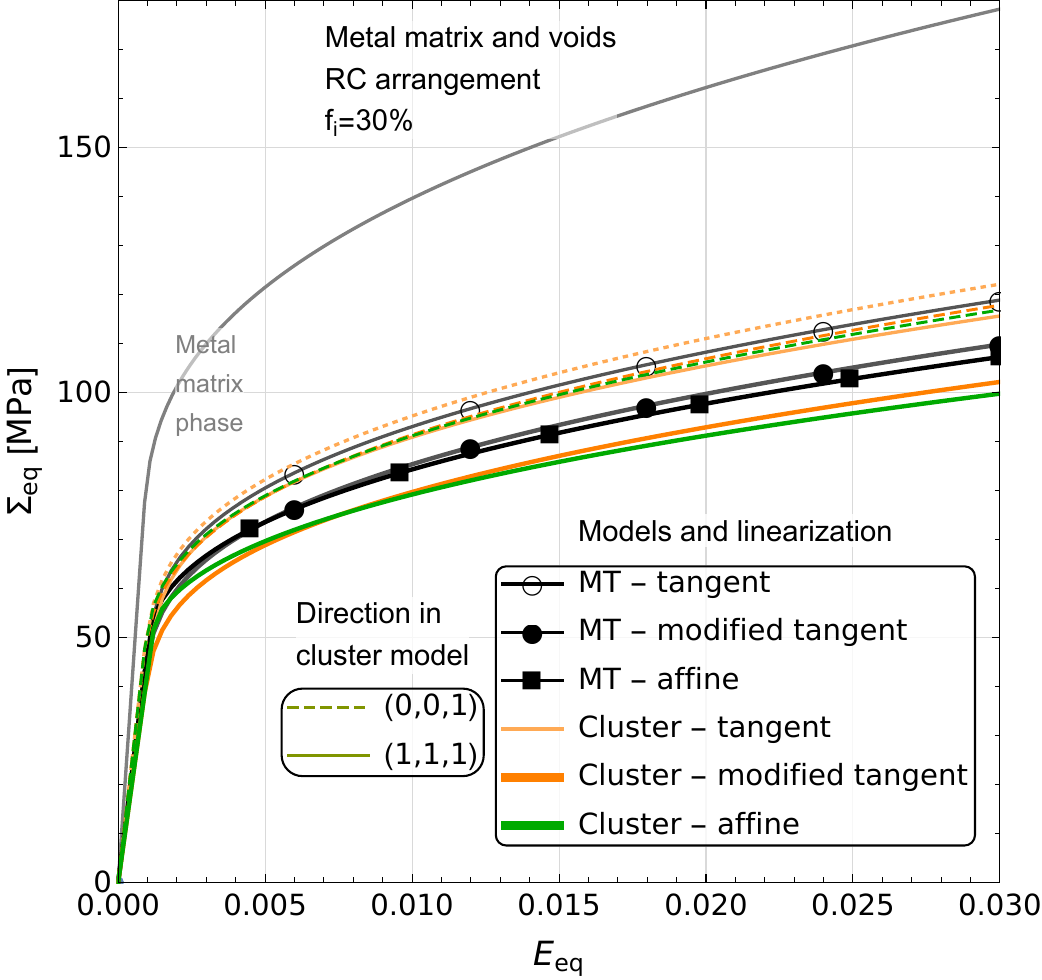}\\
		(c)&(d)\\
		\includegraphics[angle=0,width=0.45\textwidth]{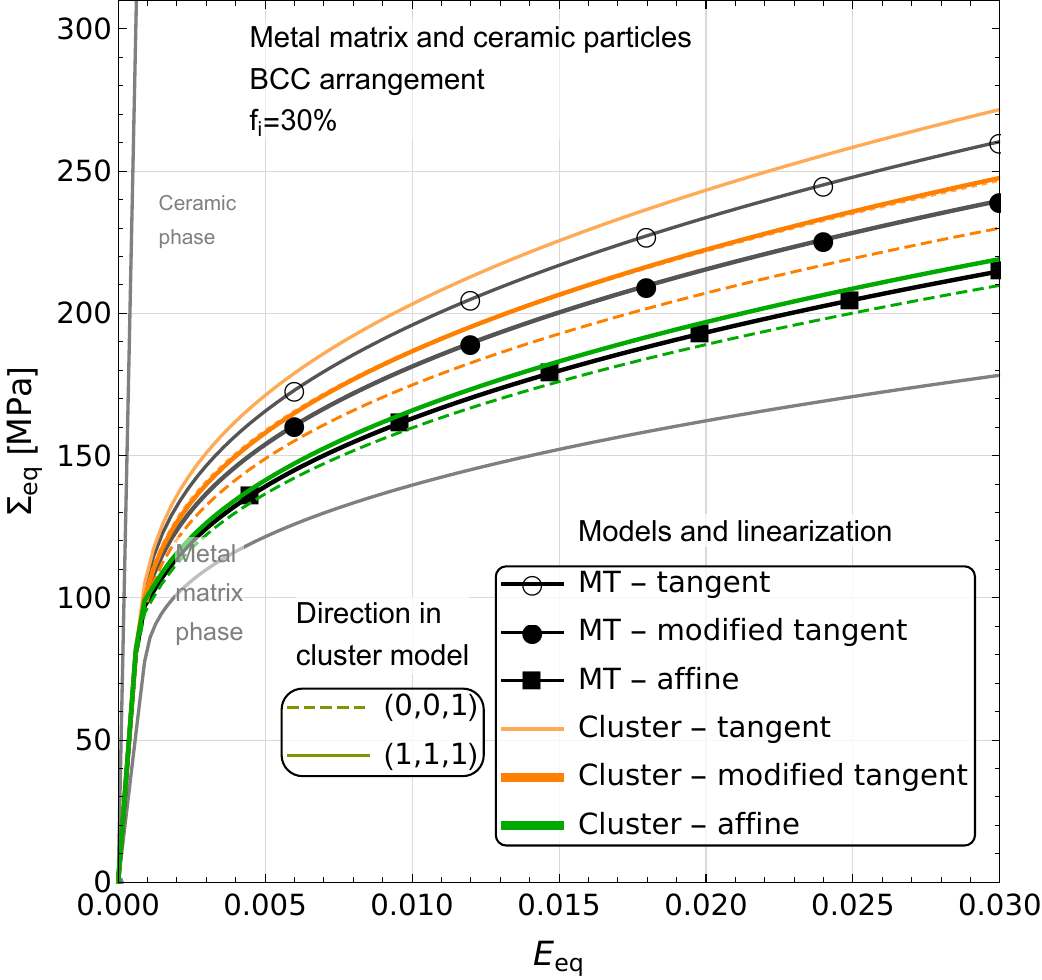}&
		\includegraphics[angle=0,width=0.45\textwidth]{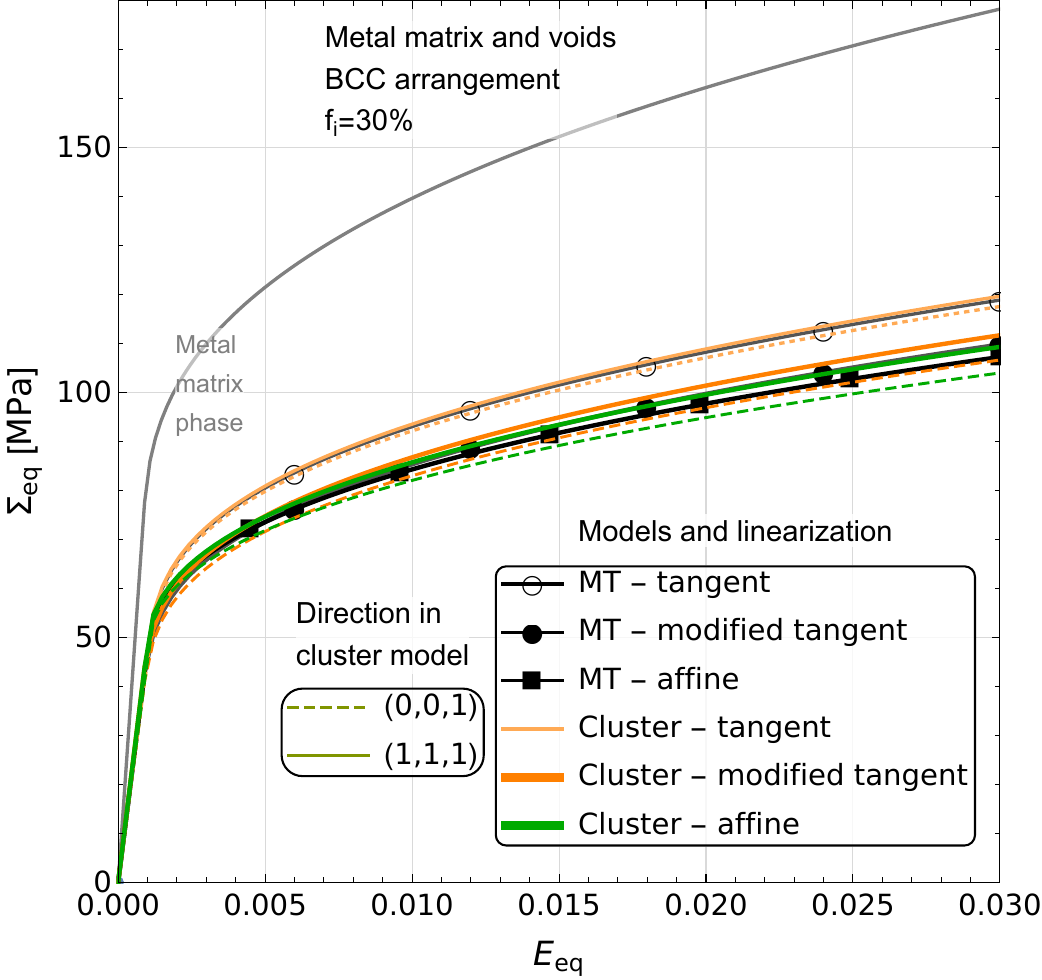}
	\end{tabular}
	\caption{
		Analytical estimations of the elastic-plastic response of the metal matrix: (a,c) reinforced with ceramic particles, (b,d) weakened by voids (Table~\ref{Tab:MaterialParameters}), for
		$f_{i}=30\%$ and (a,b) regular cubic (RC), (c,d) BCC arrangements of inclusions. 
		Macroscopic Huber-von Mises equivalent stress $\Sigma_{\eq}$ vs. equivalent strain $E_{\eq}$.
		Tangent, modified tangent and affine linearization of the micromechanical models:
		MT - the Mori-Tanaka model,
		Cluster - the cluster model presented in this paper.
		The results for the cluster model are for macroscopic isochoric extension tests in two extension directions $\mathbf{v}$: $(0,0,1)$ (dashed lines) and $(1,1,1)$ (solid lines)---Eq.~\eqref{Eq:isochoric}.
		\label{Fig:2_2_MisesStrain_SecTan_MicrMod_RC}}
\end{figure}

In Section~\ref{SubSec:Linearisation}, we presented three methods of linearization: the tangent, the modified tangent and the affine method.
Fig.~\ref{Fig:2_2_MisesStrain_SecTan_MicrMod_RC} shows the influence of the linearization scheme on the effective behaviour of the composite estimated using the classical Mori-Tanaka approach (MT) and the studied cluster model.
We investigated two composite materials (Table~\ref{Tab:MaterialParameters}): a metal matrix reinforced with ceramics (Fig.~\ref{Fig:2_2_MisesStrain_SecTan_MicrMod_RC}(a,c)) and a metal matrix with voids (Fig.~\ref{Fig:2_2_MisesStrain_SecTan_MicrMod_RC}(b,d)), undergoing an isochoric tension test. Inclusions in the materials are arranged in the RC (a,b) or BCC (c,d) lattice. As it is visible, contrary to the MT scheme, the cluster model captures the anisotropic response in two extreme directions $\mathbf{v}$ of the isochoric tension test, which are here $\mathbf{v}=(0,0,1)$ and $\mathbf{v}=(1,1,1)$ (Eq.~\eqref{Eq:isochoric}).

In the case of the metal matrix composite reinforced with ceramic particles (Fig.~\ref{Fig:2_2_MisesStrain_SecTan_MicrMod_RC}(a)), the linearization method has a strong impact on the estimated overall response of the material, especially on the cluster model's assessment for the isochoric tension along the unit cell's edge (denoted by (0,0,1) in the figure).
This is in contrast to the metal matrix with voids (Fig.~\ref{Fig:2_2_MisesStrain_SecTan_MicrMod_RC}(b)), where the linearization scheme has a much smaller visible influence.
The reason for this is the stiff ceramic particles, which play a crucial role in shaping the material's response in the isochoric tension test in the (0,0,1) direction for RC and in the (1,1,1) direction for the BCC arrangement.  It can be seen that in all cases the stiffest response is predicted by the standard tangent method, and the softest by the affine variant.
\begin{figure}[H]
	\centering
	\begin{tabular}{cc}
		(a)&(b)\\
		\includegraphics[angle=0,width=0.45\textwidth]{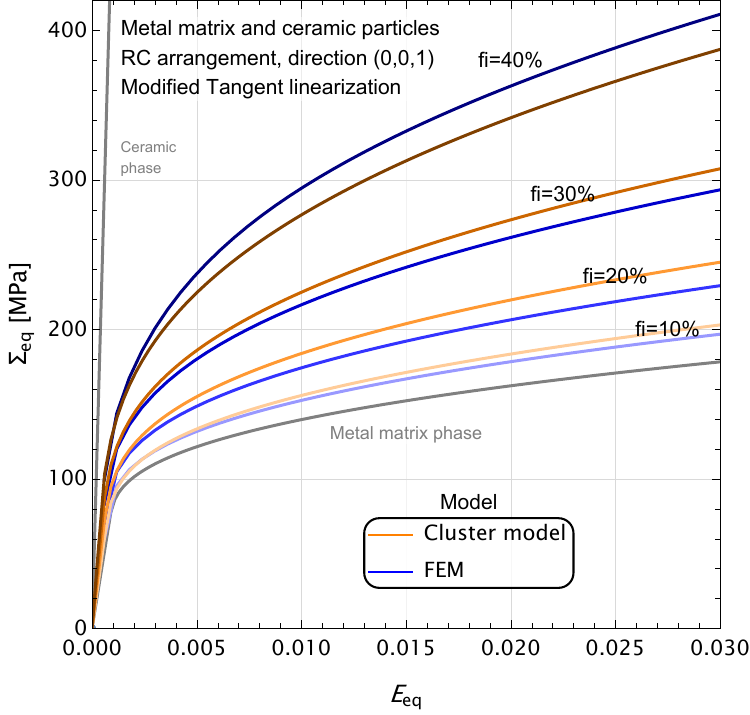}&
		\includegraphics[angle=0,width=0.45\textwidth]{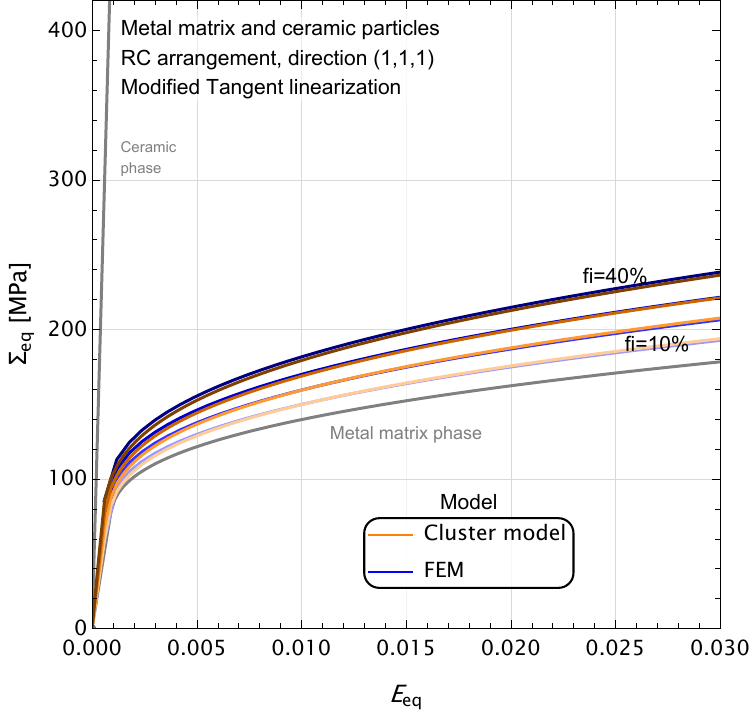}\\
		(c)&(d)\\
		\includegraphics[angle=0,width=0.45\textwidth]{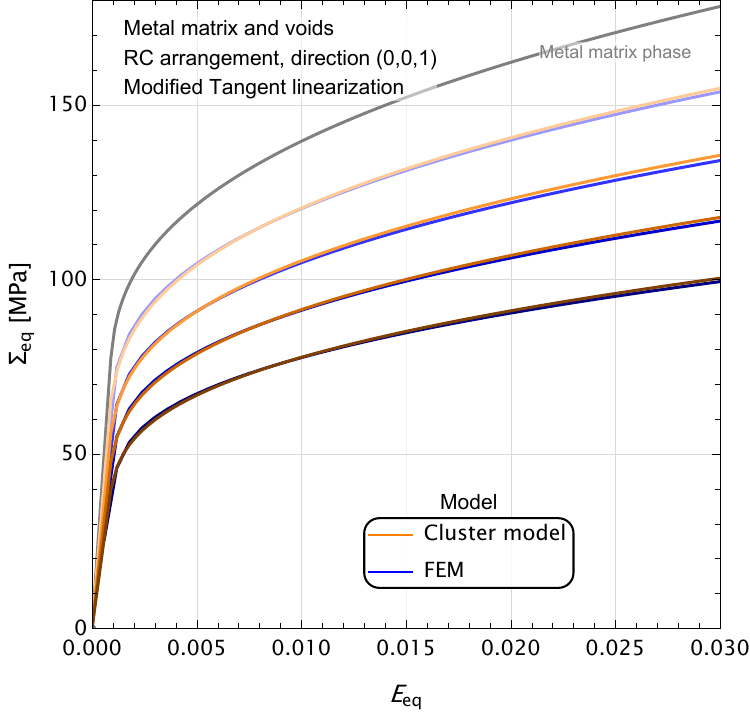}&
		\includegraphics[angle=0,width=0.45\textwidth]{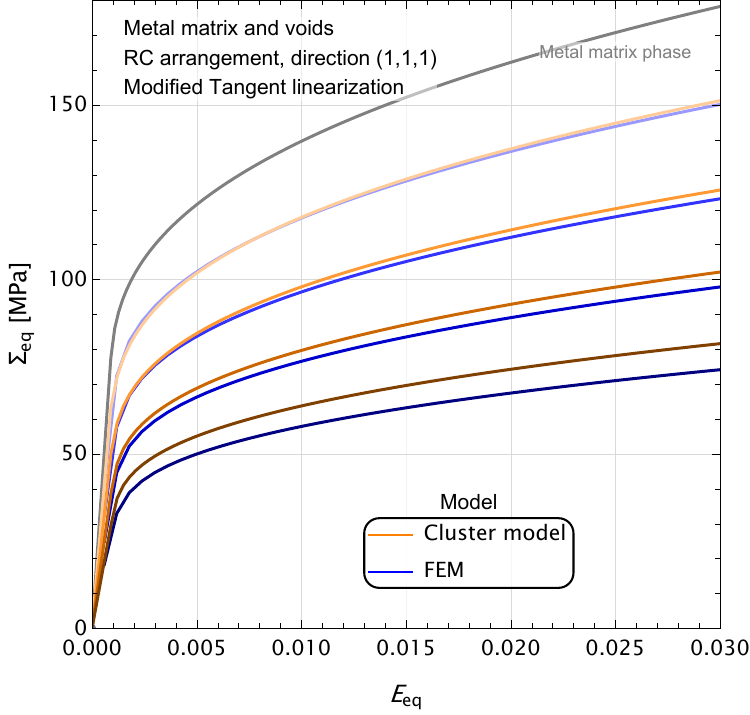}
	\end{tabular}
	\caption{
		The cluster model and numerical estimations (FEM) of the elastic-plastic response of (a,b) a metal matrix reinforced with ceramic inclusions, (c,d) a porous metal (Table~\ref{Tab:MaterialParameters}). 
		Huber-von Mises equivalent stress $\Sigma_{\eq}$ vs. equivalent strain $E_{\eq}$, obtained for
		the macroscopic isochoric extension test in two extension directions $\mathbf{v}$: (a,c) $(0,0,1)$ and (b,d) $(1,1,1)$.
		Modified tangent linearization of the cluster model was used. $f_{\rm{i}}$ equals $10\%,20\%,30\%,40\%$ and inhomogeneity arrangements are of the RC type.
		\label{Fig:4_1_Mises_Strain_fi_10_40}}
\end{figure}
As a means of verification, the results of micromechanical modelling (Cluster model with three variants of linearization) were compared with the results of numerical homogenization (FEM). We observed that the cluster model with the standard (incremental) tangent linearization and the affine method usually overestimate and underestimate the material response, respectively. Therefore, solely the results for \emph{the modified tangent linearization method} will be demonstrated in the rest of the analysis. This method has an additional advantage compared to the two remaining methods based on the first moments, as it enables one to predict matrix yielding under overall hydrostatic stress. This property will be shown in the next part of this subsection.

In Fig.~\ref{Fig:4_1_Mises_Strain_fi_10_40}, we compare the effect of inclusion volume fraction ($f_{\textup{i}}$ from 10\% to 40\%) on the effective response of the MMC and PM materials with RC inclusion/void arrangements.
Figs.~\ref{Fig:4_1_Mises_Strain_fi_10_40}(a,c) and (b,d) show the results of an isochoric tension test (Eq.~\eqref{Eq:isochoric}) along the directions (0,0,1) and (1,1,1), respectively.
\begin{figure}[H]
	\centering
	\begin{tabular}{cc}
		(a)&(b)\\
		\includegraphics[angle=0,width=0.45\textwidth]{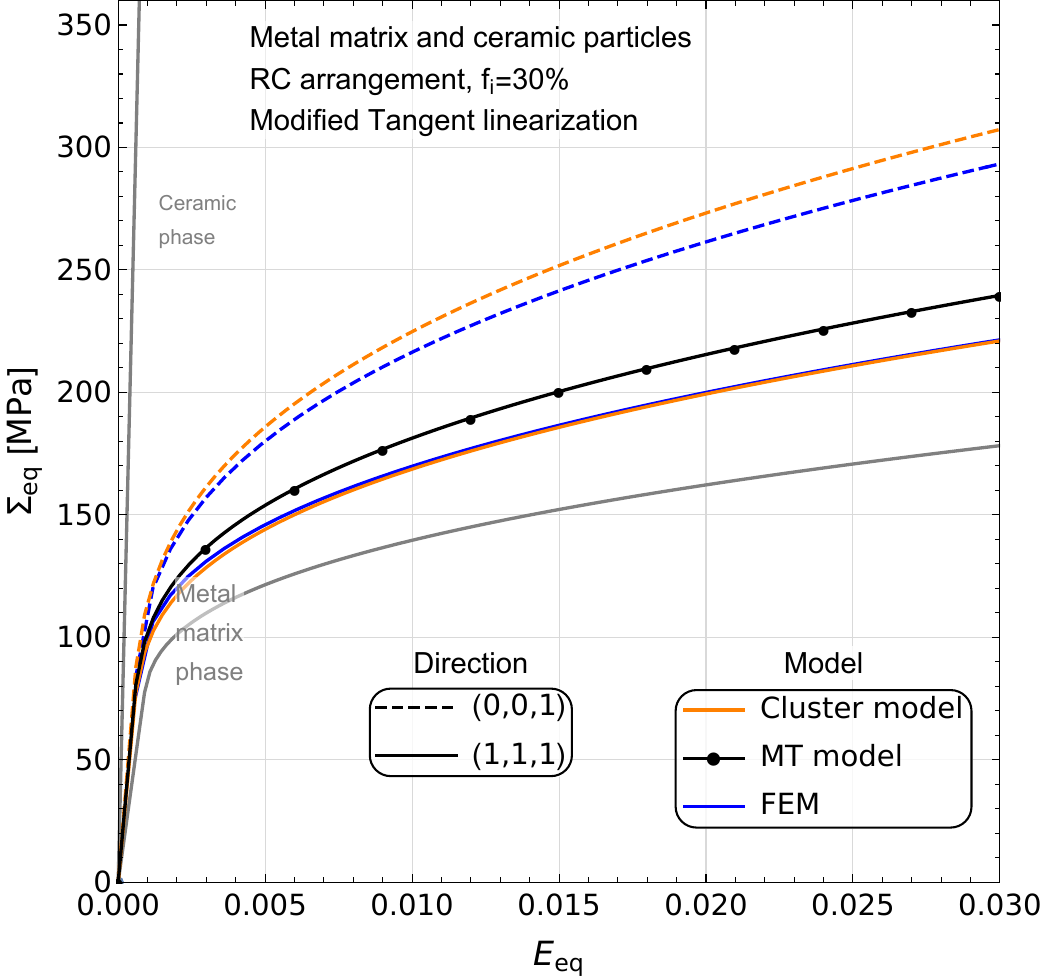}&
		\includegraphics[angle=0,width=0.45\textwidth]{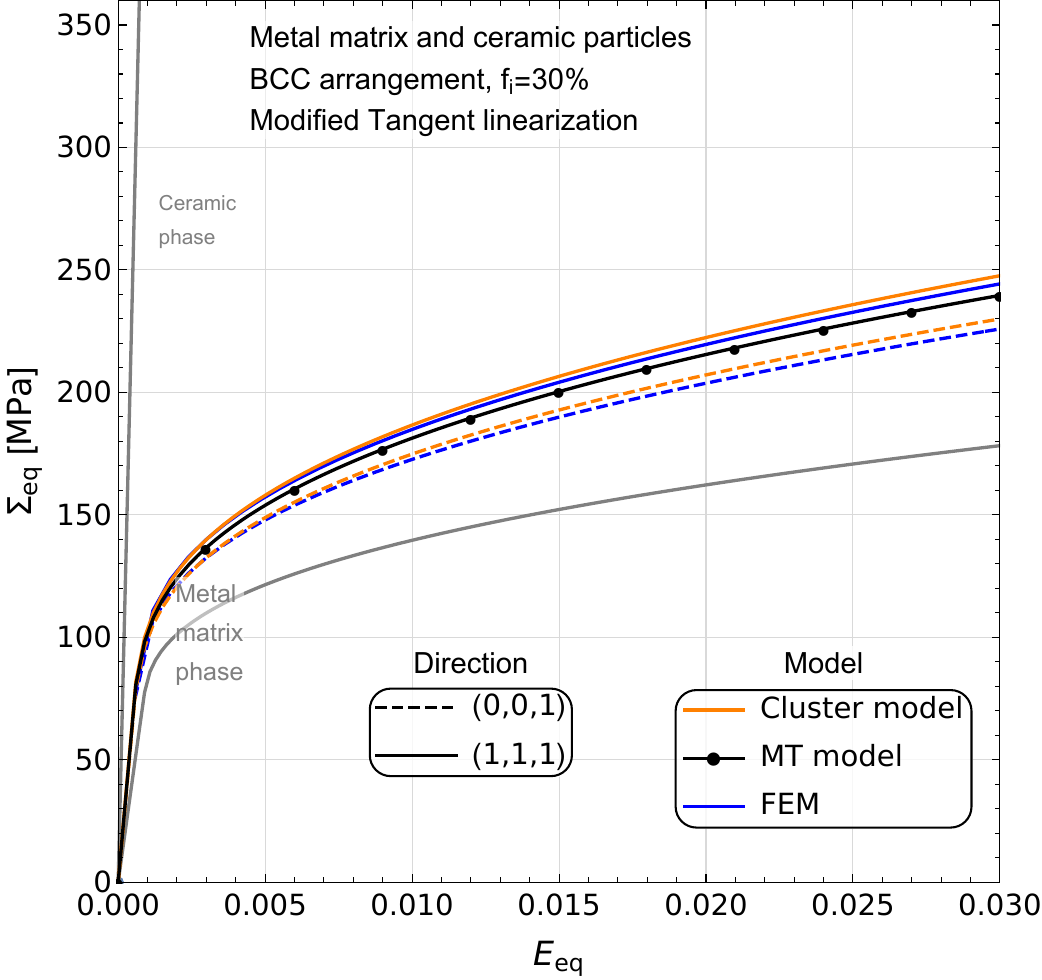}\\
		(c)&(d)\\
		\includegraphics[angle=0,width=0.45\textwidth]{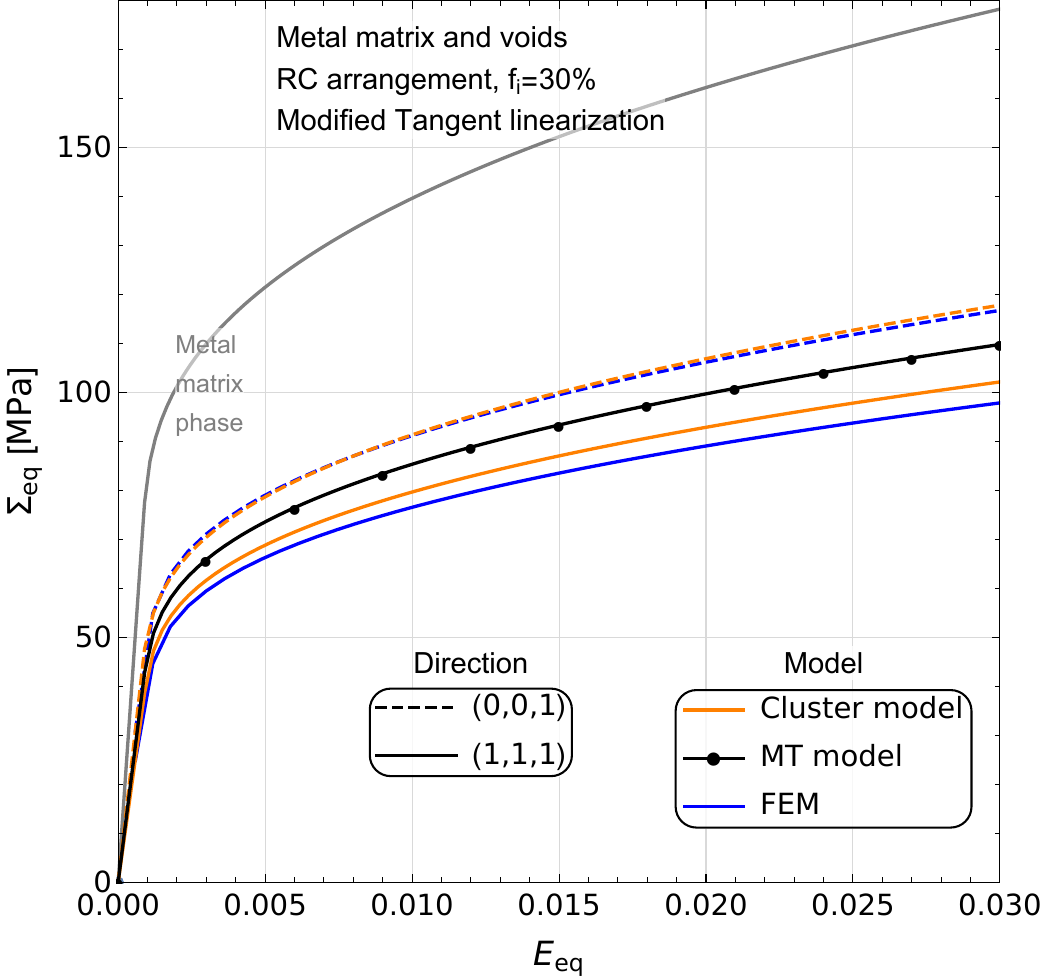}&
		\includegraphics[angle=0,width=0.45\textwidth]{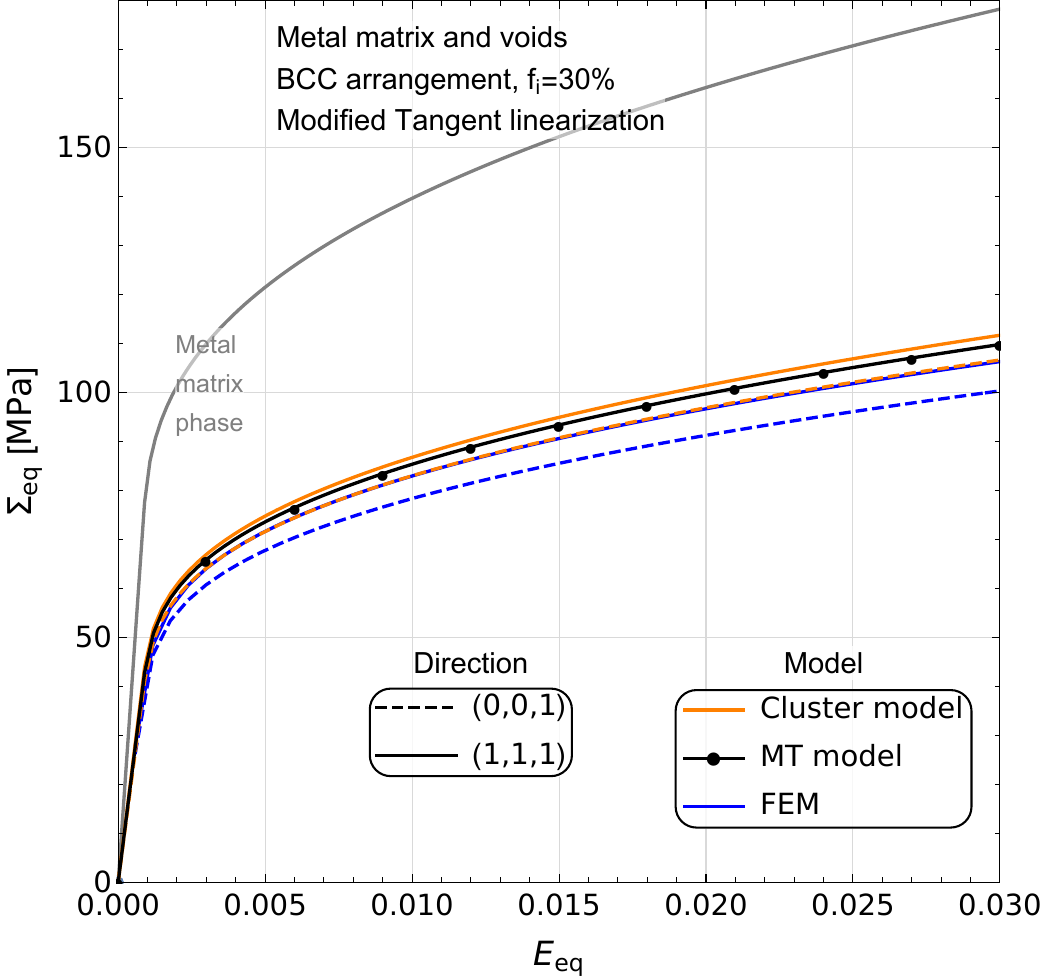}
	\end{tabular}
	\caption{
		Analytical and numerical estimations (FEM) of the elastic-plastic response of the metal matrix with: ceramic inclusions (a,b) and voids (c,d). We assumed $f_{\rm{i}}=30\%$ (Table~\ref{Tab:MaterialParameters}) and regular inclusion arrangements of type RC (a,c) and BCC (b,d). 
		Huber-von Mises equivalent stress $\Sigma_{\eq}$ vs. equivalent strain $E_{\eq}$, for the macroscopic isochoric extension test in two extension directions $\mathbf{v}$: $(0,0,1)$ and $(1,1,1)$.
		Modified tangent linearization was used for the mean-field cluster and MT  models.
		\label{Fig:4_1_Mises_Strain_fi_30_Tan_MES_MMCvoids_RCBCC}}
\end{figure}
The results of computational homogenization and the cluster mean-field model almost coincide for the direction (1,1,1) for {the MMC material}. For the stiff direction (0,0,1), the mean-field model slightly overestimates the FEM reference results for {$f_{\ii}=$10\%--30\%} (the dark-blue FEM lines are below the orange ones for the cluster model in Fig.~\ref{Fig:4_1_Mises_Strain_fi_10_40}) and underestimates for the 40\% volume fraction (the blue FEM line is over the dark-orange one).
In general, the cluster model estimates are qualitatively and quantitatively similar to the results obtained by FEM.
Both Figs.~\ref{Fig:4_1_Mises_Strain_fi_10_40}(a) and (b) have the same plot range.
By comparing the two figures, one can observe the strong impact of the tension direction; here, both methods: the cluster model and FEM, are in good agreement in estimating this anisotropic response.
Similar results are observed for the metal matrix weakened by voids in the RC arrangements demonstrated in Figs.~\ref{Fig:4_1_Mises_Strain_fi_10_40}(c,d) where an even better quantitative agreement between the cluster model and the reference FEM results is observed. Note that the weakening of the material caused by the presence of voids is more pronounced for the (1,1,1) loading direction. 

\begin{figure}[H]
	\centering
	\begin{tabular}{cc}
		(a)&(b)\\
		\includegraphics[angle=0,width=0.45\textwidth]{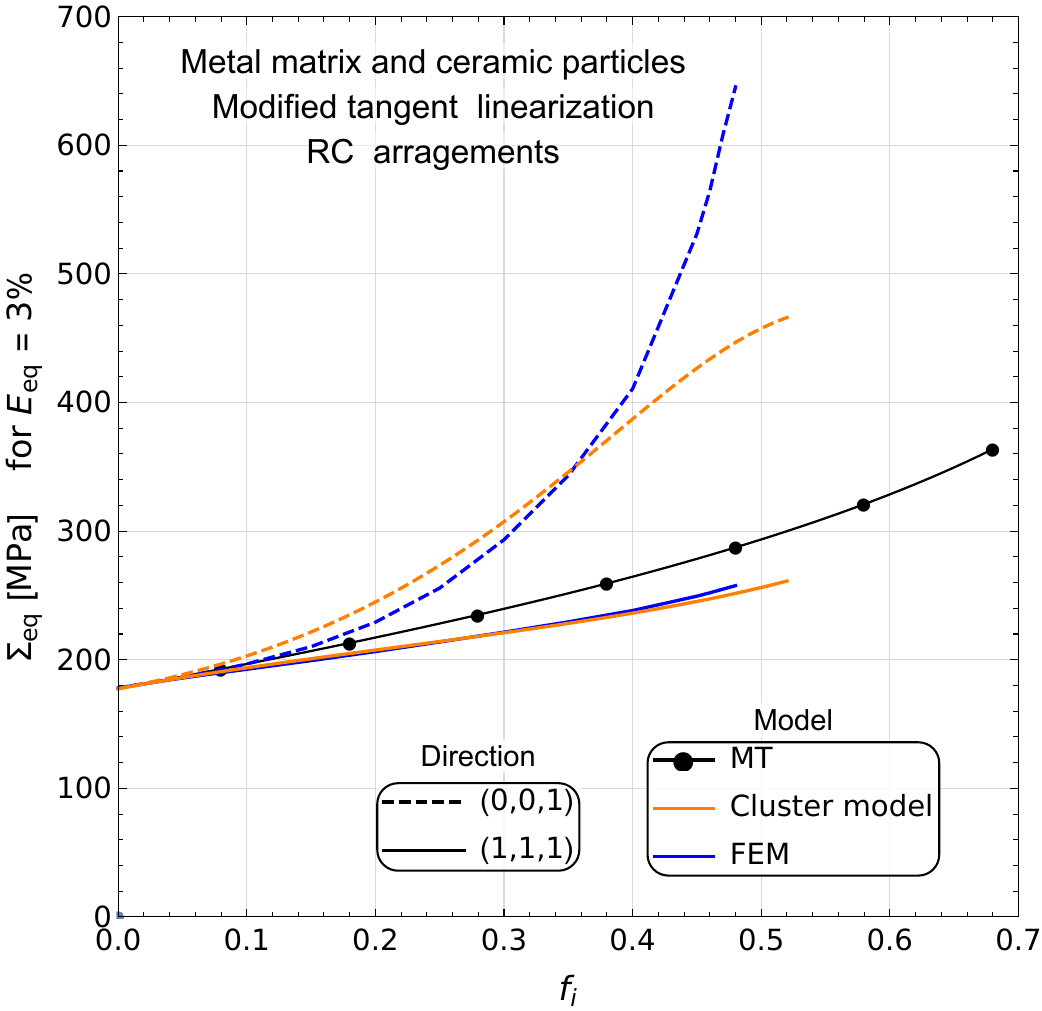}&
		\includegraphics[angle=0,width=0.45\textwidth]{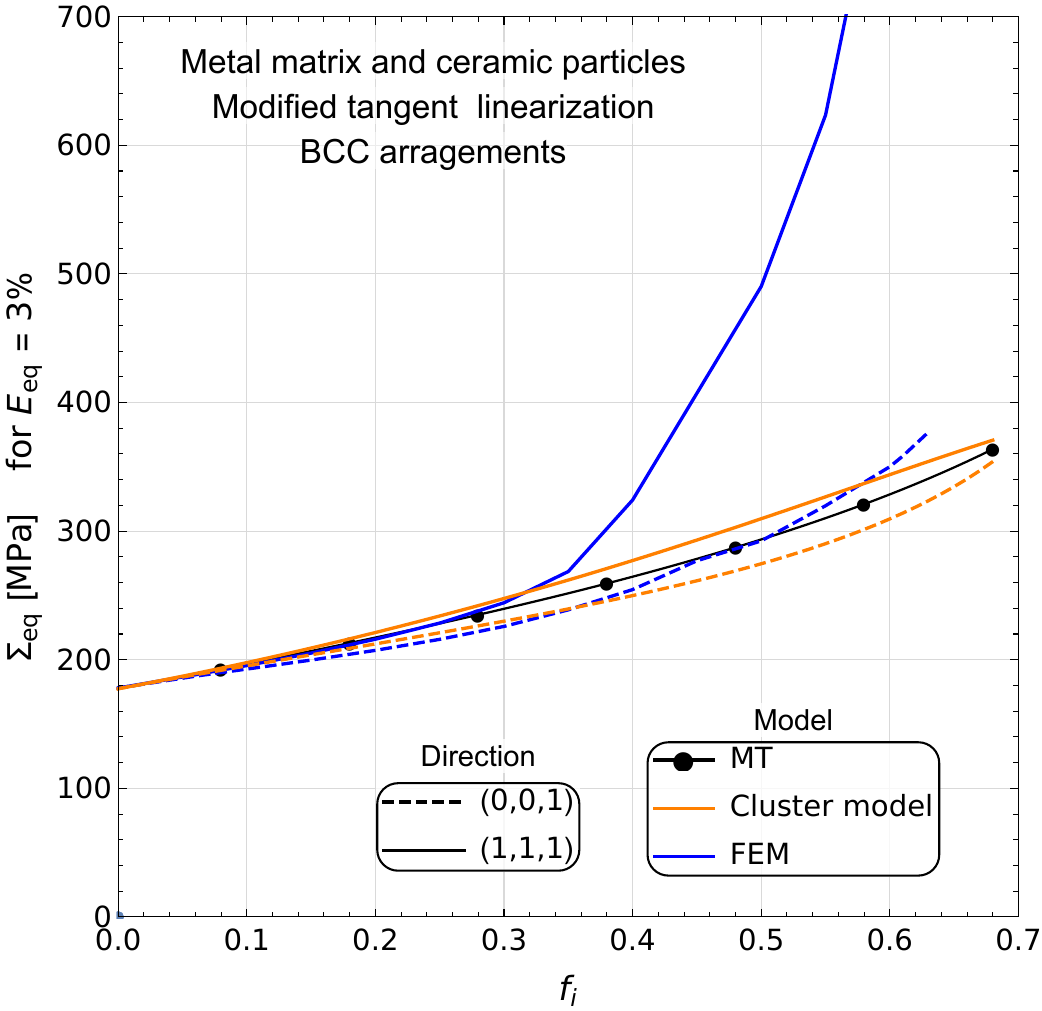}\\
		(c)&(d)\\
		\includegraphics[angle=0,width=0.45\textwidth]{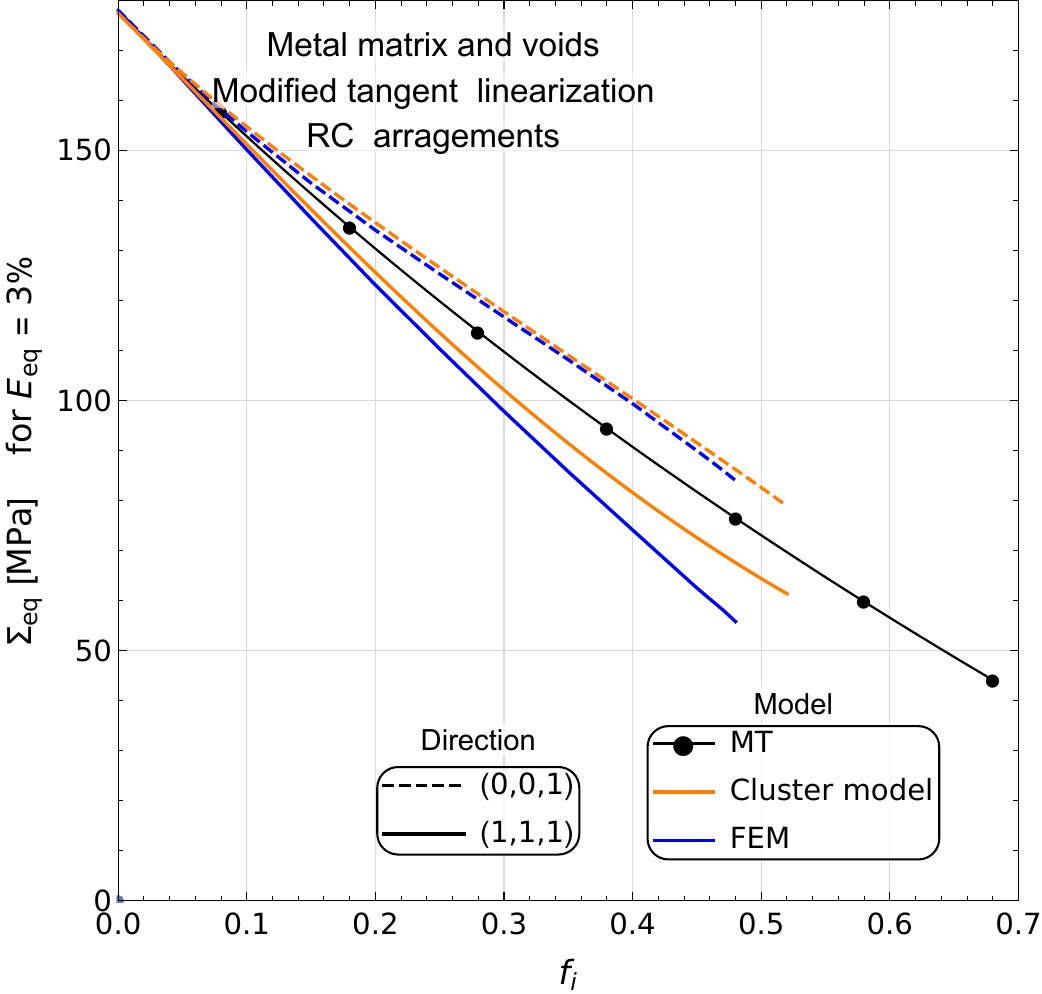}&
		\includegraphics[angle=0,width=0.45\textwidth]{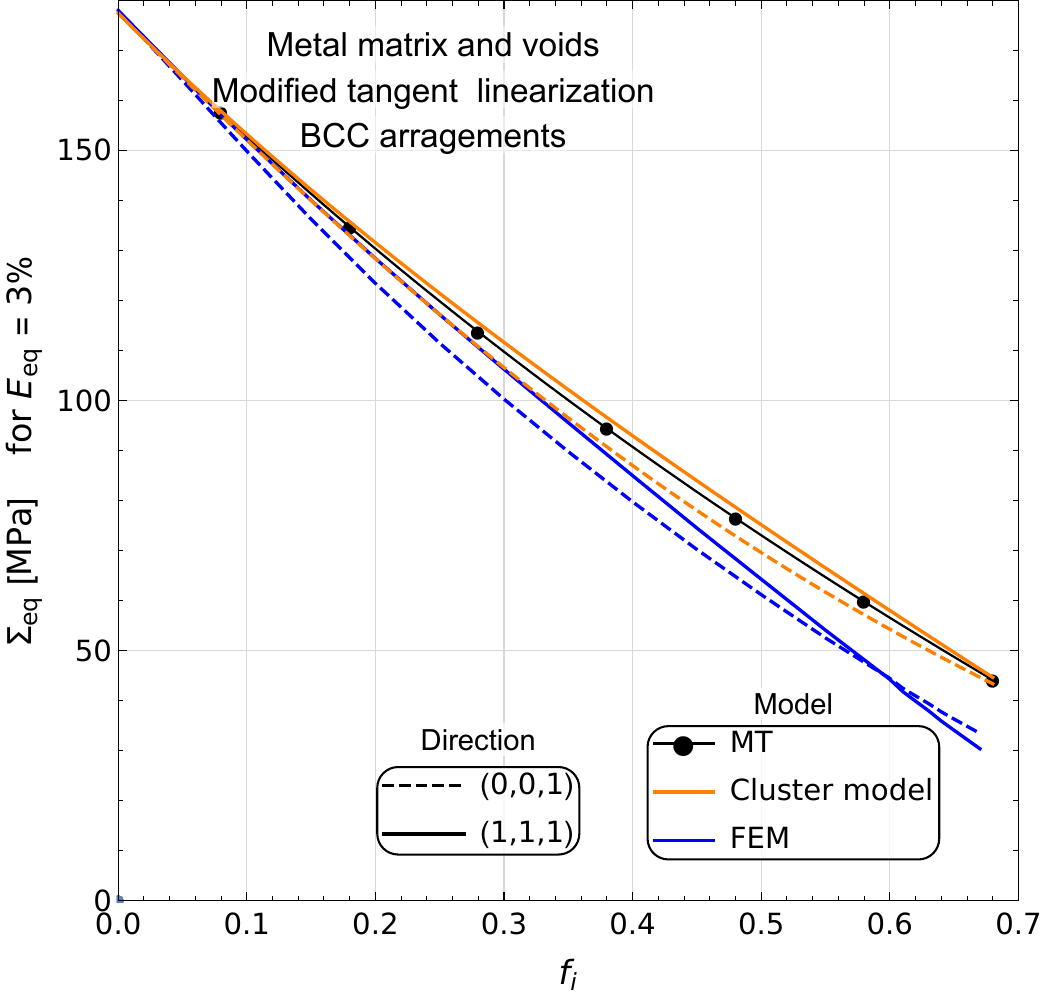}
	\end{tabular}
	\caption{
		Analytical and numerical estimations (FEM) of the elastic-plastic response of the metal matrix: (a,b) reinforced with ceramic inclusions, (c,d) weakened by voids (Table~\ref{Tab:MaterialParameters}), with regular inclusion arrangements of types RC and BCC. 
		Huber-von Mises equivalent stress $\Sigma_{\eq} \left( E_{\eq}=3\% \right)$ vs. volume fraction of inclusions $f_{\ii}$, for the macroscopic isochoric extension test in two extension directions $\mathbf{v}$: $(0,0,1)$ and $(1,1,1)$.
		Tangent linearization of the micromechanical models was used.
		MT - the Mori-Tanaka model,
		Cluster model - the cluster model.
		\label{Fig:4_1_Mises003_vs_fi_Tan}}
\end{figure}
Let us now focus on the dependence of the nonlinear response of the composite material on the stretching direction $\mathbf{v}$ (Eq.~\eqref{Eq:isochoric}).
Fig.~\ref{Fig:4_1_Mises_Strain_fi_30_Tan_MES_MMCvoids_RCBCC} shows the effective response of a composite material with ceramic inclusions (a,b) and voids (c,d).
The inclusion arrangements are: RC (Figs.~\ref{Fig:4_1_Mises_Strain_fi_30_Tan_MES_MMCvoids_RCBCC}(a,c)) and BCC (Figs.~\ref{Fig:4_1_Mises_Strain_fi_30_Tan_MES_MMCvoids_RCBCC}(b,d)).
In the RC system in Figs.~\ref{Fig:4_1_Mises_Strain_fi_30_Tan_MES_MMCvoids_RCBCC}(a,c), the influence of the stretching direction on the effective response of the material is more visible than in the case of the BCC lattice.
It is worth noting that, no matter if inhomogeneities are stiffer or softer than the matrix material, the RC system manifests a stiffer response in the direction (0,0,1) than along (1,1,1) (Figs.~\ref{Fig:4_1_Mises_Strain_fi_30_Tan_MES_MMCvoids_RCBCC}(a,c)).
In the case of the BCC system, the relationship is reversed (Figs.~\ref{Fig:4_1_Mises_Strain_fi_30_Tan_MES_MMCvoids_RCBCC}(b,d)). Those observations are true for both the FEM and the cluster model predictions. Note that the Mori-Tanaka model provides the same response for both unit cells and is not sensitive to the loading direction. The predictions of MT are always between the cluster estimates for the two extreme directions.  

To summarize the influence of the volume fraction of inclusions on the effective behaviour of the composite, Fig.~\ref{Fig:4_1_Mises003_vs_fi_Tan} presents the Mises stress $\sigma_{\textup{Mises}}$ as a function of the volume fraction of inclusions $f_{\ii}$.
The Mises stress in Fig.~\ref{Fig:4_1_Mises003_vs_fi_Tan} is the equivalent Huber-von Mises stress in the composite material measured at 0.03 equivalent strain, $\sigma_{\textup{Mises}}(E_{\eq}=3\%)$, in an isochoric tension test in two directions.
The placement of inclusions used in Figs.~\ref{Fig:4_1_Mises003_vs_fi_Tan}(a,c) and \ref{Fig:4_1_Mises003_vs_fi_Tan}(b,d) is RC and BCC, respectively.
Figs.~\ref{Fig:4_1_Mises003_vs_fi_Tan}(a,b) correspond to MMC, while Figs.~\ref{Fig:4_1_Mises003_vs_fi_Tan}(c,d) to a metal matrix with voids (PM) (Table~\ref{Tab:MaterialParameters}).
The results of the standard MT model with the modified tangent linearization have been added for comparison.
In all four graphs in Fig.~\ref{Fig:4_1_Mises003_vs_fi_Tan}, the influence of the stretching direction on the effective behaviour of the composite is visible, both in FEM simulations and mean-field modelling using the cluster model.
Therefore, the anisotropy induced by the spatial particle arrangements is an important factor which should be represented by the micromechanical model, especially when regular arrangements of inclusions are analyzed. As can be seen, the cluster model (orange line), which takes the effect of stretching direction into account, is in better compliance with FEM (blue line) than the classical  MT  model (black lines) which, as already seen in Fig.~\ref{Fig:4_1_Mises_Strain_fi_30_Tan_MES_MMCvoids_RCBCC} for 30\% volume fraction, lies between the cluster model predictions for the two considered directions. Similarly to the elastic regime, the agreement between the cluster scheme and the reference FEM results is satisfactory up to $\sim$40\% of the inclusion content. 
When the volume fraction of inclusions tends to its maximum value (around 0.52 for RC and 0.68 for BCC),
there is a significant stiffening of the MMC response, especially for the along-the-edge extension direction of the RC cell and diagonal direction of the BCC cell. 
The cluster model reflects this behaviour to some degree -- despite its systematic underestimation of the stress values at larger inclusion volume contents. 
{In the case of the metal matrix with voids in the RC arrangement, compared to MMC, a smaller influence of the extension direction on the effective response of the material is observed, both in the mean-field model's results and in FEM calculations.} Overall, a good quantitative agreement is observed between the two models, with some discrepancy seen only for the (1,1,1) direction and {larger values of} $f_{\ii}$. On the other hand, in the case of BCC the agreement between the qualitative character of the curves is preserved for the whole range of volume contents {(curves for the two directions initially diverge and then get closer again as $f_{\ii}$ approaches its maximum allowable value)}. For this case the cluster model slightly overestimates the reference FEM results.

\begin{figure}[H]
	\centering
	\includegraphics[angle=0,width=0.45\textwidth]{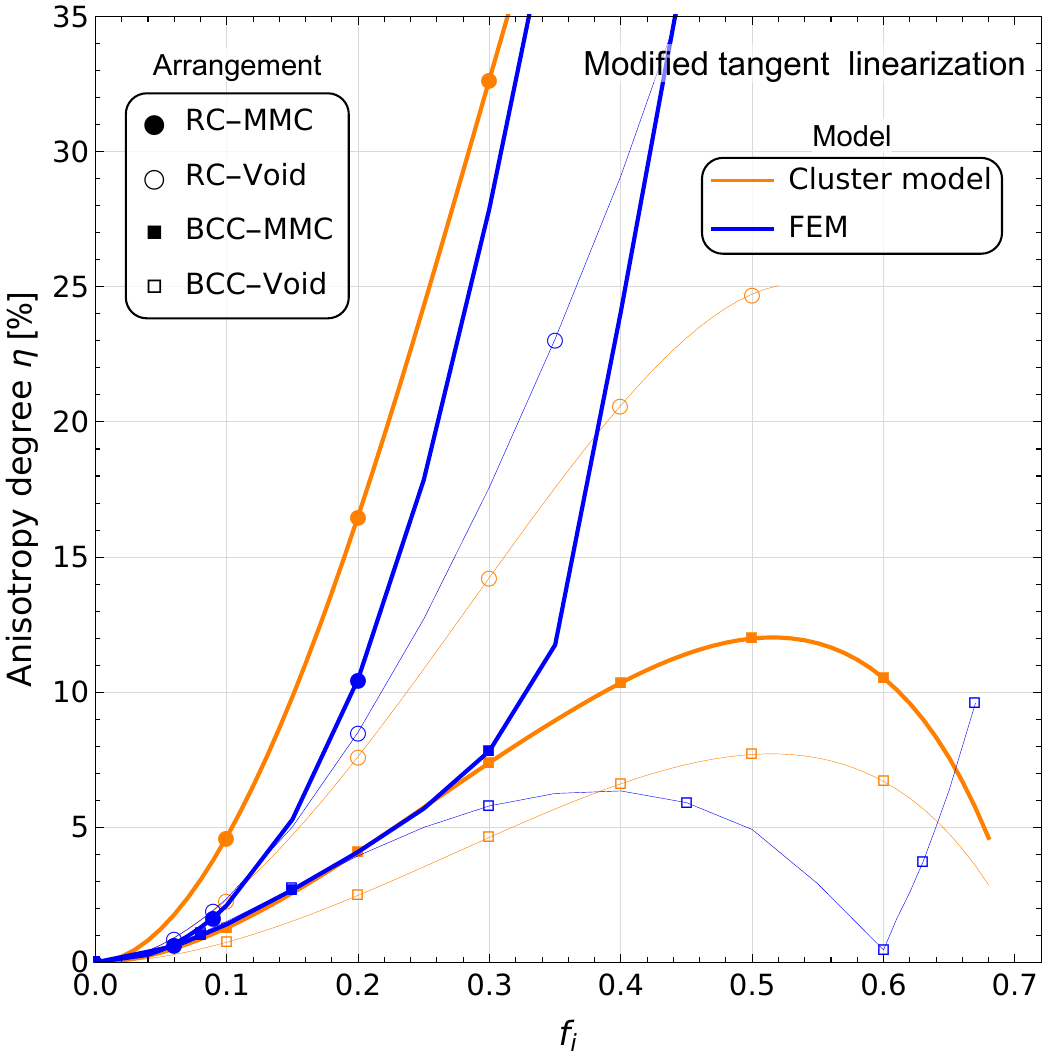}
	\caption{Anisotropy degree $\eta$ [\%], specified by Eq.~\eqref{anisotropyfactor}, of the Huber-von Mises equivalent stress $\Sigma_{\eq}$ for $E_{\eq}=3\%$ in the metal matrix composite with ceramic inclusions or voids in the RC or BCC arrangement undergoing a macroscopic isochoric tension test.
		\label{Fig:plast_anisotropy}}
\end{figure}

The presented analysis enables one also to assess the limit of validity of the isotropic approximation of the material response by the MT scheme for the case of regular particle or void arrangements. In Fig.~\ref{Fig:plast_anisotropy} we present the anisotropy factor defined as
\begin{equation}\label{anisotropyfactor}
\eta=\frac{2\left|{\Sigma}_{\eq}^{(100)}-{\Sigma}_{\eq}^{(111)}\right|}
	{{\Sigma}_{\eq}^{(100)}+{\Sigma}_{\eq}^{(111)}}100\%\,,
\end{equation}
calculated for all analyzed cases at 3\% strain in a given loading direction. The figure confirms that the RC inhomogeneity arrangement results in a higher anisotropy degree of the material than the BCC arrangement. Note that the character of the FEM curve for the BCC unit cell with voids changes when ${\Sigma}_{\eq}^{(100)}$ starts to be larger than ${\Sigma}_{\eq}^{(111)}$. Moreover, for $f_{\ii}$ below 0.1 for RC and below 0.2 for BCC, the anisotropy factor is smaller than 5\%, so the isotropic approximation is admissible for such inclusion volumes. This result is of practical importance. 
For example, the common methodology of numerical analysis of the void growth phenomenon is to use a cubic unit cell with a single cavity, under periodic boundary conditions, as a representative piece of material with a random distribution of voids, while in fact such a unit cell corresponds to their regular cubic (RC) distribution. The present analysis shows that such an approach is justified only for sufficiently small volume fractions of voids.   

\paragraph{Hydrostatic loading} In this part, the boundary conditions are described by the global deformation tensor $\bar{\boldsymbol{\varepsilon}}$ with the following representation in the basis coaxial with the unit cell's edges:
\begin{equation} \label{Eq:hydrostatic}
{\bar{\varepsilon}}_{ij}^{\rm{h}}={E}_{\rm{mean}} \, \, 
\left(\begin{array}{ccc} 1 & 0& 0 \\ 0 & 1 & 0 \\ 0 & 0 & 1 \end{array}\right)\,.
\end{equation}
Note that such a strain tensor corresponds to hydrostatic expansion $\bar{\boldsymbol{\varepsilon}}={E}_{\rm{mean}}\mathbf{I}$. For the unit cell of cubic symmetry, this strain process results in the hydrostatic overall stress, namely: $\bar{\boldsymbol{\sigma}}'=\mathbf{0}$. Such loading conditions are known to pose important difficulties for many mean-field models of metal matrix elastic-plastic composites \citep{NailiDoghri23}. In particular, those mean-field approaches which make use of mean values of stress (the first moment) to perform the `uniformization' step will predict a purely elastic response of the phases, even though plastic yielding of the metal matrix is observed in the full-field calculations. The problem of accounting for yielding initiation is overcome when the second stress moment is used in the `uniformization' step. However, the extent of plastic yielding may be not correctly estimated. This problem seems to be of prime importance when plastic yielding of {porous} metals is analyzed and when void growth phenomena are of interest.

\begin{figure}[H]
	\centering
	\begin{tabular}{cc}
		(a)&(b)\\
		\includegraphics[angle=0,width=0.45\textwidth]{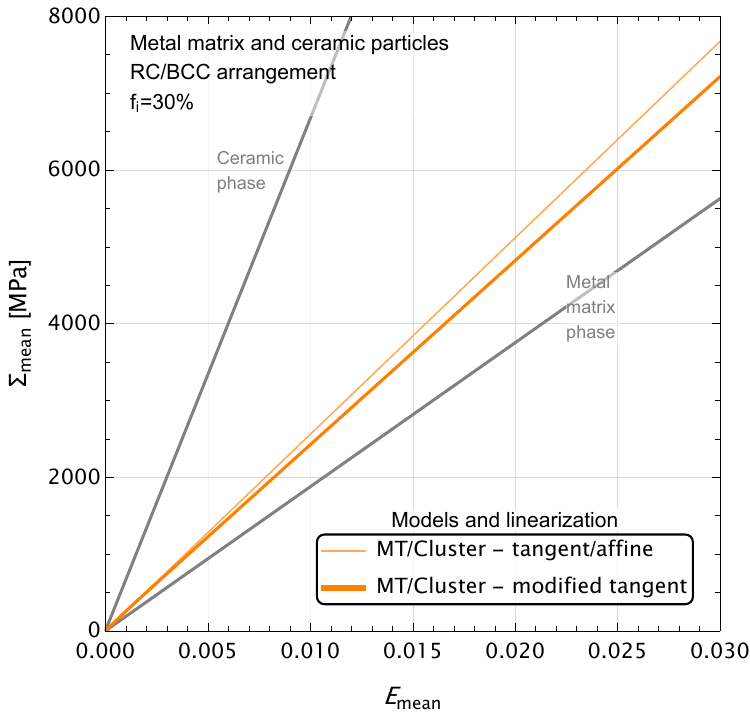}&
		\includegraphics[angle=0,width=0.45\textwidth]{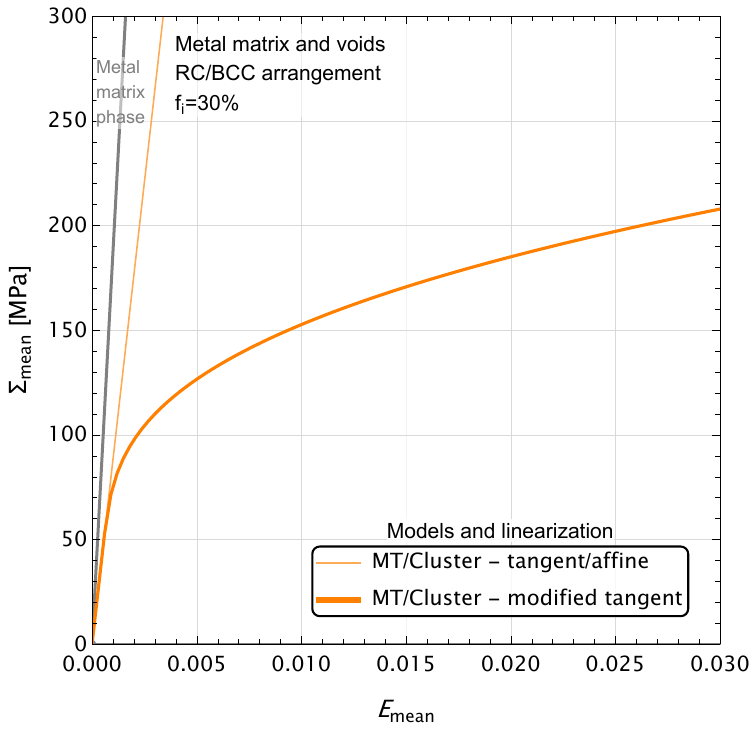}
	\end{tabular}
	\caption{
		Mean-field estimations of the elastic-plastic response of the metal matrix: (a) reinforced with ceramic particles, (b) weakened by voids (Table~\ref{Tab:MaterialParameters}), for
		$f_{\ii}=30\%$ and cubic arrangements of inclusions subject to hydrostatic extension---Eq.~\eqref{Eq:hydrostatic}. 
		The mean component of the overall stress ${\Sigma}_{\textup{mean}}=1/3\,\rm{tr}\,\bar{\boldsymbol{\sigma}}$ vs. the mean component of the overall strain ${E}_{\textup{mean}}=1/3\,\rm{tr}\,\bar{\boldsymbol{\varepsilon}}$.
		Tangent, modified tangent and affine linearization of the micromechanical models:
		MT - the Mori-Tanaka model,
		Cluster - the cluster model presented in this paper. {Predictions of the two models coincide for the same linearization method.}	
		\label{Fig:2_2_MeanStrain_SecTan_MicrMod_RC}}
\end{figure}

Fig.~\ref{Fig:2_2_MeanStrain_SecTan_MicrMod_RC} compares the predictions of the proposed variants of the cluster model for metal matrix composites and porous metals in terms of the mean component of the macroscopic stress ${\Sigma}_{\rm{mean}}=1/3\,\mathrm{tr}\,\bar{\boldsymbol{\sigma}}$ vs. the mean component of the macroscopic strain ${E}_{\rm{mean}}=1/3\,\mathrm{tr}\,\bar{\boldsymbol{\varepsilon}}$. In both cases, due to cubic symmetry of the particle or void distribution, the cluster model is insensitive to the unit cell's geometry and provides the same outcome as the Mori-Tanaka model (with the same linearization and uniformization procedure) applied to the metal-matrix constitutive response. Additionally, formulations based on the tangent or affine linearizations employing only the first moment of stress, predict a purely elastic response ($\bar{\sigma}_{\eq}=0$ in the metal phase) and thus their predictions coincide in this strain scenario. On the other hand, for the modified tangent linearization one finds $\bar{\bar{\sigma}}_{\eq}\neq 0$ in the metal matrix, so the plastic flow is initiated at some stage of the process. As seen in Fig.~\ref{Fig:2_2_MeanStrain_SecTan_MicrMod_RC}, while the resulting overall response of both classes of models is not much different for metal matrix composites (a), a huge difference is found for porous materials (b). As will be demonstrated below, the predictions of the modified tangent variant of the proposed mean-field approach are in a much better agreement with full-field simulations.

\begin{figure}[H]
	\centering
	\begin{tabular}{cc}
		(a)&(b)\\
		\includegraphics[angle=0,width=0.45\textwidth]{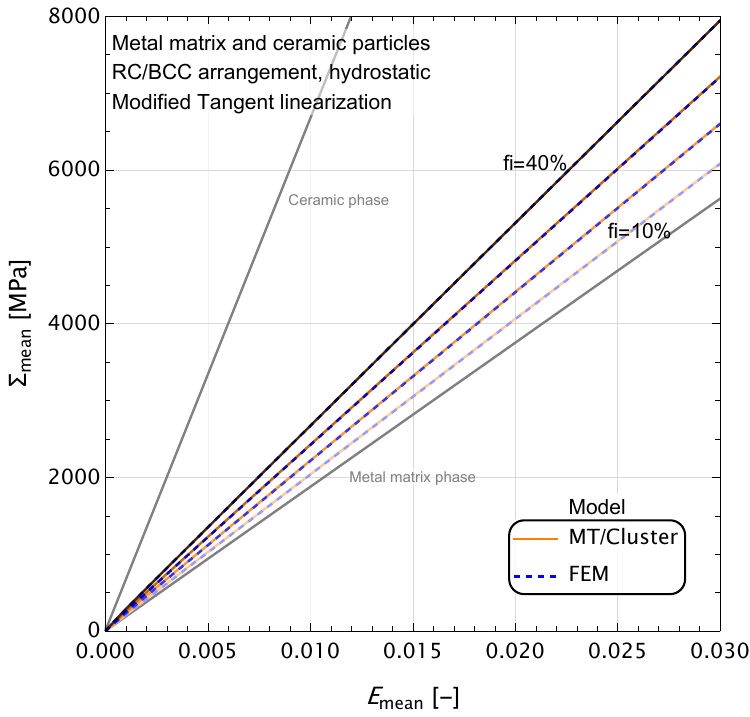}&
		\includegraphics[angle=0,width=0.45\textwidth]{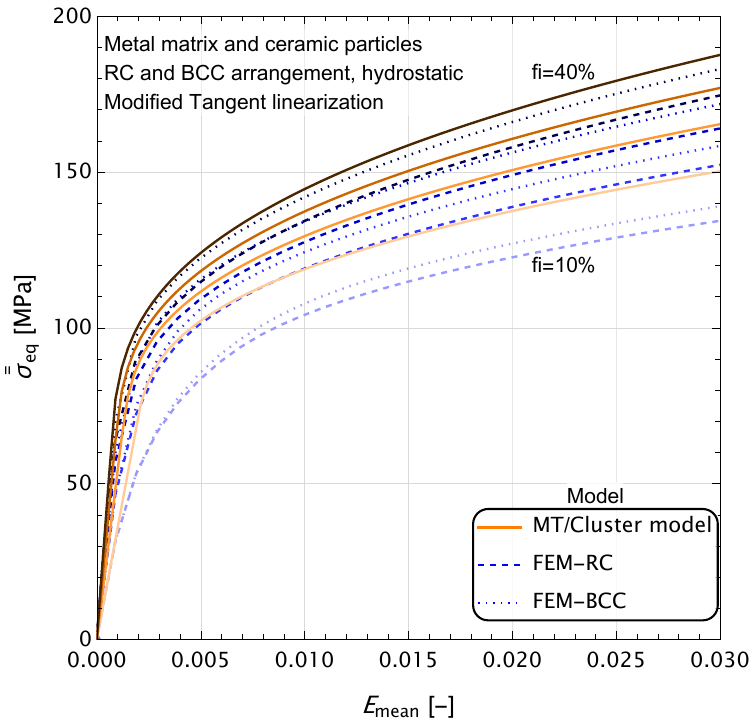}\\
		(c)&(d)\\
		\includegraphics[angle=0,width=0.45\textwidth]{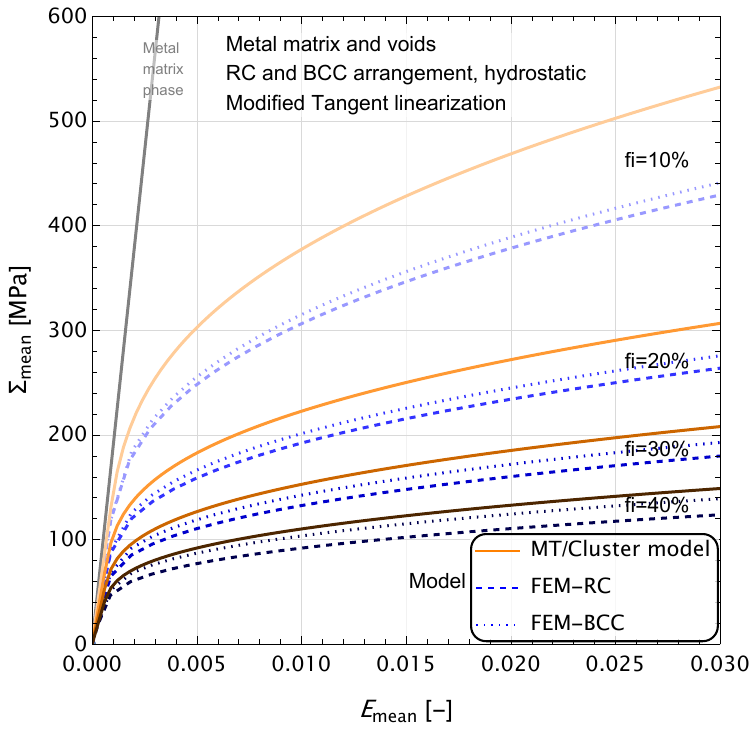}&
		\includegraphics[angle=0,width=0.45\textwidth]{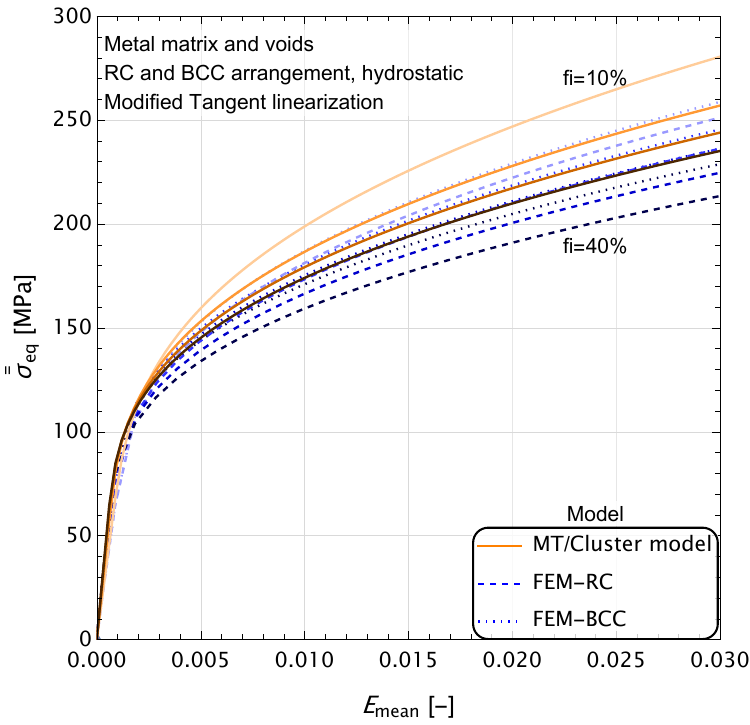}
	\end{tabular}
	\caption{
		The cluster model and numerical estimations (FEM) of the elastic-plastic response of (a,b) the metal matrix reinforced with ceramic inclusions, (c,d) the porous metal (Table~\ref{Tab:MaterialParameters}). 
	Macroscopic mean stress $\Sigma_{\textup{mean}}$ vs. macroscopic mean strain $E_{\textup{mean}}$ (a,c) and equivalent Huber-Mises stress $\bar{\bar{\sigma}}_{\textup{eq}}$ in the metal matrix vs. macroscopic mean strain $E_{\textup{mean}}$ (b,d), obtained for
	the macroscopic hydrostatic extension test.
	Modified tangent linearization of the cluster model was used. $f_{\rm{i}}$ equals $10\%,20\%,30\%,40\%$ and inhomogeneity/void arrangements are of the RC or BCC type (note that for the cluster model the same response in both cases is obtained and it is equal to the respective MT model).
		\label{Fig:4_1_Mean003_vs_strain-withFE}}
\end{figure}

In Fig.~\ref{Fig:4_1_Mean003_vs_strain-withFE}, the estimates of the proposed mean field approach for hydrostatic extension of MMC and PM are compared to the reference FEM results. As concerns the overal mean stress in MMC (Fig.~\ref{Fig:4_1_Mean003_vs_strain-withFE}(a)), the estimates are in perfect agreement with FEM calculations. It should be noted that in terms of these results there is no visible difference between the composite response for the RC and BCC particle arrangements also in FEM predictions. In the case of stress fluctuations described by the second moment of stress $\bar{\bar{\sigma}}_{\textup{eq}}$ in the matrix (Fig.~\ref{Fig:4_1_Mean003_vs_strain-withFE}(b)), the FEM values for BCC are slightly larger than those for RC. Moreover, the proposed mean-field model overpredicts the magnitude of $\bar{\bar{\sigma}}_{\textup{eq}}$, especially for RC, but the results are of the same order of magnitude. The largest difference is observed for the smallest volume fraction of reinforcements ($f_{\rm{i}}=10\%$). The macroscopic mean stress in the porous metal (Fig.~\ref{Fig:4_1_Mean003_vs_strain-withFE}(c)) is overpredicted by the modified tangent cluster approach. However, overall, the predictions are of reasonable quality compared to the classical tangent linearization. Plastic yielding of the composite is well predicted. Contrary to the mean-field approach, the FEM calculations exhibit some difference between the BCC and RC response. However, the difference is not substantial. Fig.~\ref{Fig:4_1_Mean003_vs_strain-withFE}(d) demonstrates the second moment of stress in the metal matrix. Also in this case, the cluster model overestimates the magnitude with respect to FEM outcomes. When comparing MMC with PM it is seen that, at the same level of the macroscopic mean strain, {PM exhibits larger values of $\bar{\bar{\sigma}}_{\textup{eq}}$}, so more plasticity develops in the porous material than in the metal matrix composite under hydrostatic loading.

\begin{figure}[H]
	\centering
	\begin{tabular}{cc}
		(a)&(b)\\
		\includegraphics[angle=0,width=0.45\textwidth]{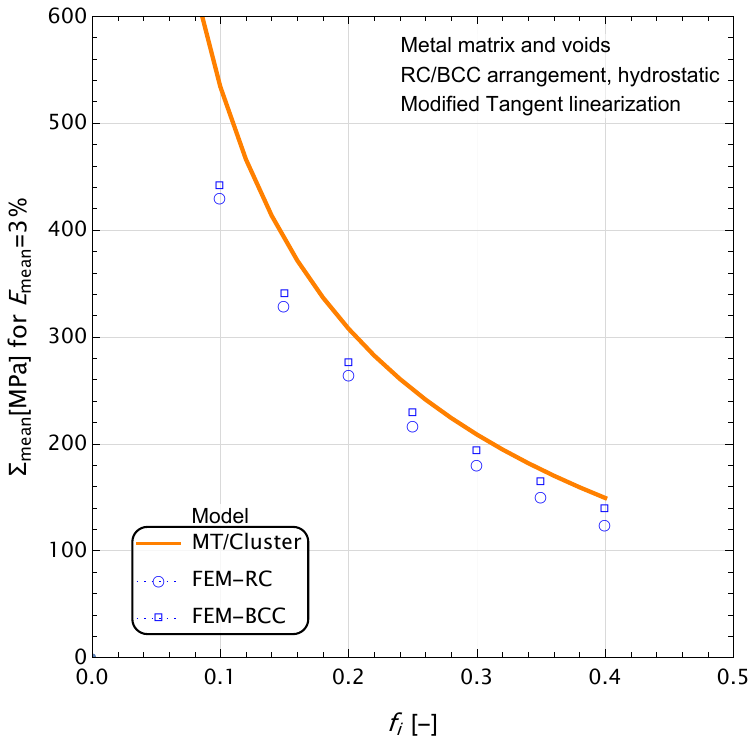}&
		\includegraphics[angle=0,width=0.45\textwidth]{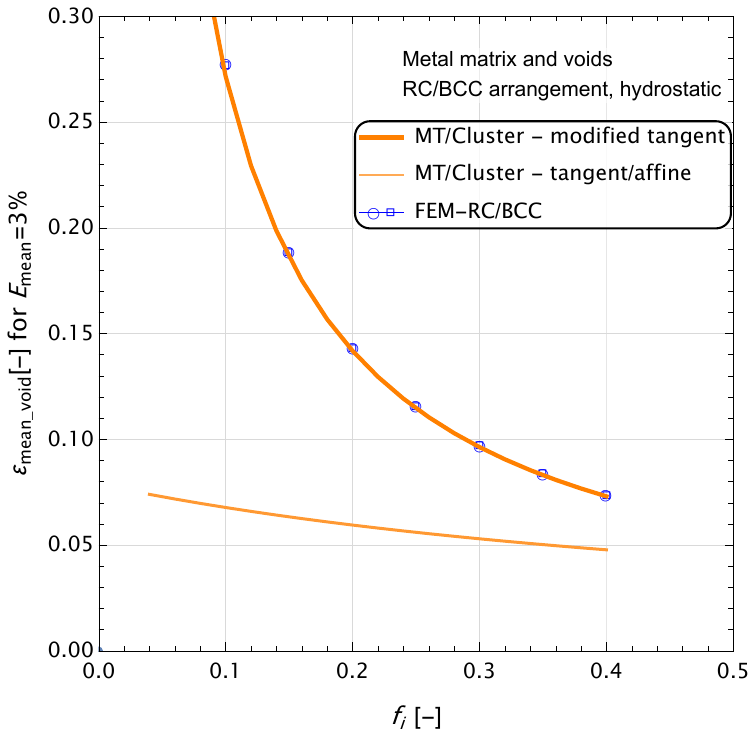}
	\end{tabular}
	\caption{
		Analytical and numerical estimations (FEM) of the elastic-plastic response of the metal matrix weakened by voids (Table~\ref{Tab:MaterialParameters}), with regular inclusion arrangements of types RC and BCC. 
		(a) Macroscopic mean stress $\Sigma_{\textup{mean}} \left( E_{\textup{mean}}=3\% \right)$ vs. volume fraction of voids $f_{\textup{i}}$. (b) Mean strain in the voids $\varepsilon_{\textup{mean\_}\rm{void}} \left( E_{\textup{mean}}=3\% \right)=1/3\,\mathrm{tr}\boldsymbol{\varepsilon}_{\rm{i}}$ vs. volume fraction of voids $f_{\textup{i}}$ for the macroscopic hydrostatic extension test.
		Tangent linearization of the micromechanical models was used.
		MT - the Mori-Tanaka model,
		Cluster - the cluster model.
		\label{Fig:4_1_Mean_vs_fi}}
\end{figure}

Fig.~\ref{Fig:4_1_Mean_vs_fi} presents a comparison of the modified cluster model and FEM results for the porous metal under hydrostatic extension in relation to void volume fraction. As it is seen in Fig.~\ref{Fig:4_1_Mean_vs_fi}(a), the value of the macroscopic mean stress at $3\%$ macrocopic hydrostatic strain is overpredicted. However, the difference between both approaches decreases with the increasing volume fraction. In Fig.~\ref{Fig:4_1_Mean_vs_fi}(b), the predicted volumetric strain of voids at the same macrocopic strain level is shown. As it can be seen, the predictions of the modified tangent cluster approach are in a very good agreement with FEM results, contrary to the classical tangent or affine formulations based solely on the first moments of stress, which would highly underestimate this value. This is an important advantage of the proposed approach as the volumetric strain is governing void growth which eventually leads to material ductile damage. It is worth observing that the difference between the modified tangent and standard tangent approaches \emph{increases} with \emph{decreasing} void volume fraction. Let us also note that similarly to this mean-field model, for this quantity no visible difference in predictions for the BCC and RC arrangements is seen in FEM simulations.  

Overall, the use of the modified version of tangent linearization {enables the proposed model to make satisfactory predictions for the case of hydrostatic loading}, contrary to other approaches of comparable computational efficiency available in the literature.

\section{Conclusions}
The goal of this paper was to propose a micromechanical mean-field model enabling reliable prediction of the \emph{anisotropic} overall response of composite or porous materials, 
both in the elastic and elastic-plastic regime. To this end, we adopted and further developed the cluster model (the mean-field interaction model), originally proposed in \citep{MolinariElMouden96} for elastic materials, which includes the effect of the morphological and spatial distribution of particles. For the cluster model to encompss the elastic-plastic regime as well, we have proposed a new modified tangent linearization method for the matrix response. In the method, the second moment of stress $\bar{\bar{\sigma}}_{\eq}$ is used to calculate the tangent elastic-plastic modulus of the metal matrix. The evolution of $\bar{\bar{\sigma}}_{\eq}$ is tracked by an additional relation stemming from the Hill-Mandel lemma. To preserve the computational efficiency of the framework, isotropization of the tangent elastic-plastic stiffness was performed. Moreover, with the same goal, at each time increment the second moment of stress was updated based on its first and second rates.

The cluster model allows one to perform calculations for different regular and irregular arrangements of inclusions, affording flexibility in adjusting inclusion sizes, their locations and the representative unit cell's proportions. In the present paper, we focused on regular arrangements of spherical particles or voids exhibiting cubic symmetry: regular cubic, body-centered cubic or face-centered cubic. We restricted ourselves to the so-called single-family situation, concentrating our attention on the overall material anisotropy induced by the spatial distribution of inhomogeneities (particles or voids). For the selected arrangement, we provided closed-form relations for the average interaction tensor $\bar{\boldsymbol{\Gamma}}$. In the framework of the cluster model, this tensor transfers the information on the particle distribution to the estimated overall properties.  

Throughout the paper, full-field FEM calculations were used as benchmarks for checking the soundness of the cluster model predictions.
These computations were performed on microstructural unit cells with applied periodic displacement boundary conditions. In the case of elastic properties, the cluster model estimates were also compared with other analytical solutions available in the literature. 

In the case of the elastic regime, effective bulk and shear moduli estimations for two material scenarios were considered: one with ultra-hard inclusions and one with ultra-soft inclusions.
The effective bulk modulus estimated by the mean-field interaction model is in agreement with computational homogenization in the whole range of volume fractions for the ultra-soft inclusions. The estimated value is the same as the one predicted by the standard Mori-Tanaka model. 
On the other hand, for the ultra-hard inclusions, there is good agreement between the two methods until the volume fraction approaches its limits specified by particle overlapping. For these large values of the volume fraction, the numerical results are stiffer than the cluster model predictions.
Because of the cubic symmetry of the arrangements, the anisotropic effective properties of the composites are described by two shear moduli. 
The strongest anisotropy, as defined by the Zener anisotropy ratio of the two moduli, is visible for the regular cubic (RC) arrangement.
The body-centered cubic (BCC) and face-centered cubic (FCC) arrangements behave in a very similar way, with a lower, but still visible, difference between the two shear moduli. Moreover, in the case of RC the Zener anisotropy ratio $\bar{G}_2/\bar{G}_1$ is smaller than one, while for BCC and FCC it is larger than one. It should be stressed that composite
anisotropy is not predicted by the classical Mori-Tanaka and self-consistent models whose predictions account only for the volume content and shape of inclusions; when inclusions are spherical, the predicted effective stiffness is isotropic.

{In the elastic-plastic regime, the considered heterogeneous materials also} consisted of two phases with a high contrast: the elastic-plastic metal binder and the filler (elastic ceramic particles for MMC and voids for PM, respectively).
{We investigated isochoric extension of these materials in two directions: along the unit cell's edges and along the cell's main diagonals, as well as their volumetric (hydrostatic) extension.}
Contrary to the MT scheme, the cluster model accounts for the anisotropic non-linear response for different directions of the isochoric tension test. 
The results of computational homogenization are in good agreement with the cluster model predictions for volume fractions up to around 40\%.
{We observed a strongly increasing role of the tension direction when the volume content of inhomogeneities rose above 5\%,} with the cluster model and the finite element method being in good agreement in estimating this phenomenon. Thus, a side conclusion of the paper is that validation of mean-field approaches dedicated to random composites by means of the analysis of a unit cell containing a single particle should be avoided, especially in the case of an increased volume content of particles. {The analysis of hydrostatic loading cases has shown that the use of the modified version of tangent linearization enables the model to provide satisfactory predictions for the case of hydrostatic loading as well, which sets the present approach apart from other approaches of similar complexity available in the literature.}  The problem of accounting for yielding initiation under volumetric extension has been overcome by the use of the second stress moment within the `uniformization' {step. This allowed the extent of plastic yielding to be satisfactorily estimated for both the MMC and PM cases.} In particular, very good agreement was obtained in the assessment of the volumetric strain of the void phase---a quantity crucial in the description of the void growth phenomenon. It should be remarked that similarly to the bulk modulus in the elastic regime, the response under hydrostatic loading predicted by the proposed mean-field model is insensitive to the particular {arrangement of inhomogeneities if only it is of cubic symmetry, whereas some differences between arrangements are detected by full-field numerical homogenization.} 

As mentioned before, effective properties of random composites can be predicted using the multi-family variant of the cluster model. It enables one to study differences between mean strains in subsequent inclusions in the representative volume stemming from the specificity of their locations with respect to other inclusions. This will be the subject of our future studies. {As the proposed modified tangent linearization imposes no restrictions on the deformation process, thus permitting a change of the deformation path, cyclic and non-radial processes are planned to be investigated as well, for both regular and random inhomogeneity arrangements, using the proposed framework. This will allow its farther validation.}

Overall, the ability of the proposed cluster model to capture topological effects opens interesting perspectives, e.g., regarding 3-D printed materials.
In particular, the model could be used as an efficient tool for adjusting the internal structure of composite materials with non-linear matrix properties in order to obtain desirable effective properties.

\section*{Acknowledgements}
The research was partially supported by the {project} 2021/41/B/ST8/03345 of the National Science Centre, Poland.

\appendix
\section{Algorithm for the cluster model for the single-family case \label{SubSec:AppStepByStep}}
The present appendix contains a step-by-step algorithm for the cluster model as applied to a two-phase composite with isotropic phases, from the beginning (setting input parameters) to the end (obtaining effective values for the composite). It allows the reader to implement it and try it using examples and parameters of their choice. The procedure consists of the following steps:

0. ``Step zero'' is to create input data for the model. These data consist of material parameters of all phases, as well as the geometry of the sample:
\begin{itemize} 
\item elastic constants for both phases (the bulk and shear moduli and Poisson's ratio: $K_{r},G_{r},\nu_{r}$, $r=\ii,\mm$), 
\item coordinates $(x,y,z)$ of the centers {and radii} of all inclusions within the unit cell; the number of inclusions will be denoted by $M$ and we are assuming here that they are all symmetrically equivalent {-- in particular, their radii are equal},
\item radius $R_{\rm{c}}$ of the cluster {(we assume for simplicity that $R_{\rm{c}}$ is a multiple of the unit cell size $d$)}. 
\end{itemize} 

\begin{enumerate}	
\item Find the centermost inclusion $I$ in the unit cell and its coordinates $(x^I,y^I,z^I)$. It will serve as the base inclusion of reference.
\item {Construct a cluster of radius $R_{\rm{c}}$ by repeating the unit cell periodically in three dimensions until {the assembly of cells} accommodates a sphere of radius $R_{\rm{c}}$ centered at the base inclusion. {Then select from the assembly those inclusions which are within distance $R_{\rm{c}}$ (measured center to center) from the base inclusion.} The number of inclusions in the cluster is denoted by $N$.}  	
\begin{figure}[ht]
	\centering
	\includegraphics[width=0.4\textwidth]{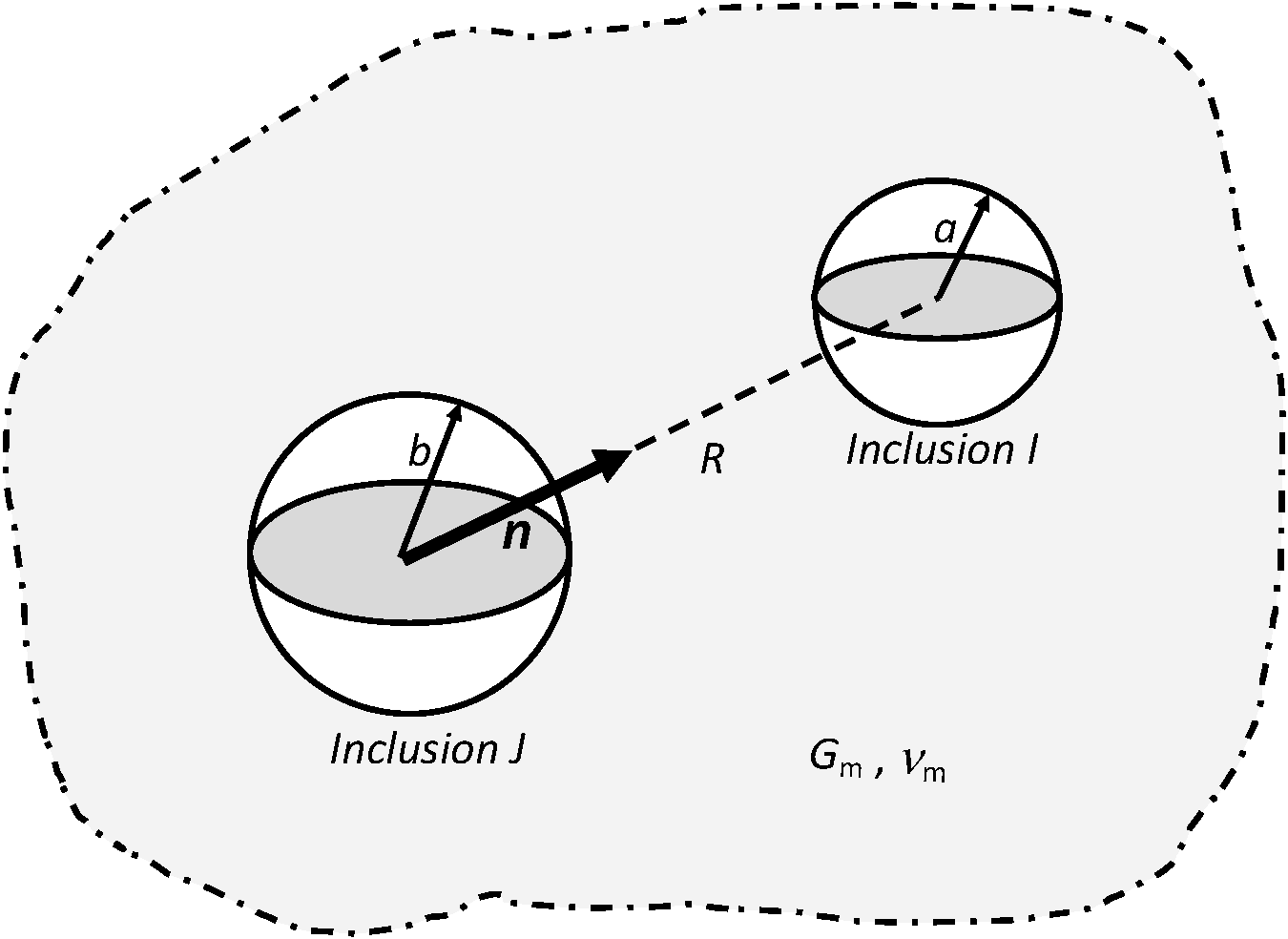}
	\caption{Interaction tensor $\boldsymbol{\Gamma}^{IJ}$ is calculated for each pair $I$-$J$ separately.}
	\label{Fig:cluster_interaction}
\end{figure}
\item Calculate the interaction tensor $\boldsymbol{\Gamma^{IJ}}$ for each pair of inclusions: the base inclusion $I$ and any other inclusion $J$ (Fig.~\ref{Fig:cluster_interaction}). There should be $N-1$ pairs and thus $N-1$ tensors $\boldsymbol{\Gamma^{IJ}}$. The components of the interaction tensor are calculated as follows:
\begin{equation}\label{Eq:GIJ1111}
\Gamma^{IJ}_{1111}=\Gamma^{IJ}_{2222}
=
\frac{b^{3}}{12 R^{3}}\frac{-1} {G_{\mm}\left(1-\nu_{\mm}\right)}
\left(1-4\nu_{\mm}+\frac{9}{5}\rho^2\right)\,,
\end{equation}

\begin{equation}
\Gamma^{IJ}_{1122}=\Gamma^{IJ}_{2211}
=
\frac{b^{3}}{12 R^{3}}\frac{-1} {G_{\mm}\left(1-\nu_{\mm}\right)}
\left(-1+\frac{3}{5}\rho^2\right)\,,
\end{equation}

\begin{equation}
\Gamma^{IJ}_{1133}=\Gamma^{IJ}_{2233}=\Gamma^{IJ}_{3311}=\Gamma^{IJ}_{3322}
=
\frac{b^{3}}{12 R^{3}}\frac{-1} {G_{\mm}\left(1-\nu_{\mm}\right)}
\left(2-\frac{12}{5}\rho^2\right)\,,
\end{equation}

\begin{equation}
\Gamma^{IJ}_{1212}=\Gamma^{IJ}_{1221}=\Gamma^{IJ}_{2121}=\Gamma^{IJ}_{2112}
=
\frac{b^{3}}{12 R^{3}}\frac{-1} {G_{\mm}\left(1-\nu_{\mm}\right)}
\left(1-2\nu_{\mm}+\frac{3}{5}\rho^2\right)\,,
\end{equation}

\begin{equation}
\Gamma^{IJ}_{1313}=\Gamma^{IJ}_{1331}=\Gamma^{IJ}_{3113}=\Gamma^{IJ}_{3131}
=
\frac{b^{3}}{12 R^{3}}\frac{-1} {G_{\mm}\left(1-\nu_{\mm}\right)}
\left(1+\nu_{\mm}-\frac{12}{5}\rho^2\right)\,,
\end{equation}

\begin{equation}
\Gamma^{IJ}_{2323}=\Gamma^{IJ}_{2332}=\Gamma^{IJ}_{3223}=\Gamma^{IJ}_{3232}
=
\frac{b^{3}}{12 R^{3}}\frac{-1} {G_{\mm}\left(1-\nu_{\mm}\right)}
\left(1+\nu_{\mm}-\frac{12}{5}\rho^2\right)\,,
\end{equation}

\begin{equation}\label{Eq:GIJ3333}
\Gamma^{IJ}_{3333}
=
\frac{b^{3}}{12 R^{3}}\frac{-1} {G_{\mm}\left(1-\nu_{\mm}\right)}
\left(-8+8\nu_{\mm}+\frac{24}{5}\rho^2\right)\,,
\end{equation}

where 

\begin{equation}\label{Eq:GIJrho2}
\rho^2=\left(a^2+b^2\right)/R^2
\end{equation}
and $G_{\mm}$ is the shear modulus of the matrix, $\nu_{\mm}$ is the Poisson's ratio of the matrix, $V_J$ is the volume of inclusion $J$, $R$ is the distance between the center of the base inclusion $I$ and that of inclusion $J$, $a$ is the radius of the base inclusion $I$ and $b$ is the radius of inclusion $J$. The components are expressed in the orthonormal basis whose direction 3 is oriented along the line connecting the centers of inclusions $I$ and $J$. {It should be noted that in Fig.~\ref{Fig:cluster_interaction} and Eqs.~(\ref{Eq:GIJ1111}--\ref{Eq:GIJrho2}) the general case is presented of inclusions $I$ and $J$ having possibly different radii $a$ and $b$, respectively. In the considered one-family case of all symmetrically-equivalent inclusions, we simply set $a=b=r$.}

\item Sum up all $N-1$ interaction tensors $\boldsymbol{\Gamma}^{IJ}$. This gives the global tensor of interaction $\bar{\boldsymbol{\Gamma}}$:
\begin{equation}
\bar{\boldsymbol{\Gamma}}
=
\sum_{J\neq{I}} \boldsymbol{\Gamma}^{IJ}\,.
\end{equation}
Note that before summation the components (\ref{Eq:GIJ1111}-\ref{Eq:GIJ3333}) of subsequent $\boldsymbol{\Gamma}^{IJ}$ tensors must be transformed into the common reference frame. Otherwise, the basis-free formula (\ref{Eq:GIJ-Trans}) can be used.

\item Calculate the polarization tensor $\mathbb{P}_0$:
\begin{equation}
\mathbb{P}_0=h^{\rm{P}}_{\rm{polar}}\mathbb{I}^{\rm{P}}+h^{\rm{D}}_{\rm{polar}}\mathbb{I}^{\rm{D}}\,,
\end{equation}
where:
\begin{equation}
h^{\rm{P}}_{\rm{polar}}=\frac{1}{4G_{\rm{m}}+3K_{\rm{m}}}\,,\quad h^{\rm{D}}_{\rm{polar}}=\frac{3}{5G_{\rm{m}}}\frac{K_{\rm{m}}+2G_{\rm{m}}}
{3K_{\rm{m}}+4G_{\rm{m}}}
\end{equation}
and
\begin{equation} \label{eq:IPD}
\mathbb{I}^{\rm{P}}=\frac{1}{3}\mathbf{I}\otimes\mathbf{I}\quad\left(I^{\rm{P}}_{ijkl}=\frac{1}{3}\delta_{ij}\delta_{kl}\right)\,,\quad \mathbb{I}^{\rm{D}}=\mathbb{I}-\mathbb{I}^{\rm{P}}\quad\left(I_{ijkl}=\frac{1}{2}(\delta_{ik}\delta_{jl}+\delta_{il}\delta_{jk})\right)\,,
\end{equation}
with $\mathbb{I}$ being the fourth-order identity tensor.

\item Compose the stiffness tensors for the matrix and inclusions:
\begin{equation}
\mathbb{C}_{\ii/\mm}=3K_{\ii/\mm}\mathbb{I}^{\rm{P}}+2G_{\ii/\mm}\mathbb{I}^{\rm{D}}\,.
\end{equation}

\item Calculate localization tensors for the matrix and inclusions using Eq.~(\ref{Eq:Ai}) and (\ref{Eq:Am}).

\item Calculate the effective stiffness tensor for the entire sample using Eq.~(\ref{Eq:C}) or (\ref{Eq:C1F}).
The effective stiffness tensor is one of the output values -- an effective parameter of the composite that was the subject of mathematical homogenization. 

\item If the external strain (or stress) applied to the sample was a part of the problem, it is possible to calculate the effective stress (or strain) from the overall relation:
\begin{equation}
\bar{\boldsymbol{\sigma}}=
\bar{\mathbb{C}} \cdot \bar{\boldsymbol{\varepsilon}}\,.
\end{equation}

\item Stresses and strains for each phase $r$  are calculated as follows:
\begin{equation}
\boldsymbol{\varepsilon}_r=\mathbb{A}_r\cdot\bar{\boldsymbol{\varepsilon}}\,,\quad
\boldsymbol{\sigma}_r=\mathbb{B}_r\cdot\bar{\boldsymbol{\sigma}}\,,
\end{equation}
where $\mathbb{B}_r=\mathbb{C}_r\mathbb{A}_r\bar{\mathbb{C}}^{-1}$ $(r=\ii,\mm)$ are stress localization tensors.
\end{enumerate}

Note that the above algorithm is written for the elastic analysis. In the case of the elastic-plastic analysis the algorithm needs to be executed at each time step and, in the case of tangent linearization, formulas in steps 9 and 10 concern stress and strain rates. It is clear that in the case of the elastic-plastic analysis the availability of closed-form relations (\ref{Eq:GammaClosed}--\ref{Eq:XFCC}) speeds up calculations even more substantially since steps 1--3 are then not necessary and in step 4 the tensor $\bar{\boldsymbol{\Gamma}}$ is calculated directly from Eqs.~(\ref{Eq:GammaClosed}--\ref{Eq:XFCC}). 

\section{Properties of the interaction tensor $\mathbf{\Gamma}^{IJ}$\label{Sec:PropGammaIJ}}

The interaction tensor $\boldsymbol{\Gamma}^{IJ}$ with components given by Eqs.~(\ref{Eq:GIJ1111}--\ref{Eq:GIJ3333}) is a tensor of transverse isotropy for which the isotropic part is equal to zero (i.e. $\Gamma^{IJ}_{iijj}=0$ and $\Gamma^{IJ}_{ijij}=0$). Denoting by $\mathbf{n}$ the unit vector along the line connecting the centers of inclusions $J$ and $I$ (see Fig.~\ref{Fig:cluster_interaction}), the tensor can be written as follows in a basis-free format (compare \cite{Kowalczyk20}): 
\begin{equation}\label{Eq:GIJ-Trans}
\boldsymbol{\Gamma}^{IJ}\left(\mathbf{n},\frac{R}{b},\frac{a}{b}\right)=\frac{1}{3G_{\rm m}(1-\nu_{\rm m})(R/b)^3}\left(\Gamma^{IJ}_{1}\mathbb{P}_{12}(\mathbf{n})+2(\Gamma^{IJ}_3+\Gamma^{IJ}_4)\mathbb{P}_2(\mathbf{n})-\Gamma^{IJ}_3\mathbb{P}_3(\mathbf{n})-\Gamma_4^{IJ}\mathbb{P}_4(\mathbf{n})\right)\,,
\end{equation}
where
\begin{eqnarray}
\label{Eq:P1P2}
\mathbb{P}_{12}(\mathbf{n})&=&\frac{\sqrt{2}}{6}\left((3\mathbf{N}-\mathbf{I})\otimes\mathbf{I}+\mathbf{I}\otimes(3\mathbf{N}-\mathbf{I})\right)\,,\\
\mathbb{P}_2(\mathbf{n})&=&\frac{1}{6}(3\mathbf{N}-\mathbf{I})\otimes(3\mathbf{N}-\mathbf{I})\,,\\
\label{Eq:P3}
\mathbb{P}_3(\mathbf{n})&=&\frac{1}{2}\left(\left[(\mathbf{I}-\mathbf{N})\otimes(\mathbf{I}-\mathbf{N})\right]^{T(23)+T(24)}\!-\!
(\mathbf{I}-\mathbf{N})\otimes(\mathbf{I}-\mathbf{N})\right)\,,\\ \label{Eq:P4}
\mathbb{P}_4(\mathbf{n})&=&\frac{1}{2}
\left[\mathbf{N}\otimes(\mathbf{I}-\mathbf{N})+(\mathbf{I}-\mathbf{N})\otimes\mathbf{N}\right]^{T(23)+T(24)}\,,
\end{eqnarray}
with $(\mathbb{A}^{T(23)+T(24)})_{ijkl}\equiv(\mathbb{A})_{ikjl}\!+\!(\mathbb{A})_{ilkj}$ and $\mathbf{N}=\mathbf{n}\otimes\mathbf{n}$. The three scalar coefficients {$\Gamma_k^{IJ}=\Gamma_k^{IJ}(R/b,a/b)$}, $k=1,3,4$, 
are specified by the following formulas:
\begin{equation}\label{Eq:GammaCoeff}
\Gamma_1^{IJ}=\frac{1}{\sqrt{2}}(1-2\nu_{\rm m})\,,\quad
\Gamma_3^{IJ}=\frac{1}{10}(5-10\nu_{\rm m}+3\rho^2)\,,\quad
\Gamma_4^{IJ}=\frac{1}{10}(5+5\nu_{\rm m}-12\rho^2)\,.
\end{equation}
As can be seen when $R/b\rightarrow\infty$, that is when the distance between the inclusions is growing, {the fraction in Eq.~\eqref{Eq:GIJ-Trans} tends to zero}, and so does the whole tensor $\boldsymbol{\Gamma}^{IJ}$. This {motivates} the approximation introduced by the cluster model with a finite cluster radius.

It is worth noting that, since the decomposition of the fourth-order tensor into the isotropic and anisotropic parts is linear \citep{Forte96,Rychlewski00}, the isotropic part of the tensor $\bar{\boldsymbol{\Gamma}}$ is zero as well  (i.e. $\bar{\Gamma}_{iijj}=0$ and $\bar{\Gamma}_{ijij}=0$). A direct consequence of this fact is that the overall bulk modulus does not depend on the spatial distribution of inclusions in the case of arrangements of cubic symmetry such as RC, BCC or FCC and coincides with the estimate of the Mori-Tanaka method.

\bibliography{references}
\end{document}